\documentclass[12pt,psf,epsf]{article}
\usepackage[centertags]{amsmath}
\usepackage{amssymb}
\usepackage{graphicx}
\usepackage{epsfig}
\usepackage{ulem}
\usepackage[english]{babel}
\usepackage{array}
\usepackage{amsthm}
\usepackage{latexsym}
\usepackage[mathcal]{euscript}
\pdfoutput=1
\usepackage{epsfig}
 \usepackage{jheppub}
\usepackage{mathdots}
\usepackage{MnSymbol}
\usepackage{multirow}

\newcommand{\nn}{\nonumber}
\newcommand{\be}{\begin{equation}}
\newcommand{\ee}{\end{equation}}
\newcommand{\bea}{\begin{eqnarray}}
\newcommand{\eea}{\end{eqnarray}}

\newcommand{\e}{\mathrm{e}}
\newcommand{\tr}{\text{tr}}
\newcommand{\Tr}{\text{Tr}}
\newcommand{\ZZ}{\mathbb{Z}}

\newcommand{\RR}{\mathbb{R}}
\newcommand{\dd}{\partial}
\newcommand{\sgn}{\text{sgn}}
\newcommand{\Det}{\text{Det}}

\newcommand{\mI}{\mathcal{I}}
\newcommand{\mG}{\mathcal{G}}

\newcommand{\mO}{\mathcal{O}}
\newcommand{\mL}{\mathcal{L}}

\newcommand{\bj}{{\bf j}}

\newcommand{\tS}{\tilde{\Sigma}}

\def\Re{\text{Re}}
\def\Im{\text{Im}}
\def\Pf{\text{Pf}}
\def\diff{\text{diff}}
\def\const{\text{const}}

\newcommand{\mC}{\mathcal{C}}

\newcommand{\fk}{\mathfrak{k}}

\newcommand{\fl}{\mathfrak{l}}

\newcommand{\fm}{\mathfrak{m}}

\newcommand{\fs}{\mathfrak{s}}

\newcommand{\fT}{\mathfrak{T}}

\newcommand{\fJ}{\mathfrak{J}}

\newcommand{\fG}{\mathfrak{G}}
\newcommand{\fK}{\mathfrak{K}}

\newcommand{\fD}{\mathfrak{D}}

\newtheorem*{Lemma}{Lemma}

\definecolor{dgreen}{rgb}{0.,0.6,0.}

\title{Replica-nondiagonal solutions in the SYK model }
\author{Irina Aref'eva,}
\author{Mikhail Khramtsov,}
\author{Maria Tikhanovskaya and}
\author{Igor Volovich}

\affiliation{Steklov Mathematical Institute, Russian Academy of Sciences,\\Gubkina str. 8, 119991, Moscow, Russia}

\abstract{We study the SYK model in the large $N$ limit beyond the replica-diagonal approximation. First we show that there are exact replica-nondiagonal solutions of the saddle point equations for $q=2$ for any finite replica number $M$. In the interacting $q=4$ case we are able to construct the numerical solutions, which are in one-to-one correspondence to the analytic solutions of the quadratic model. These solutions are singular in the $M \to 0$ limit in both quadratic and quartic interaction cases. The calculations of the on-shell action at finite integer $M$ show that the nondiagonal replica-symmetric saddles are subleading in both quadratic and quartic cases. 

We also study replica-nondiagonal solutions of the SYK in the strong coupling limit. For arbitrary $q$ we show that besides the usual solutions of the replica-diagonal saddle point equations in the conformal limit, there are also replica-nondiagonal  solutions for any value of $M$ (including zero). The specific configurations that we study, have factorized time and replica dependencies. The corresponding saddle point equations are separable at strong coupling, and can be solved using the Parisi ansatz from spin glass theory. We construct the solutions which correspond to the replica-symmetric case and to one-step replica symmetry breaking. We compute the regularizized free energy on these solutions in the limit of zero replicas. It is observed that there are nondiagonal solutions  with the regularized free energy lower than that of the standard diagonal conformal solution. }

\emailAdd{arefeva@mi-ras.ru}
\emailAdd{khramtsov@mi-ras.ru}
\emailAdd{tikhanovskaya@mi-ras.ru}
\emailAdd{volovich@mi-ras.ru}

\begin{document}
\maketitle

\section{Introduction}

The Sachdev-Ye-Kitaev model \cite{Sachdev92,Kitaev,MScomments,Kitaev17} is a quantum mechanical model of $N$ Majorana fermions with disordered interactions that is solvable at large $N$ in the strong coupling limit. It was proposed \cite{Kitaev} as a solvable toy model for holographic description of quantum gravity in the AdS$_2$ spacetime (see \cite{Sarosi17} for a review). This idea is justified by the fact that the SYK model displays emergent approximate conformal symmetry in the strong coupling regime \cite{Kitaev,MScomments,Sachdev15} and that it exhibits maximal quantum chaos \cite{Kitaev,MScomments,Polchinski16} at strong coupling. The Goldstone mode corresponding to the conformal symmetry is connected to the gravitational mode in the effective description of the Jackiw-Teitelboim gravity in the AdS$_2$ bulk \cite{MScomments,Maldacena16,Kitaev17,Jevicki16,Jensen16,Engelsoy16}. Its dynamics in the leading order in inverse coupling turned out to be completely solvable \cite{Engelsoy16,Bagrets16,Stanford17,Mertens17}, and in the leading order in $1/N$ all correlation functions of operators dual to the matter fields in the bulk were computed as well \cite{Gross17}. While the precise bulk dual theory is still unknown, the SYK model has already allowed to obtain significant insight in the physics of black holes and wormholes \cite{Cotler16,Maldacena18}. 

The defining characteristic of the SYK model is the quenched disorder which randomizes the couplings between sites. Under the assumption that the system is self-averaging, one can perform averaging over the disorder by introducing replicas, and obtain the path integral in terms of auxiliary fields $G_{\alpha\beta} (\tau_1, \tau_2)$ and $\Sigma_{\alpha\beta} (\tau_1, \tau_2)$, where $\alpha, \beta$ are the replica indices \cite{Kitaev,Kitaev17,Bagrets16}. The results for SYK regarding the large $N$ solution and thermodynamics, are obtained under the assumption that the auxiliary fields are diagonal in replicas. This assumption is justified by the exact diagonalization numerics \cite{MScomments,Bagrets16,Garcia-garcia16,Cotler16}, and by a physical qualitative argument\footnote{We discuss this argument and its applicability in detail in Appendix \ref{sec:Correlators}.} \cite{Georges00,Fu16,Polchinski16} that prohibits realization of replica-nondiagonal behavior. Besides that, the work which studied spin glass phases (which are usually realized as particular replica-nondiagonal saddle points of the free energy path integral) \cite{Sachdev92,Georges00,Fu16,Ye18,Gur-Ari18,Caracciolo18} found no numerical evidence of glassy behavior in the fermionic SYK model and provided several analytic arguments of why there should not be spin glass physics. However, all these considerations do not conclusively exclude the existence of \textit{general} replica-nondiagonal saddle points of the path integral at either finite or zero replicas (in the case of the free energy). Meanwhile, recent work \cite{Cotler16,Saad18,Harlow18} hints that replica-nondiagonal saddle points in annealed quantities, such as the spectral form factor, are responsible for manifestations of quantum chaotic behavior in black holes and possible recovery of information from the black hole, which is lost at the semiclassical level, via holographic duality. 

Motivated by these points, in the present work we study the replica-nondiagonal large $N$ saddle points of of the SYK model at general replica number. We start off with the $q=2$ variant of the model (where $q$ is the degree of the interacting Hamiltonian). We obtain a family of analytic replica-nondiagonal solutions of the saddle point equations of the disorder-averaged partition function for $M$ replicas of the SYK chain, where $M$ can be understood as an arbitrary real non-negative number. These solutions have an important property of being singular at $M \to 0$. This means that these saddles do not contribute to the free energy. Computing the on-shell action on these solutions at finite replica number, we show that for analytically continued $0 < M < 1$ there is a nontrivial phase structure of the path integral, however for $M > 1$ the standard diagonal solution always dominates. As a next step, we use these analytic solutions in the $q=2$ model to construct nondiagonal numerical solutions to the exact saddle point equations in the interacting $q=4$ variant of the model. These numerical solutions also exhibit the singular behavior at $M \to 0$ and $J=0$. In the $q=4$ case the replica-nondiagonal saddles are also subleading in the replica partition function at $M >1$. 

In the second part of the work, we focus on the strong coupling, or IR limit of the SYK model. In this limit one can find analytic replica-nondaigonal solutions in the interacting model either at finite replica number or in the limit of zero replicas. We study the class of solutions, for which the time dependence and replica dependence are factorized. In this case the saddle point equations separate in the strong coupling limit. The solution is then constructed by solving the equation for the temporal part in the same way as in the replica-diagonal case, and by solving an algebraic equation for the replica part. We perform the latter by using the Parisi ansatz \cite{Mezard91,ParisiBook}. The algebraic equation for the replica dependence in the zero replica limit transforms into an integral equation. To solve it, we restrict ourselves to the step-function ansatz, which corresponds to the one-step replica symmetry breaking. We study the solutions at $M =0$ and compute the regularized free energy on the corresponding saddle points.  

The paper is organized as follows. In the section \ref{sec:Setup} we briefly review the main features of the replica-diagonal saddle-point of SYK, and discuss the non-perturbativity of the exact replica-nondiagonal saddles. We start the study of the replica-nondiagonal solutions in the section \ref{sec:q=2}, which is devoted to the quadratic variation of the model. The exact nondiagonal solutions and their properties are discussed. The next section \ref{sec:Numerics} contains a description and the results of the numerical study of exact saddles in the interacting $q=4$ model, and also a general remark on the large replica number limit. In the section \ref{sec:Factorized} we switch gears to the study of replica-nondiagonal solutions in the strong coupling limit. Assuming the factorized ansatz, we derive the reduced saddle point equations and explain in detail the general strategy for constructing the solutions and computing the regularized free energy in the zero replicas limit. Subsequently, in the section \ref{sec:RSB1} we construct the solutions in the one-step replica symmetry breaking ansatz and compute the leading contribution to the regularized free energy in the strong coupling limit. In the next section \ref{sec:Implications} we make some comments about generating other solutions using the reparametrization symmetry, possible holographic interpretation in particular cases and about solutions beyond the strong coupling limit. We discuss our results and unanswered questions in the section \ref{sec:Discus}. The appendix \ref{sec:Parisi} provides a brief introduction to Parisi matrices and derivations of a few formulas used in the main text, and the appendix \ref{sec:appB} contains a few general formulae regarding the computation of on-shell action in different cases. In the appendix \ref{sec:Correlators} we present some general considerations of the disordered correlation functions and other observables and their relation to our results.

\section{Setup}
\label{sec:Setup}

The object of our study is the Sachdev-Ye-Kitaev model \cite{Kitaev,MScomments,Kitaev17}, which is a theory of $N \gg 1$ interacting Majorana fermions in $0+1$ dimensions\footnote{In this paper we work in the Euclidean time, unless mentioned otherwise.}. The Hamiltonian is given by 
\be
H = \frac{i^{q/2}}{q!} \sum_{i_1, i_2,\dots,i_q=1}^N j_{i_1 i_2\dots i_q} \psi_{i_1} \psi_{i_2} \dots \psi_{i_q}\,. \label{Hsyk} 
\ee
Here $\psi_i$ are the Majorana fermions, and $ j_{i_1\dots i_q}$ are totally antisymmetric couplings randomized via the Gaussian distribution: 
\be
P(j_{i_1\dots i_q}) = \sqrt{\frac{N^{q-1}}{2 (q-1)!\pi J^2}}
\ \e^{-\frac{N^{q-1} j_{i_1 \dots i_q}^2}{2 (q-1)! J^2}}\,. \label{Gaussian}
\ee
To calculate a physical quantity in this model, one has to average over the disorder using the rules which follow from the distribution (\ref{Gaussian}): 
\be
\overline{j_{i_1 \dots i_q}} = 0\,, \qquad \overline{j_{i_1 \dots i_q}j_{i_1 \dots i_q}} =  \frac{(q-1)! J^2}{N^{q-1}} \quad \text{(no sum)}\,. \label{disorder2point}
\ee
To study physically meaningful quantities, one usually has to average over all realizations of the disorder. The free energy of the model with quenched disorder is given by
\be
F = -\frac{1}{\beta} \overline{\log Z}\,, \label{F}
\ee 
where $Z = \Tr\ \e^{-\beta H}$. 
To simplify evaluating the disorder average, one employs the replica trick, which we write in the form \cite{SpinGlassBook}: 
\be\label{ReplicaTrick}
\overline{\ln Z} = \lim_{M \to 0} \frac{\ln \overline{Z^M}}{M}\,.
\ee
In the case of integer $M$ on the right hand side there is a path integral over $M$ copies of SYK, so the fields now carry an additional index $\alpha = 1, \dots, M$. We study the case of integer values of $M$ separately. The limit $M \to 0$ requires an analytic continuation, which will be considered on a particular ansatz. 

One can calculate the disorder average for the replica partition function and rewrite it in terms of the path integral over $O(N)$-invariant auxiliary fields with replica indices. The derivation was presented in detail in \cite{Bagrets16,Kitaev17}. After this procedure, one obtains the following  expression for the replica partition function: 
\bea
&& \overline{Z(\beta)^M}= \int DG D\Sigma\ \Pf[\delta_{\alpha\beta} \dd_\tau - \hat{\Sigma}_{\alpha\beta}]^{N}\times \nn\\&& \exp\left[-\frac{N}{2}\int_0^\beta \int_0^\beta d\tau_1 d \tau_2  \left(\Sigma_{\alpha\beta}(\tau_1, \tau_2) G_{\alpha\beta}(\tau_1, \tau_2) - \frac{J^2}{q} G_{\alpha\beta}(\tau_1, \tau_2)^q\right)\right]\,, \label{Zreplica}
\eea
where $G$ is the bilocal field which has the meaning of the Majorana fermion two-point function, and $\Sigma$ is the auxiliary bilocal field\footnote{Here and henceforth the hat denotes the integral operator corresponding to the kernel given by the bilocal field. } which has the meaning of the fermion self-energy. These bilocal fields are supposed to satisfy the antisymmetry condition
\be
G_{\alpha\beta}(\tau_1, \tau_2) = - G_{\beta\alpha}(\tau_2, \tau_1)\,; \qquad \Sigma_{\alpha\beta}(\tau_1, \tau_2) = - \Sigma_{\beta\alpha}(\tau_2, \tau_1)\,.
\ee
In the present work we essentially study the saddle points of (\ref{Zreplica}) for different values of $M$. The saddle points of the path integral are defined by the following equations: 
\bea
 \dd_\tau G_{\alpha\gamma}(\tau, \tau'')-\int d\tau' G_{\alpha\beta}(\tau, \tau') \Sigma_{\beta\gamma}(\tau', \tau'') &=&  \delta_{\alpha\gamma}\delta(\tau-\tau'')\,; \label{saddle-point-1}\\
 \Sigma_{\alpha\beta}(\tau, \tau') &=& J^2 G_{\alpha\beta}(\tau, \tau')^{q-1}\,. \label{saddle-point-2}
\eea

\subsection{Review of the replica-diagonal solution}

The replica partition function (\ref{Zreplica}) has a family of replica-diagonal saddle points, which have been extensively studied in the literature \cite{Kitaev,MScomments,Kitaev17,Polchinski16,Bagrets16,Gross17}. One assumes the ansatz
\be
G_{\alpha\beta} (\tau, \tau')= G (\tau, \tau') \delta_{\alpha\beta}\,;\qquad  \Sigma_{\alpha\beta}(\tau, \tau') = \Sigma (\tau, \tau') \delta_{\alpha\beta}\,.
\ee
The quenched average in this case coincides with the annealed average up to subleading orders in $1/N$ expansion \cite{Kitaev17,Gu16}: 
\be
\underbrace{\overline{ \log Z_{\text{RD}}}}_{\text{quenched}} = \underbrace{\log  \overline{Z_{\text{RD}}}}_{\text{annealed}} + O \left(\frac{1}{N^{q-2}}\right)\,,
\ee
which means that up to subleading orders in $1/N$ one can take off the replica indices in the path integral (\ref{Zreplica}):
\be
\overline{Z_{\text{RD}}}= \int D G D \Sigma\ \Pf[\dd_\tau - \hat{\Sigma}]^{N} \exp\left[-\frac{N}{2}\int_0^\beta\!\!\!\! \int_0^\beta d\tau_1 d \tau_2  \left(\Sigma (\tau_1, \tau_2)G (\tau_1, \tau_2) - \frac{J^2}{q} G (\tau_1, \tau_2)^q\right)\right]\,. \label{ZMS}
\ee
At large $N$ the asymptotic of the RHS of \eqref{ZMS} is given by saddle points contributions
\be
\overline{Z_{\text{RD}}}=\exp\{- S_{RD}\}
\ee
where the saddle points of (\ref{ZMS}) are given by the Schwinger-Dyson equations for melonic diagrams \cite{Kitaev,MScomments}
\bea
&& \frac{1}{\dd_\tau - \hat{\Sigma}} = \hat{G}\,;\label{RD-eq-1}\\
&& \Sigma (\tau, \tau') = J^2 G(\tau, \tau')^{q-1}\,,\label{RD-eq-2}
\eea
where in the first equation hats denote the integral operators with the kernels defined by the corresponding bilocal fields. 

The solutions of these equations in general case can be constructed numerically, which was done in the previous work \cite{MScomments,Kitaev17,Cotler16,Bagrets16}, and analytic solution is known in the IR/strong coupling limit $\beta J \gg 1$. 

\subsubsection{Strong coupling limit}

Substituting (\ref{RD-eq-2}) into (\ref{RD-eq-1}) and taking $\dd_\tau \to 0$, one obtains the equation
\be
J^2 \int_0^\beta d\tau' G(\tau, \tau') G(\tau', \tau'')^{q-1} = -\delta(\tau-\tau'')\,. \label{saddle-point-RS}
\ee
The solution for $G$ of the equation \eqref{saddle-point-RS} has the form of the conformal propagator on the circle: 
\be
 g_{c,\beta} (\tau,\tau')= b \left( \frac{\pi}{\beta}\right)^{2\Delta} \frac{\sgn (\tau-\tau')}{\left| \sin \frac{\pi}{\beta} (\tau-\tau') \right|^{2\Delta}}\,, \label{Gconf}
\ee
where $\Delta = \frac{1}{q}$ is the conformal dimension of the Majorana fermion and 
\be
b^q = \frac{(q-2) \tan \frac{\pi}{q}}{2\pi q J^2}\,. \label{b}
\ee

In the frequency space the conformal propagator on a circle
has a form \cite{Gurau17,Jensen16}: 
\begin{align}\label{gcomega}
 g_{c,\beta}(\omega_n)  
  =  -i2b \left(  \frac{2\pi}{\beta} \right)^{2\Delta-1} \cos(\pi \Delta)  
\frac{ \Gamma\left(  \frac{\beta}{2\pi} \omega_n +  \Delta   \right) \Gamma(1-2\Delta)  }{ \Gamma\left(  \frac{\beta}{2\pi} \omega_n +1 -  \Delta   \right)  }    \;,
\end{align}
where $\omega_n$ are Matsubara frequencies:
\be
\omega_n = \frac{2\pi}{\beta} \left(n + \frac12\right)\,, \quad n \in \ZZ\,. \label{Matsubara}
\ee
At zero temperature \eqref{Gconf} and \eqref{gcomega} reduce to 
\bea
g_c(\tau)&=&\frac{b}{|\tau|^{2\Delta}}\sgn(\tau)\label{gct}\\
g_c(\omega)&=&=i b \,  2^{ 1-2\Delta } \sqrt{\pi }
   { \Gamma( 1 - \Delta ) \over \Gamma( { 1 \over 2 } + \Delta ) } 
 |\omega|^{ { 2 \Delta  } - 1}   \sgn(\omega)\,,\label{gco}
\eea
where $b$ is the  dimensional constant fixed from \eqref{b}. 
The equations (\ref{RD-eq-1}), (\ref{RD-eq-2}) are invariant under time reparametrizations $\tau \to f(\tau)$ in the strong coupling limit, provided the bilocal fields transform as follows (we assume without loss of generality that $f$ is a monotonically increasing function): 
\bea
&& G(\tau_1, \tau_2) = f'(\tau_1)^{\Delta} f'(\tau_2)^{\Delta} G(f(\tau_1), f(\tau_2)) \,; \label{repG-RD}\\
&& \Sigma(\tau_1, \tau_2) = f'(\tau_1)^{1-\Delta} f'(\tau_2)^{1-\Delta} \Sigma(f(\tau_1), f(\tau_2))\,. \label{repS-RD}
\eea
Acting with these transformations on the solution (\ref{Gconf}), one can generate the full infinite-dimensional manifold $\text{diff}(S^1)/SL(2,\RR)$ of the replica-diagonal saddle points.

\subsubsection{On-shell action for replica-diagonal solution}
The replica-diagonal on-shell action is given by 
\be
\frac{2}{N} S_{RD}=\fs_1+\fs_2,\ee
where 
\bea
\fs_1&=&- \Tr \log \left(\dd_\tau- \hat{\Sigma}\right);\\
\fs_2&=&\int_0^\beta\!\!\!\int_0^\beta d\tau_1 d \tau_2\, \left(G(\tau_1, \tau_2) \Sigma(\tau_1, \tau_2)- \frac{J^2}{q}G(\tau_1, \tau_2)^q\right)\nn\\&=&\left(1-\frac{1}{q}\right) J^2\int_0^\beta\!\!\!\int_0^\beta d\tau_1 d\tau_2 \,G(\tau_1, \tau_2)^q= \left(1-\frac{1}{q}\right) \Tr({\bf 1} - \dd_\tau \hat{G} )\,,
\eea
where $G$ and $\Sigma$ solve \eqref{RD-eq-1}
and \eqref{RD-eq-2}.

The conformal limit is obtained by neglecting the time derivative. The $s_1$ and $s_2$ can be simplified and we can rewrite them in the equivalent forms
\bea
\fs_{1,c} &=& - \Tr \log \left(- \hat{\Sigma}_c\right)=- \Tr \log \left(-J^2 \hat{g}_c^{q-1}\right)\label{s1c}\\
\fs_{2,c} &=& \left(1-\frac{1}{q}\right) J^2 \int d\tau_1 d\tau_2 g_c(\tau_1, \tau_2)^q= \left(1-\frac{1}{q}\right) \Tr\,\,{\bf 1} \label{s2c}.
\eea
In the conformal limit both these pieces in the on-shell action diverge and have to be regularized. It is expected that the renormalization can be performed pertrurbatively (in $1/\beta J$) by reinstating the time derivative in $\fs_1$ and evaluating the corresponding counterterms, and the resulting renormalized action equals to the on-shell action, evaluated on the solution of exact saddle point equations \cite{MScomments}.

\subsection{Nonperturbative nature of the replica-nondiagonal correlators}
\label{sec:weak-coupling}

Here we will show that one cannot obtain a replica-nondiagonal large $N$ solution in perturbation theory over the free fermionic theory. We will work in the frequency space, and we also fix $q=4$. In the free case we have $J = 0$, and the saddle point equations (\ref{saddle-point-1})-(\ref{saddle-point-2}) are solved by 
\be
G_{\alpha\beta}(\omega) = G_f(\omega) \delta_{\alpha\beta}\,; \label{Gfree}
\ee
where $G_f$ is defined from 
\be
-i \omega G_f(\omega) = 1\,.
\ee
We introduce the dimensionless parameter $\lambda$ and look for the solution by perturbing the free UV fixed point: 
\be
G_{\alpha\beta}(\omega) = G_f(\omega) \delta_{\alpha\beta} + \lambda g^{(1)}_{\alpha\beta}(\omega) + \lambda^2 g^{(2)}_{\alpha\beta}(\omega) + \dots,.
\ee
The field $\Sigma$ in this case is expanded as follows: 
\bea
\Sigma_{\alpha\beta}(\omega) &=& 3J^2 \lambda (G_f * G_f * g^{(1)}_{\alpha\beta})(\omega) \delta_{\alpha\beta} + 3 J^2 \lambda^2 (G_f * g^{(1)}_{\alpha\beta} * g^{(1)}_{\alpha\beta})(\omega) \delta_{\alpha\beta} \\&+& 3 J^2 \lambda^2 (G_f * G_f * g^{(2)}_{\alpha\beta})(\omega) \delta_{\alpha\beta} + J^2 \lambda  (g^{(1)}_{\alpha\beta} * g^{(1)}_{\alpha\beta} * g^{(1)}_{\alpha\beta})(\omega) + \dots\,,
\nn\eea
where the star denotes the functional contraction in the frequency space, and the replica matrices are always multiplied a-la Hadamard, i.e. component-wise. Note that the leading possible replica-nondiagonal contribution to $\Sigma$ is of order $\lambda^{q-1}$. Substituting all this into the saddle point equation (\ref{saddle-point-1}) and equating the powers of $\lambda$, we arrive at an infinite system of linear inhomogeneous integral equations for $g^{(k)}$. For example, for $\lambda^1$ we obtain
\be
i \omega g^{(1)}_{\alpha\beta}(\omega) =3 J^2 G_f(\omega)\ (G_f * G_f * g^{(1)}_{\alpha\beta})(\omega) \delta_{\alpha\beta}\,.
\ee
It is clear that the solution for $g^{(1)}$ of this equation is replica-diagonal. Having solved this equation, one can substitute the solution into the $\lambda^2$ equation, which would then allow to solve for $g^{(2)}$. However, since everything in the equation will be replica-diagonal, the solution for $g^{(2)}$ will also be replica-diagonal. Using this expansion one can construct the solution up to any finite order in $\lambda$, and it will thus always remain replica-diagonal, ultimately because the free fixed point $G_f$ is replica-diagonal\footnote{Note that this argument does not rule out the replica-diagonal solutions with broken replica symmetry, i.e. different values of $g_{\alpha\alpha}$ for distinct $\alpha$.}. 

This means that we cannot obtain a replica-nondiagonal solution as a low-energy limit of a replica-diagonal solution, and any replica-nondiagonal large $N$ solution would be nonperturbative in $(\beta J)^{-1}$.

\section{Nondiagonal saddles in the $q=2$ model}
\label{sec:q=2}

In this section we consider the $q=2$ variant of the SYK model. We present a class of simple exact replica-nondiagonal solutions and study their properties. 

\subsection{The solutions}

Let us consider the $q=2$ case and show that there are replica-nondiagonal solutions of the saddle point equation. The saddle point equations (\ref{saddle-point-1})-(\ref{saddle-point-2}) in the $q=2$ case condense to a single equation, which in terms of replica matrices is written as 
\bea
\label{Gm-kappa-q2}
G(\omega)\cdot \Big(-i\omega\ I- J^2 G(\omega)\Big)&=& I\,.
\eea
Here $G(\omega)=(G(\omega))_{\alpha\beta}$ is a $M\times M$ matrix, where $\omega$ is the Matsubara frequency, and $I$ is the unit matrix in the replica space.  We assume the replica-symmetric ansatz
\be\label{NDS}
G_{\alpha\alpha}=G_{0},\,\,\,\,\,G_{\alpha\beta}=G_{1},\,\,for\,\,\alpha\neq\beta,\,\,\,\,\,\alpha,\beta=1,...M
\ee
The equation \eqref{Gm-kappa-q2} turns into the pair of equations for $G_0$ and $G_1$:
\bea\label{eq1}
-i\omega G_0-J^2\Big(G_0^2+
   (M-1)G_1^2\Big) &=&1\\\label{eq2}-i \omega G_1 -J^2\Big(2 G_0
  G_1+G_1^2 (M-2)\Big)&=&0
   \eea
In order to obtain the fermionic solutions, we have to impose the antisymmetry condition
 \bea   G_{\alpha\beta}(\omega)&=& -G_{\beta\alpha}(-\omega)\,,\label{AsComega}\eea
which in terms of the replica-symmetric ansatz simply means that $G_0$ and $G_1$ have to be odd functions in frequency and time domains.

The equations (\ref{eq1})-(\ref{eq2}) are readily solved. There are two replica-diagonal solutions:
       \bea
       G_{0}^{(1)}(\omega)&=& \frac{
  -i \omega+\,i\sgn(\omega)\sqrt{4 J^2+\omega^2}}{2J^2}\,; \label{G01}\\
  G_{0}^{(2)}(\omega)&=& \frac{
  -i \omega-\,i\sgn(\omega)\sqrt{4 J^2+\omega^2}}{2J^2} \,; \label{G02}\\
    G_{1}^{(j)}&=&0\,\,\,\,\,\mbox{for } \,\,j=1,2, \label{G112}
\eea
and two replica-nondiagonal solutions
\bea
 G_{0}^{(3)}(\omega)&=&\frac{-i  \omega+ \,i\,\sgn(\omega)\sqrt{4
   J^2+\omega^2}\left(1-\frac2M\right)}{2J^2}\,; \label{G03}\\
  G_{1}^{(3)}(\omega)&=& -\,i\,\sgn(\omega)\frac{\sqrt{4 J^2+\omega^2}}{J^2 M}
  \,;\label{G13}\\
  G_{0}^{(4)}(\omega)&=&\frac{-i  \omega- \,i\,\sgn(\omega)\sqrt{4
   J^2+\omega^2}\left(1-\frac2M\right)}{2J^2}\,; \label{G04}\\
  G_{1}^{(4)}(\omega)&=& i\,\sgn(\omega)\frac{\sqrt{4 J^2+\omega^2}}{J^2 M}\,. \label{G14}
\eea
 Let us make a few remarks about these solutions. The first solution coincides with the solution presented in \cite{MScomments}. It admits the $J=0$ limit, where its leading asymptotic is given by the free correlator $\frac{1}{-i \omega}$. The second solution is the one, which gives subleading saddles, discussed by Cotler et al in \cite{Cotler16}.
All solutions are pure imaginary and 
 \bea
 \Re\ G_{k}^{(j)}(\omega) &=&0= \Re\ G_{k}^{(j)}{(\omega)}\,;\\
 G_{k}^{(j)}(-\omega)&=&-G_{k}^{(j)}(\omega),\,\,\,
  G_{k}^{(j)*}(\omega)=-G_{k}^{(j)}(\omega)=G_{k}^{(j)}(-\omega),\,\,\,\,\,k=0,1
 \eea
There are also the relations
\bea
 G_{0}^{(3)}-G_{1}^{(3)}&=& G_{0}^{(1)}\,;\\
 G_{0}^{(4)}-G_{1}^{(4)}&=& G_{0}^{(2)}\,.\eea
In regards to the replica-nondiagonal third and fourth solutions, the important property worth pointing out here is that they are singular in the $M \to 0$ limit. Also, one can check that these solutions are singular in the free limit $J \to 0$, which confirms the nonperturbative nature of these solutions, discussed in the sec. \ref{sec:weak-coupling}. We will see that these properties remain for the class of numerical $q=4$ solutions that we studied.

We conclude this subsection with a comment about other solutions of the equation (\ref{Gm-kappa-q2}). This is a quadratic matrix equation, which means that in principle one can find and classify all of the solutions at any fixed $M$. Their general form can be found by rewriting the equation (\ref{Gm-kappa-q2}) as follows\footnote{We thank Andrey Mikhailov for pointing this out.}: 
\begin{eqnarray}
\left(J G(\omega) +\frac{i\omega}{2J}
I\right)^2=-\left(1+\frac{\omega^2}{4J^2}\right) I\,.
\end{eqnarray}
The general solution will have a form 
\begin{eqnarray}
G =-\frac{i\omega}{2J^2} I \pm i\frac{\sqrt{4J^2+\omega^2}}{2J^2} X\,; \label{Ggen}
\end{eqnarray}
where $X$ is a matrix which parametrizes a particular solution, such that:
\be X \cdot X = I\,.\label{XX=I} \ee
There is no explicit dependence on $M$ in (\ref{Ggen}), however the diagonal component of equation (\ref{XX=I}) has a sum of $M$ terms equal to $1$. That means that the individual non-trivial components of $X$ should contain the $1/M$ dependence to compensate. Because of this argument, we expect that every solution will have singularity at $M \to 0$. 

\subsection{On-shell action}
\label{sec:action-q=2}

\begin{figure}[t]
\centering
\includegraphics[scale=0.3]{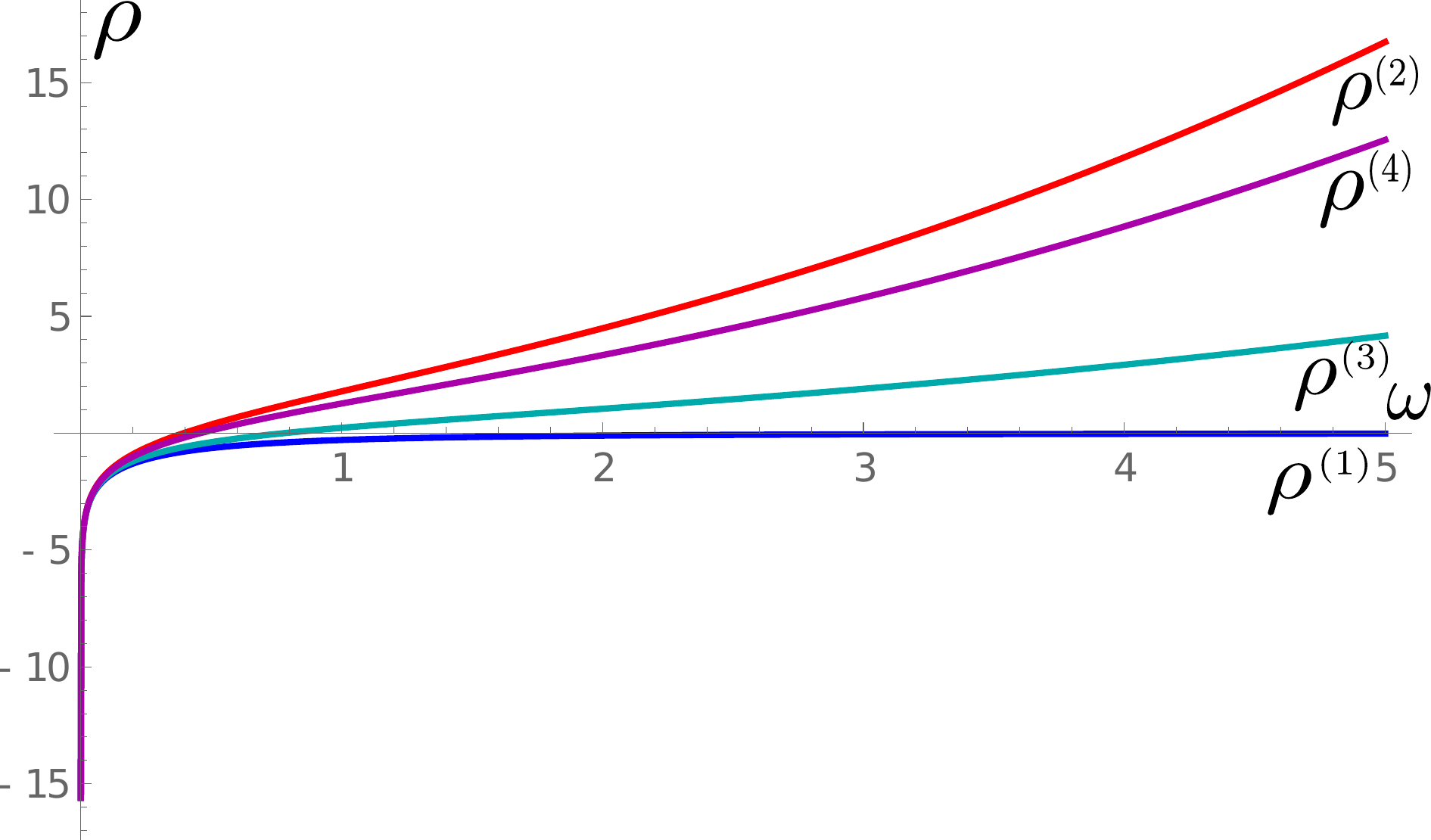} $\,\,\,${\bf A.}$\,\,\,\,\,\,\,\,\,\,$
\includegraphics[scale=0.3]{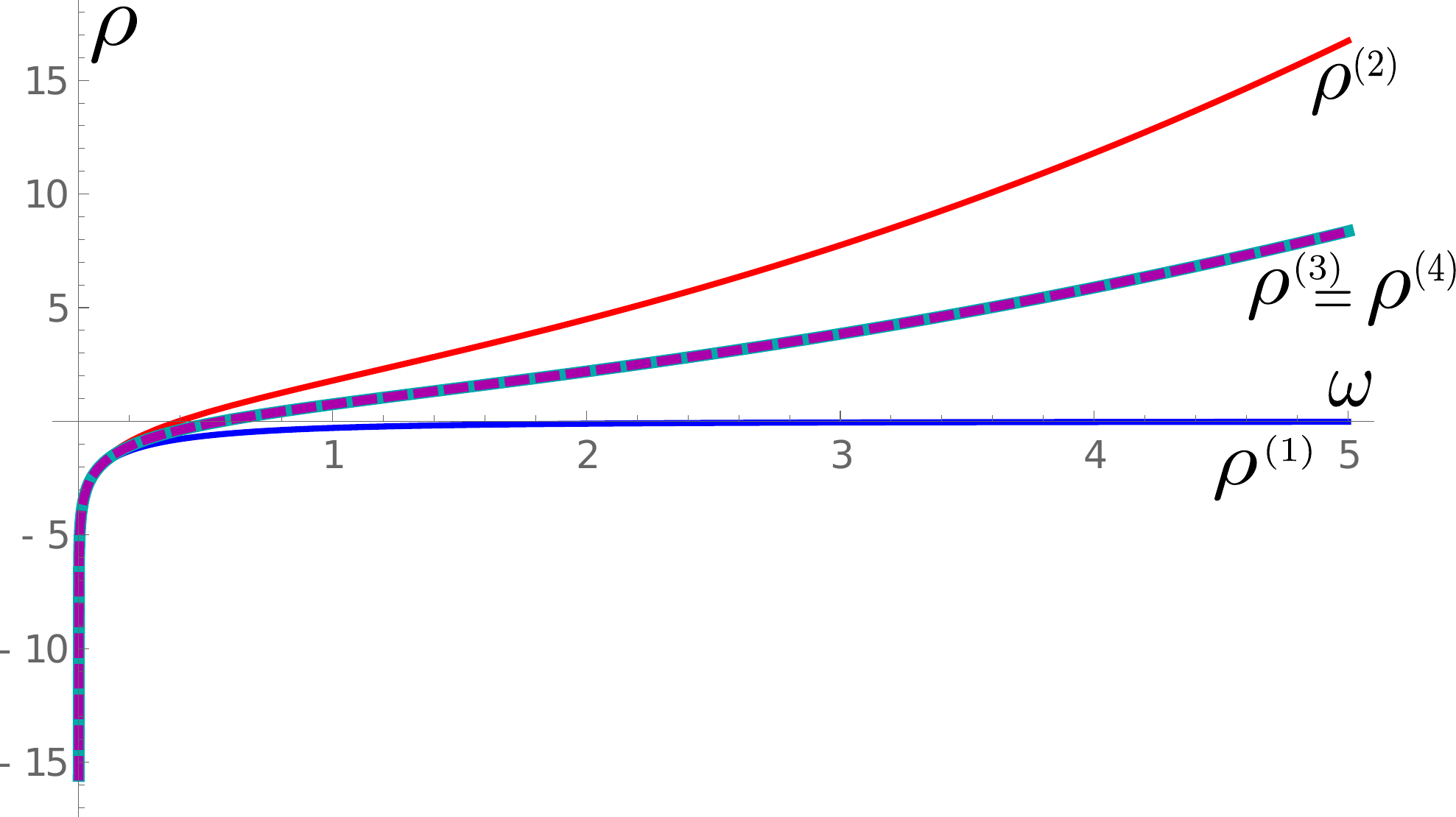}$\,\,\,${\bf B.} 
\includegraphics[scale=0.3]{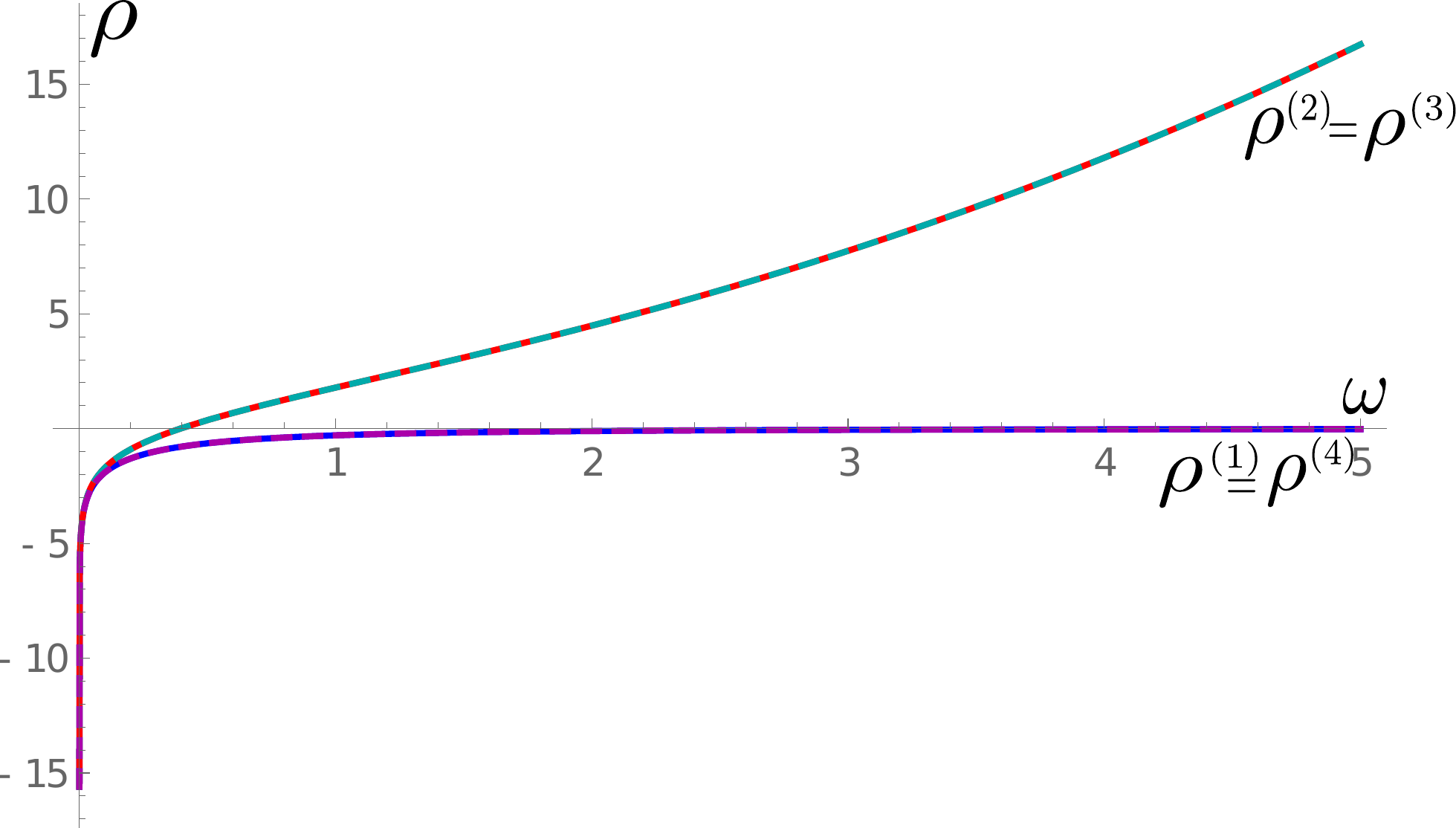}$\,\,\,${\bf C.} $\,\,\,\,\,\,\,\,\,\,$
\includegraphics[scale=0.3]{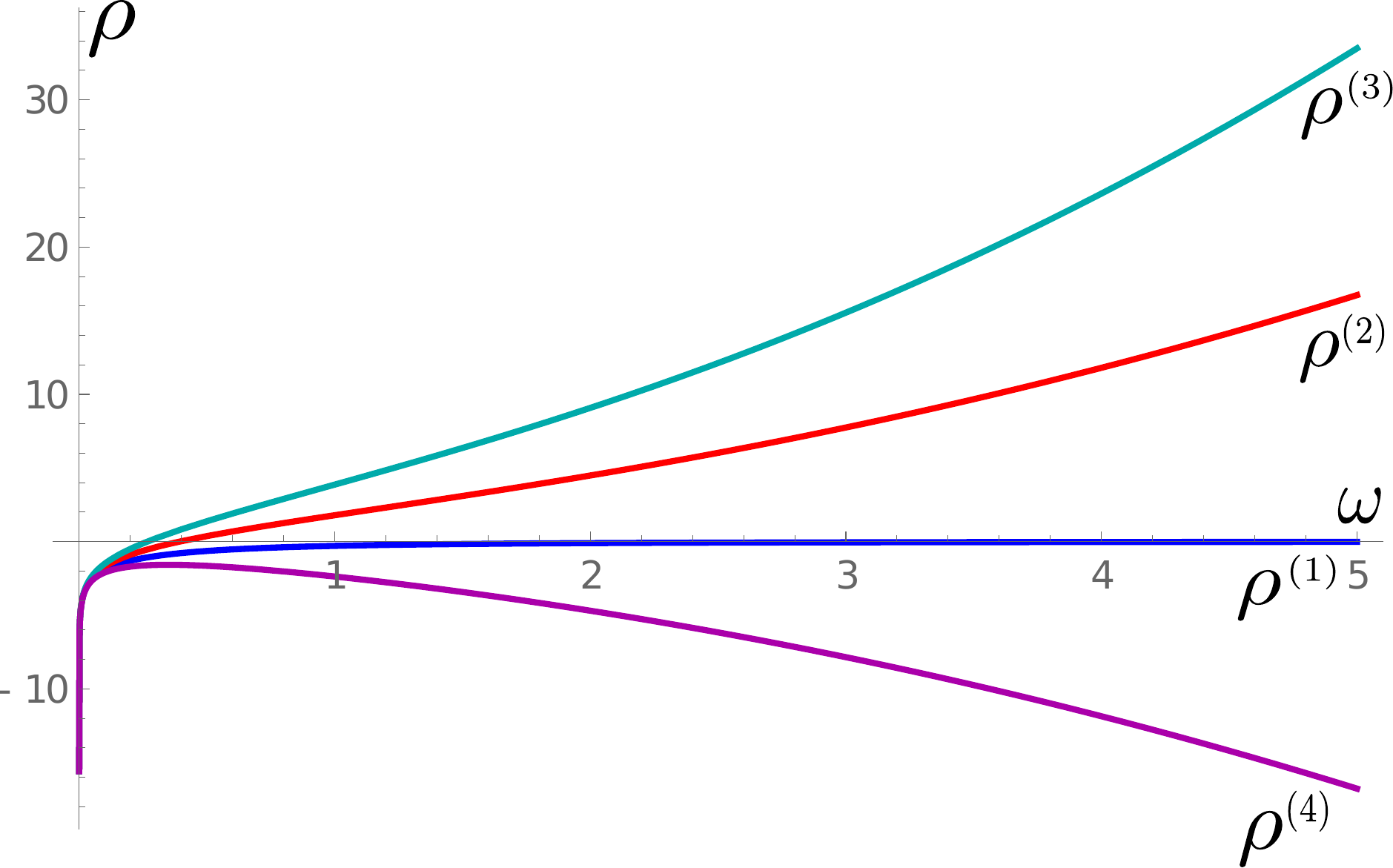}$\,\,\,${\bf D.} 
\caption{Action densities \eqref{action1-4} on 4 roots. Here $J = 1$. 
\textbf{A}. $M=4$. \textbf{B}. $M=2$. In this case $\rho^{(3)} = \rho^{(4)}$. \textbf{C}. $M=1$. In this case $\rho^{(1)}=\rho^{(4)}$ and $\rho^{(2)}=\rho^{(3)}$. \textbf{D}. $M=0.5$.}
\label{fig:DDD}
\end{figure}
Now we turn to the study of contributions of the nondiagonal saddle points, described above, to the replica partition function. Since the saddle point equations for different frequency modes decouple, we can consider the density of the action, which we denote by $\rho$ and define at zero temperature as follows: 
\bea
\frac{4\pi}{NM{\cal V}}S_{M}&=&\int d\omega\ \rho(\omega,J,M)\,.\label{Ss}
\eea
Here ${\cal V}$ is the regularized volume. At finite temperature the definition is generalized by setting ${\cal V} = 2\pi$ and $\int d\omega \to \sum_{\omega_n}$. 
The decoupling of the saddle point equations means that for every allowed frequency one can in principle choose any of the four solutions obtained above. The question, in which we are interested here in this section, is whether there are any saddle points that would dominate over the replica diagonal saddle. Because of the frequency decoupling, to check this fact it is enough to compare the action density $\rho$ evaluated on different roots. Using the formulae (\ref{fs1-RS}),(\ref{fs_2-RS}) at $q=2$, the action density is written as 
\bea
 \rho&=&-\fl_1-\frac{1}{M}\,\fl_2+\frac{J^2}2\,\fm\label{sj}\,;\\
 \fl_1&=& \log\left(1+J^2\frac{G_0(\omega) - G_1(\omega)}{i\omega}\right)\,;\\
  \fl_2&=& \log\left(1 +M\frac{J^2G_1}{i\omega+J^2(G_0(\omega) -G_1(\omega))}\right)\,;\\
\fm &=&  |G_0(\omega)|^2
+(M-1)|G_1(\omega)|^2\,.
\eea
We denote the contributions from different solutions (\ref{G01})-(\ref{G14}) as
\bea
\rho^{(j)}(\omega,J,M)&=&-\fl_{1}^{(j)}-\frac{1}{M}\fl_{2}^{(j)}+\frac{J^2}2\fm^{(j)},\,\,\,\,j=1,2,3,4
\label{action1-4}\eea
\begin{figure}[t]
\centering
\includegraphics[scale=0.21]{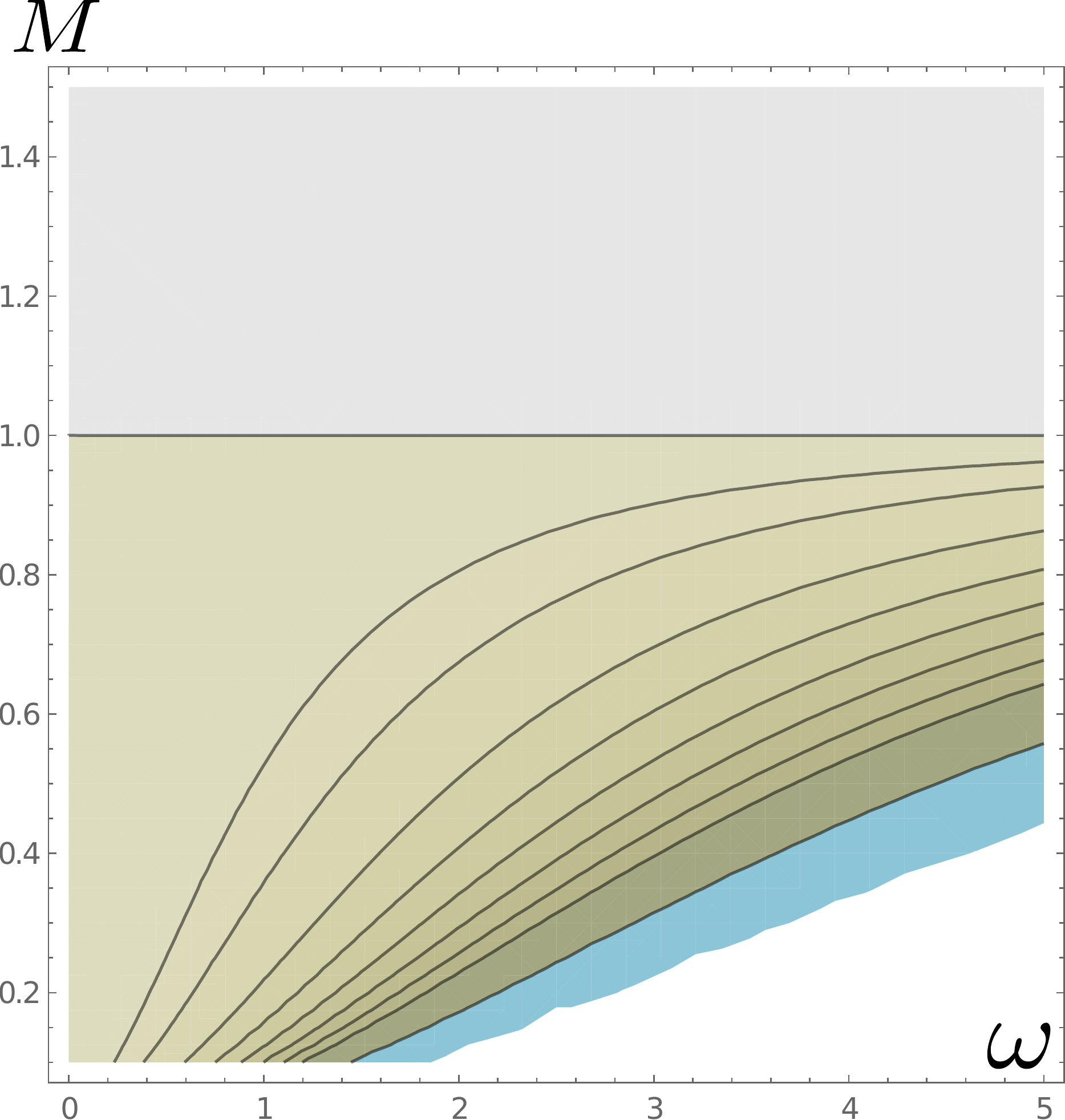}
\includegraphics[scale=0.2]{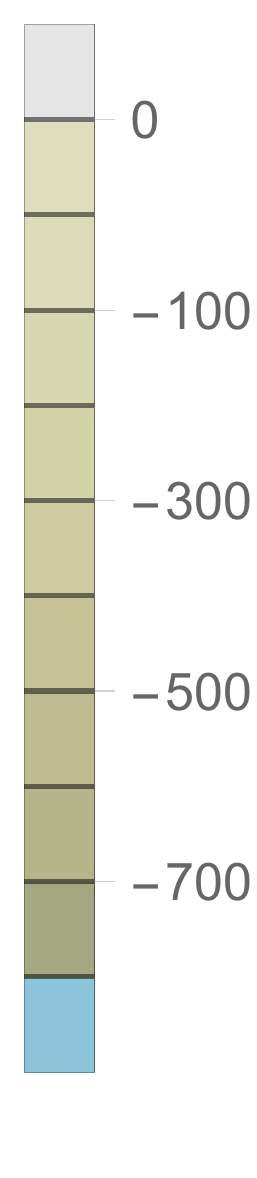}A.
\includegraphics[scale=0.21]{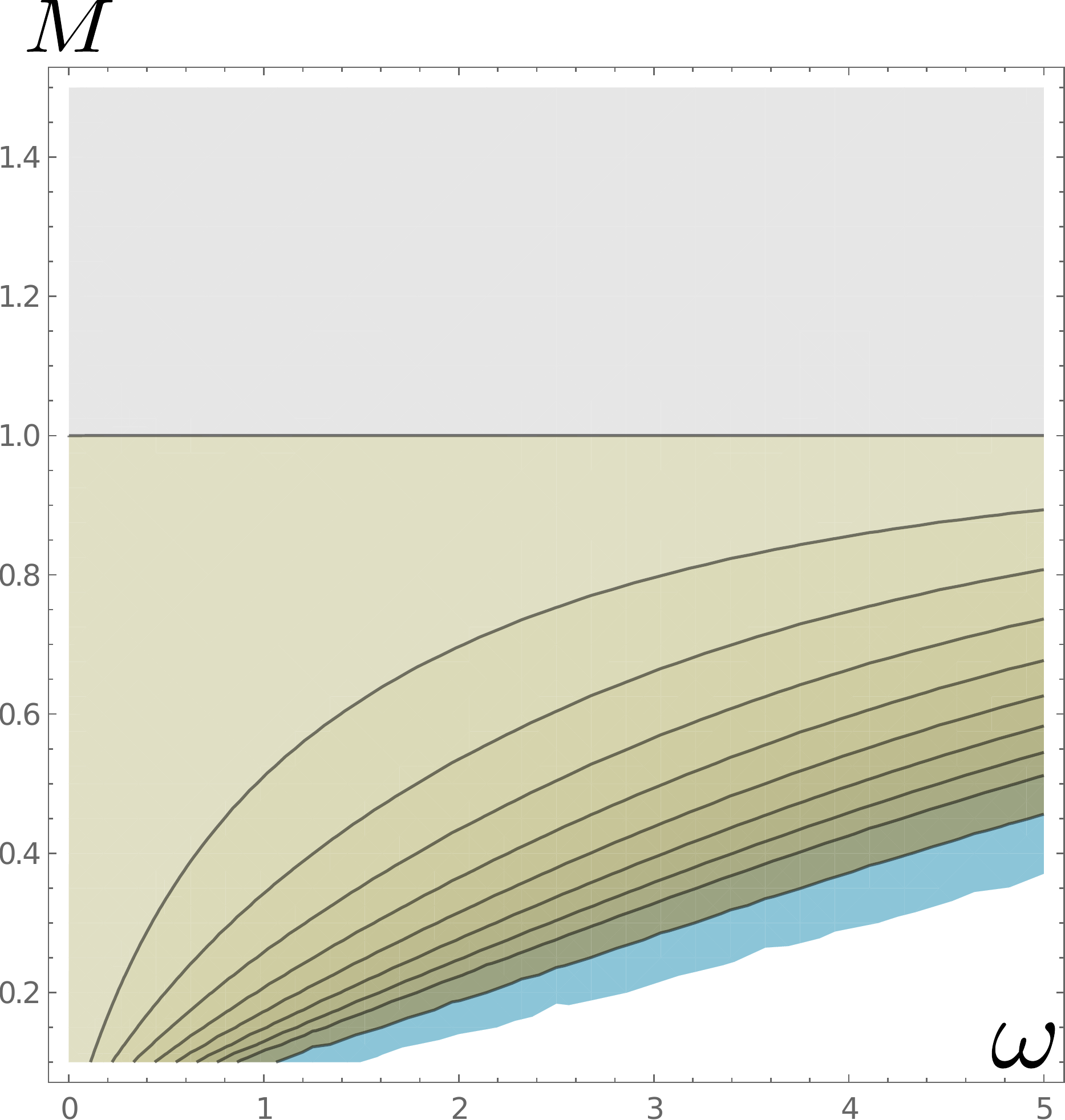} 
\includegraphics[scale=0.2]{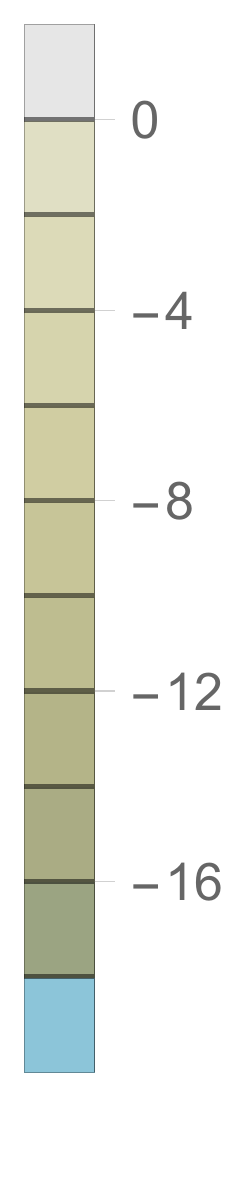}B.
\includegraphics[scale=0.21]{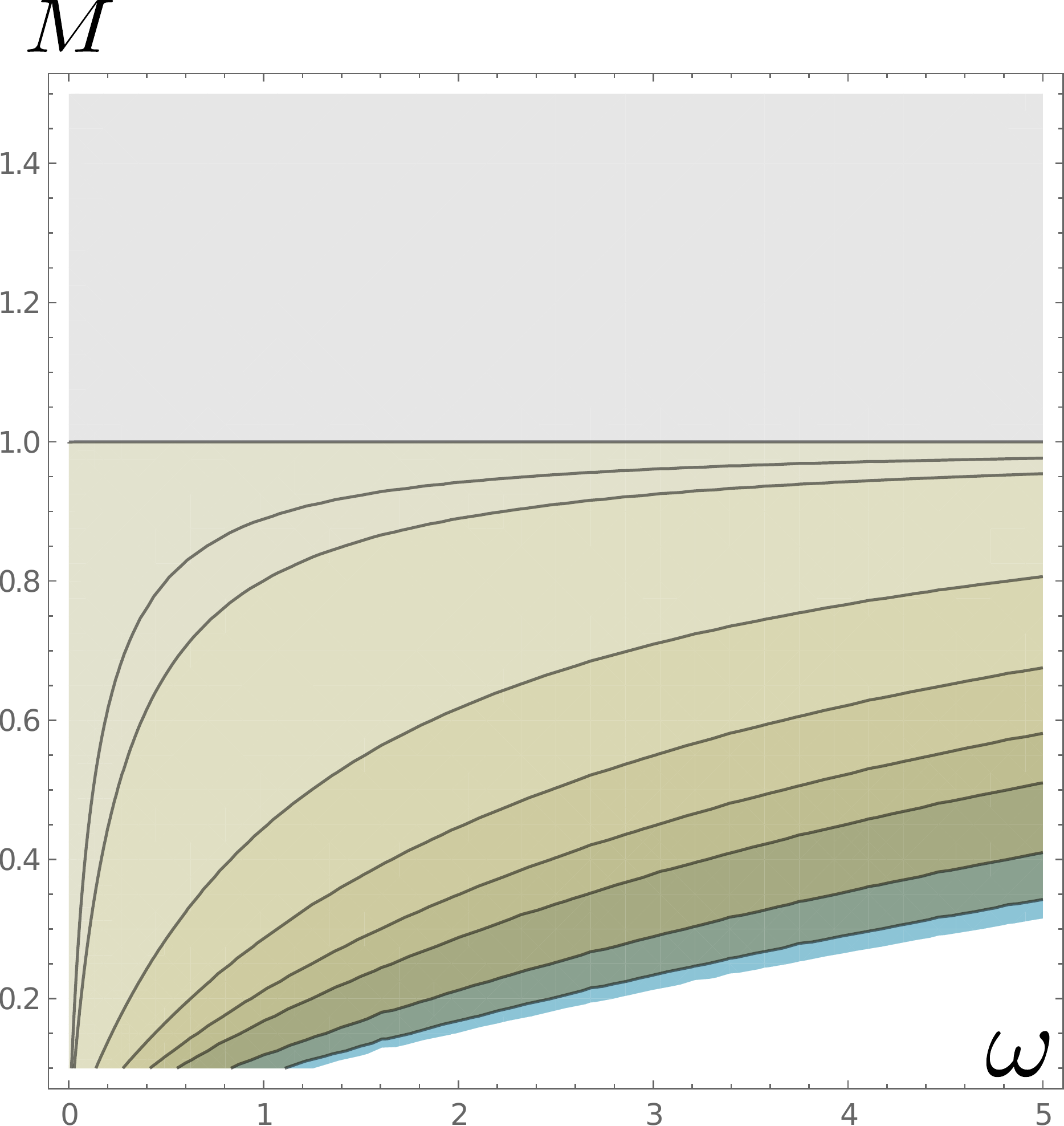} 
\includegraphics[scale=0.2]{figures/M-Delta-J1-leg-corr2.pdf}C.\\
$\,$\\
\includegraphics[scale=0.21]{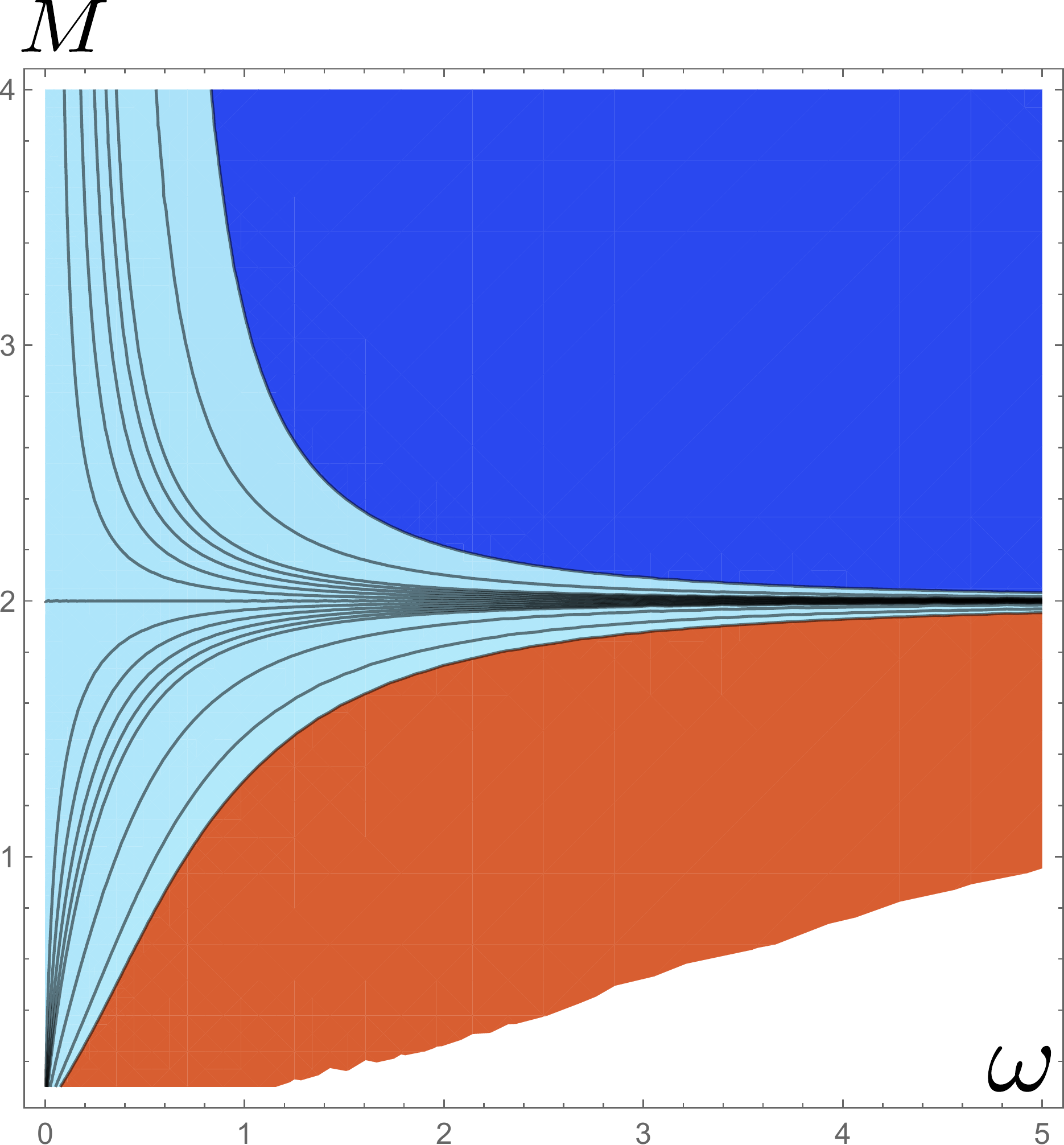}
\includegraphics[scale=0.2]{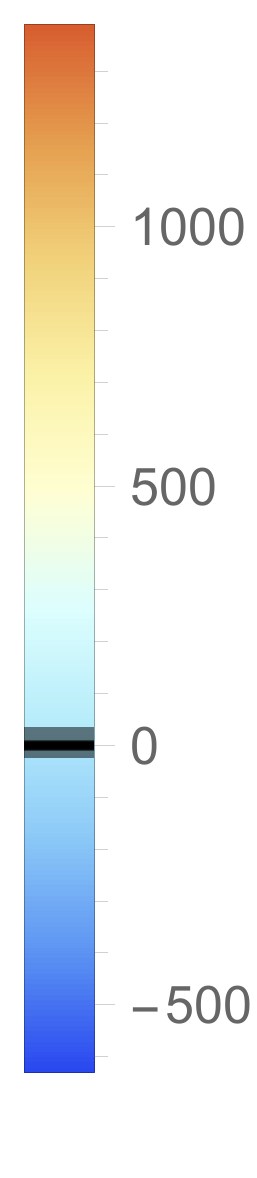}D. 
\includegraphics[scale=0.21]{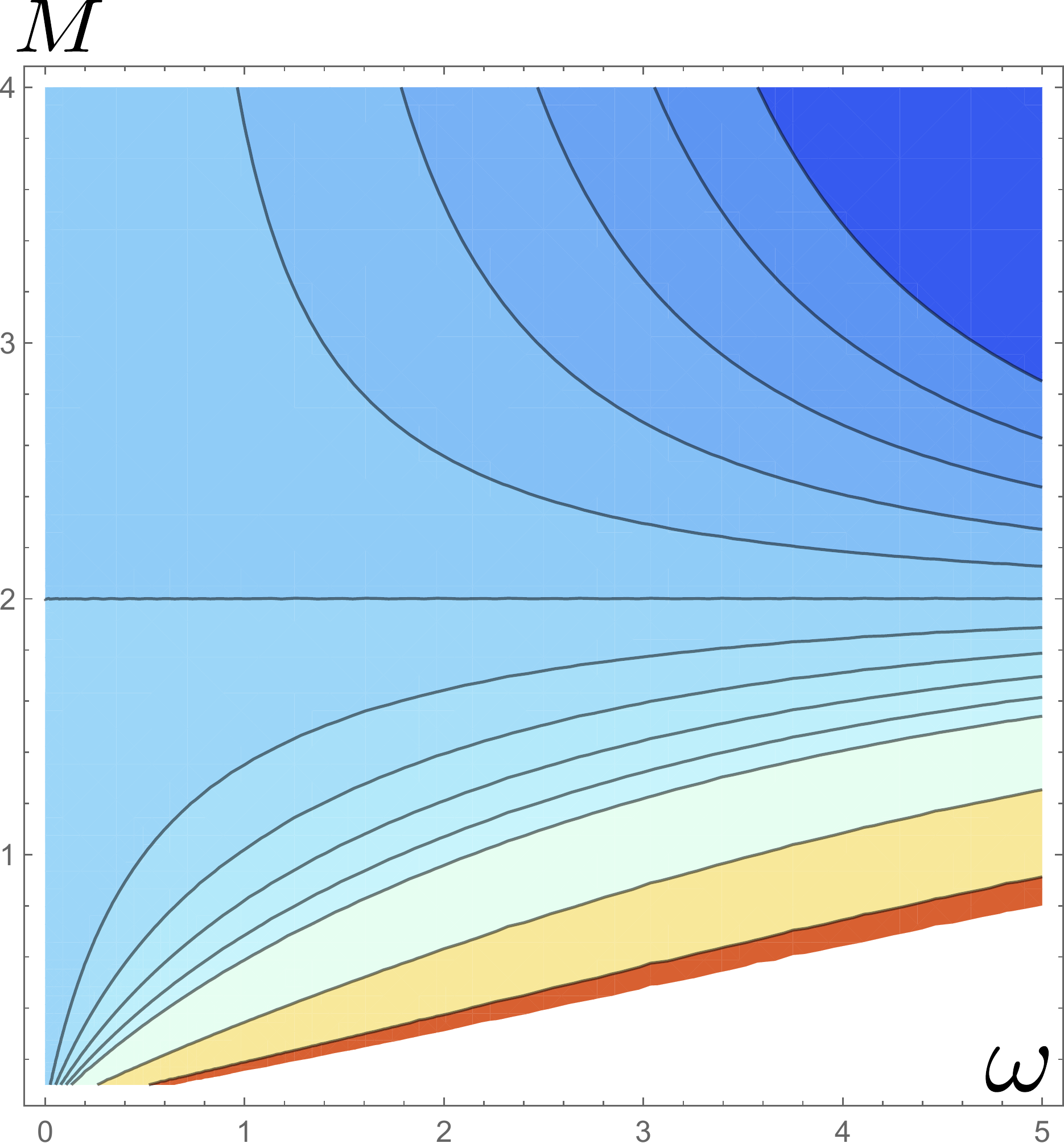}
\includegraphics[scale=0.2]{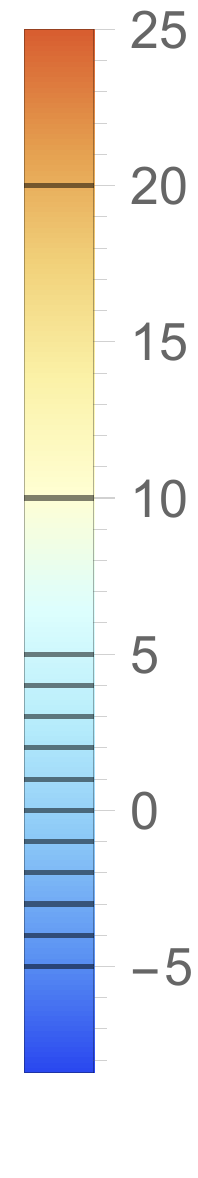}E.
\includegraphics[scale=0.21]{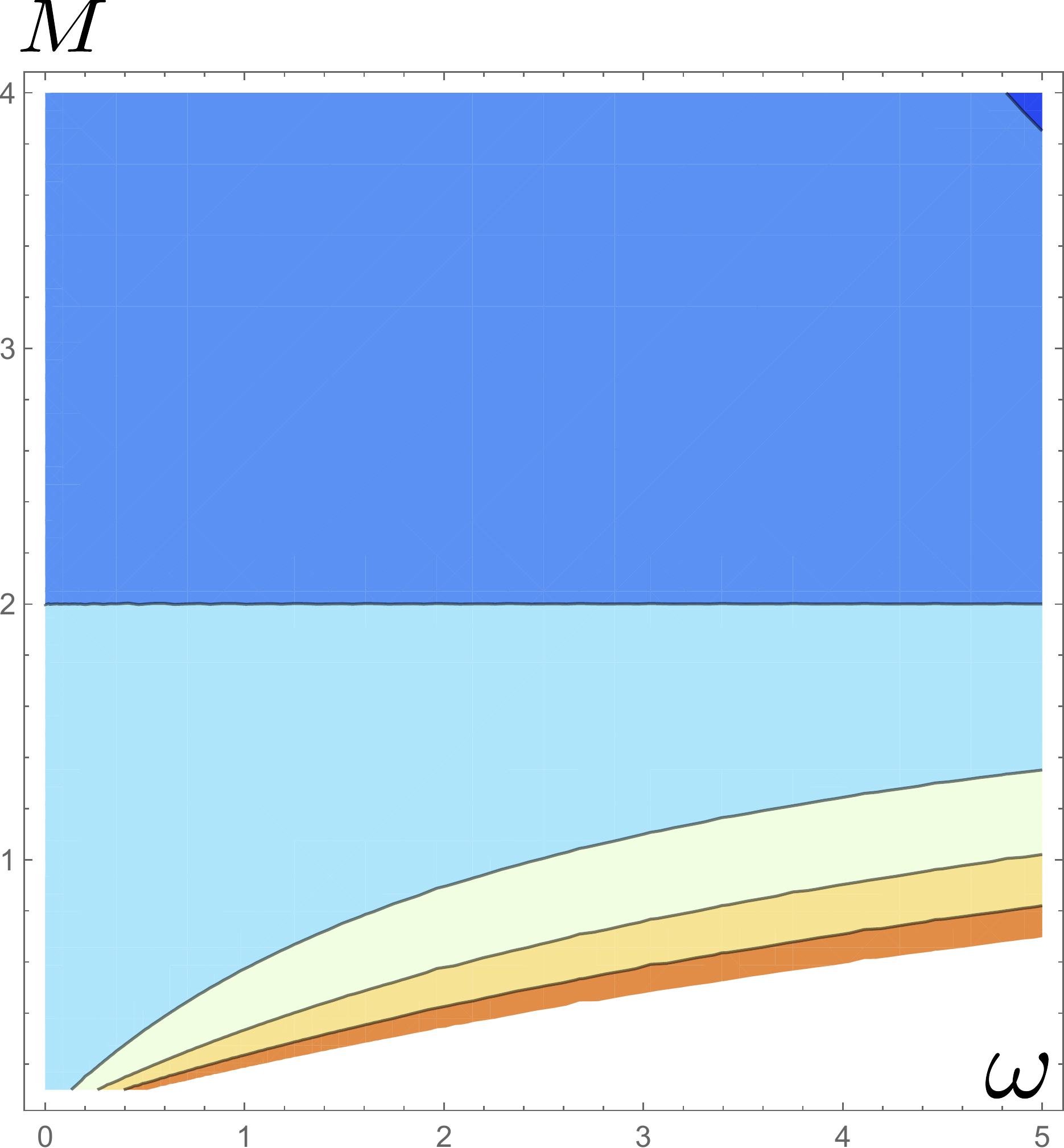}
\includegraphics[scale=0.3]{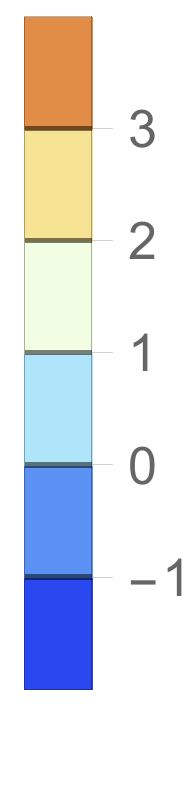}F.\\
$J=0.1\,\,\,\,\,\,\,\,\,\,\,\,\,\,\,\,\,\,\,\,$
$\,\,\,\,\,\,\,\,\,\,\,\,\,\,\,\,$
$\,\,\,\,\,\,\,\,\,\qquad\qquad J=1\,\,\,\,\,\,\,\,\,\,\,\,\,\,\,\,\,\,\,\,\,\,\,\,$ $\,\,\,\,\,\,\,\qquad \qquad   J=5\,\,\,\,\,\,\,\,\,\,\,\,\,\,\,\,\,\,\,\,\,$
\caption{First line: the difference $\rho(\text{non-diag})-\rho^{(1)}$ as a function of $\omega$ and $M$. The regions of dominance of the diagonal solution are shown by gray color. Second line: the difference $\rho^{(3)}-\rho^{(4)}$ as a function of $\omega$ and $M$.
}
\label{fig:RPA}
\end{figure}
If we see an interval where $\rho^{(3)} < \rho^{(1)}$ ($\rho^{(4)} < \rho^{(1)}$), then that means that there are solutions that have action lower than the replica-diagonal. One such solution can be constructed by selecting the $3$-rd ($4$-th) root for the frequencies from the interval where the above inequality is true, and choosing the $1$-st root for the rest of the frequencies. Thus we focus on the study of the function $\rho(\omega)$ in the remainder of this subsection. 

Let us first estimate the contributions of the first two diagonal solutions to the on-shell action. For these roots $\fl_2=0$ and:
\bea
\rho^{(1,2)}&=&-\fl_{1}^{(1,2)}+\frac{J^2}2\fm^{(1,2)},\nn\\
\fl_{1}^{(1,2)}&=&\log\left(1+J^2\frac{G^{(1,2)}_0}{i \omega} \right)=
   \log\left(\frac{
   1\pm \sgn(\omega)\sqrt{\frac{4 J^2}{\omega^2}+1}}{2}\right)\,;\\
   \fm^{(1,2)}
&=& |G^{(1,2)}_0(\omega)|^2=  \frac{
  4}{\left( 
   \omega\,\pm\,\sgn(\omega)\sqrt{4 J^2+\omega^2}\right)^2}\nn
\eea   
We can expand in terms of small $\omega>0 $. In this case we get 
\be
\rho^{(1,2)} = \frac12 + \log \frac{\omega}{J} \mp \frac{\omega}{J} + O(\omega^2)\,.\label{12s}
\ee
Let us now present the contributions to the action density the last two solutions. 
   \bea
&\,&\rho^{(3,4)}=-\fl_{1}^{(3,4)}-\frac{1}{M}\fl_{2}^{(3,4)}+\frac{J^2}2\fm^{(3,4)}\\
&=&-\log \left(\frac{1}{2} \left(\pm\sqrt{\frac{4 J^2}{\omega^2}+1}\ 
   \sgn(\omega)+1\right)\right)
   -
   \frac1M \log \left(\frac{\omega\mp\sqrt{4 J^2+w^2}
   \text{sgn}(\omega)}{\omega\pm\sqrt{4 J^2+\omega^2}
   \text{sgn}(\omega)}\right)\nn \\
   &+& \frac{\mp  (M-2) M \left| \omega\right|  \sqrt{4
   J^2+\omega^2}+M^2
  \omega^2}{4 J^2 M^2} +\frac12\,.
\eea
For small frequencies $\omega\to 0$ ($\omega>0$), we get
\bea
\rho^{(3,4)}= \frac12 + \log \frac{\omega}{J}- \frac{i \pi}{M} \mp \left(1-\frac{2}{M}\right) \frac{\omega}{J} + O(\omega^2)\label{w03}\,.\eea
Comparing \eqref{w03} with  \eqref{12s} we see in the leading order the asymptotic coincides, except for phase contribution, which would be proportional to $2\pi i$ in the total action and thus inconsequential. However, in the subleading $\omega^1$ order and higher there is difference. As hinted by this asymptotic and confirmed by the plots of the exact expressions on Fig.\ref{fig:DDD}, the $M=1$ is a threshold value which distinguishes between two different types of behavior of saddle points: 
\begin{itemize}
\item $M > 1$. In this case $1-2/M > -1$, so that the diagonal $1$-st solution dominates, see Fig.\ref{fig:DDD}A,B. 
\item $M < 1$. In this case $1-2/M < -1$, and consequently nondiagonal solutions dominate over the diagonal $1$-st solution, see Fig.\ref{fig:DDD}D. 
\end{itemize}
The $M=1$ case is degenerate, where the $1$-st and $4$-th solutions, as well as the $2$-nd and $3$-rd solutions give pairwise equal contributions to the action, see Fig.\ref{fig:DDD}C.  Also note the peculiar case of $M=2$, where the two nondiagonal solutions give the same action density, as shown on Fig.\ref{fig:DDD}C. For $M>2$ we have $\rho^{(3)}<\rho^{(4)}$, and for $M<2$ we have $\rho^{(3)}>\rho^{(4)}$. This is also illustrated on the plots Fig.\ref{fig:RPA}D-F we plot the difference between the contributions of two nondiagonal solutions $\rho^{(3)}-\rho^{(4)}$ as a function of $\omega$ and $M$. 

On Fig.\ref{fig:RPA}A-C we plot the difference $\rho(\text{non-diag})-\rho^{(1)}$\footnote{Here $\rho(\text{non-diag})$ is the contribution of 3-rd root for $M > 2$ and 4-th root for $M < 2$ (which is on the plot), since they exchange dominance relative to each other at that point.}. Besides the observations mentioned above, from these density plots it is evident that the IR region seems to be more robust in the singular $M \to 0$ limit, rather than the UV. One can interpret this as a hint towards the fact that the singular behavior in $M \to 0$ limit of nondiagonal solutions is the UV effect in the SYK model. We explain more evidence for this in other sections of the paper. 

\section{Exact nondiagonal saddles in $q=4$ SYK: numerical study}
\label{sec:Numerics}

Having found nondiagonal solutions in the $q=2$ model, we now turn to study the interacting $q=4$ model. In this case the saddle point equations (\ref{saddle-point-1})-(\ref{saddle-point-2}) cannot be solved analytically in general, so we construct the solutions numerically. As was done in the previous section, we assume the replica-symmetric ansatz (\ref{NDS}). In terms of independent variables the saddle point equations read
\bea
&& -i \omega G_0(\omega) - G_0(\omega) \Sigma_0(\omega) -(M-1) G_1(\omega) \Sigma_1(\omega) =1\,;\label{SD-RS-1-diag}\\
&& -i \omega G_1(\omega) - G_1(\omega) \Sigma_0(\omega)- G_0(\omega) \Sigma_1(\omega)- (M-2) G_1(\omega) \Sigma_1(\omega) =0\,;\label{SD-RS-1-offdiag}\\
&& \Sigma_{0,1}(\tau, \tau') = J^2 G_{0,1}(\tau, \tau')^{q-1}\,.\label{SD-RS-2}
\eea
We solve the equations numerically at finite temperature, with $\beta = 2\pi$. 

\subsection{Comments on the method}

We solve the system of integral equations (\ref{SD-RS-1-diag})-(\ref{SD-RS-2}) by iterating them. We use the approach employed in \cite{Cotler16} in studies of subleading replica-diagonal saddles. The main idea is to start iterations with $q=2$, using a particular solution of the $q=2$ model as a trial functions, and gradually increase $q$ from $2$ to $4$ during the procedure. Let us know discuss the procedure in more detail. 

\paragraph{Initial condition.} The trial functions for $G_0$ and $G_1$ are constructed by choosing one of the four solutions (\ref{G01})-(\ref{G14}) for every Matsubara frequency. We want the resulting solution in the interacting model to have the asymptotic behavior in the UV region that would correspond to the free theory, so we only consider the $q=2$ trial functions for which $\exists \bar{n}$ such that $G_{0,1}(\omega_n) = -G_{0,1}(-\omega_n) = G^{(1)}_{0,1}(\omega_n)$  $\forall n \geq \bar{n}$.
In this case for any $n < \bar{n}$ we can choose $G_{0,1}(\omega_n) = -G_{0,1}(-\omega_n)$ to be equal to any of the four solutions (\ref{G01})-(\ref{G14}).

\paragraph{Iteration procedure.} We divide the iterations into two stages. 
\begin{itemize}

\item[1.] We start first stage of iterations at $q=2$. At each iteration, $q$ is increased by some small amount. At every step $\Sigma_0$ and $\Sigma_1$ are computed in the position space using fast Fourier transform for $G$ and equation (\ref{SD-RS-2}), and then the inverse fast Fourier transform is performed on $\Sigma_{0,1}$. Then  $G_0(\omega)$ and $G_1(\omega)$ are updated according to the weighted rule (as also used in \cite{MScomments}) of the form  
\be
G_{0,1}^{\text{new}} = (1-x)\ G_{0,1} + x\ \tilde{G}_{0,1}\,,
\ee
where $0<x<1$ is the weighting coefficient and $\tilde{G}$ is defined by solving the equations (\ref{SD-RS-1-diag})-(\ref{SD-RS-1-offdiag}) in terms of $G_0$ and $G_1$: 
\bea
\tilde{G}_0 &=& \frac{1}{-i \omega_n - \Sigma_0 + \Sigma_1} + \frac{\Sigma_1}{(i \omega_n + \Sigma_0 - \Sigma_1)(i \omega_n + \Sigma_0 + (M-1) \Sigma_1)}\,;\\
\tilde{G}_1 &=& \frac{\Sigma_1}{(i \omega_n + \Sigma_0 - \Sigma_1)(i \omega_n + \Sigma_0 + (M-1) \Sigma_1)}\,.
\eea
At this stage we keep the weight fixed, and the procedure is finished when $q$ reaches $4$. 

\item[2.] The second stage of iterations is performed at fixed $q=4$. Its purpose is to tune the solutions, obtained in the previous stage, to the desired precision. In this procedure we take the approach of \cite{MScomments} and control the $\mathcal{L}_2$-norm of the solutions between successive steps 
\be 
|| \Delta G_{0,1} ||^2 = \int_0^\beta d\tau |G_{0,1}(\tau)^{\text{new}}-G_{0,1}(\tau)^{\text{old}}|^2\,,
\ee
decreasing the weight $x$ every time the $|| \Delta G_{0,1} ||^2$ starts increasing. The procedure is completed once $|| \Delta G_{0,1} ||^2$ reaches zero (up to desired numerical accuracy). 
\end{itemize}

The main limitation of our approach is that only real-valued stable numerical solutions can be obtained. This puts limitations on making connections with the analytic solutions in the conformal limit, which we discuss in the section \ref{sec:Factorized}) and thereafter. 

\begin{figure}[t]
\centering
\includegraphics[scale=0.16]{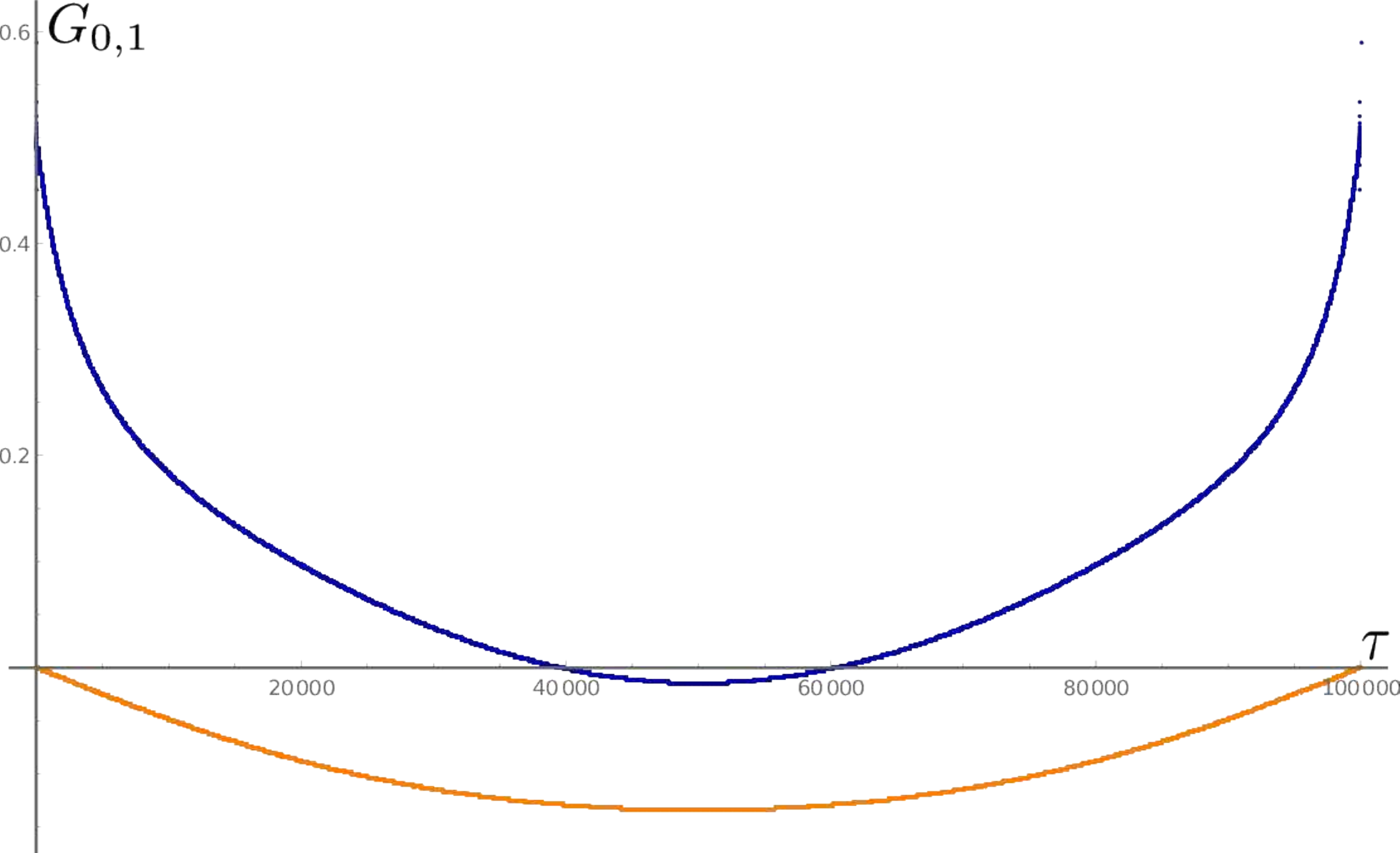}A.
\includegraphics[scale=0.16]{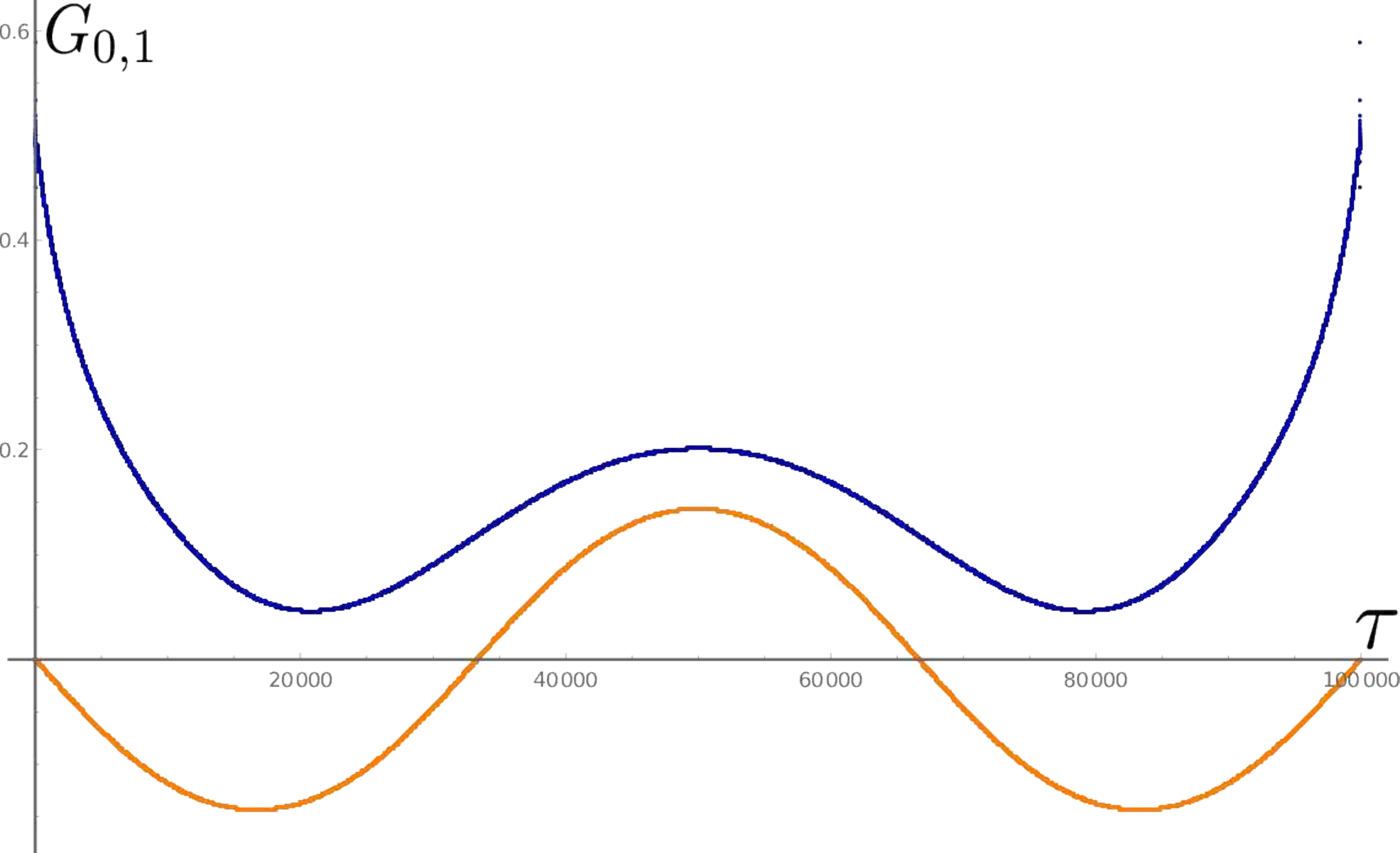}B.
\includegraphics[scale=0.16]{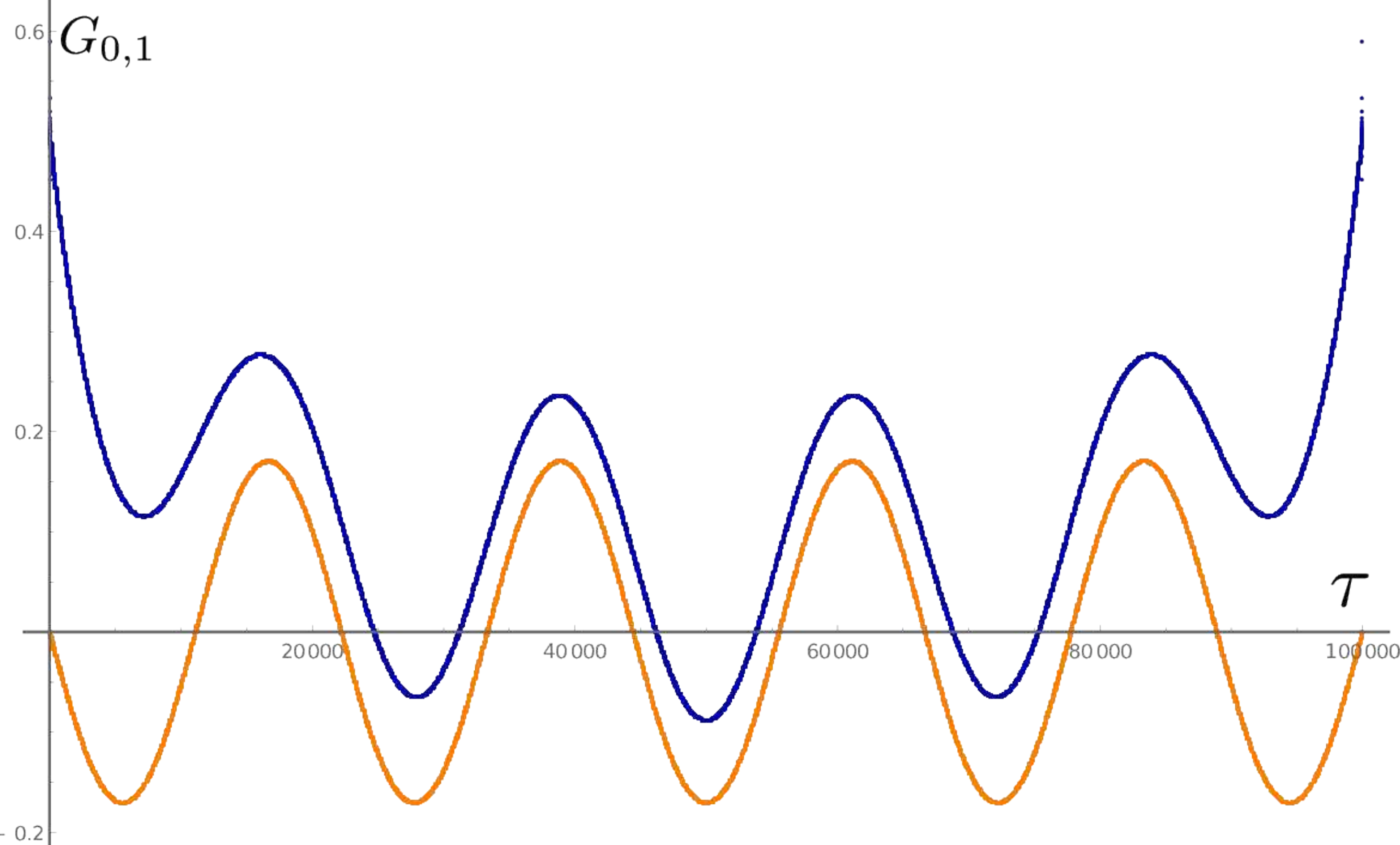}C.\\

\includegraphics[scale=0.17]{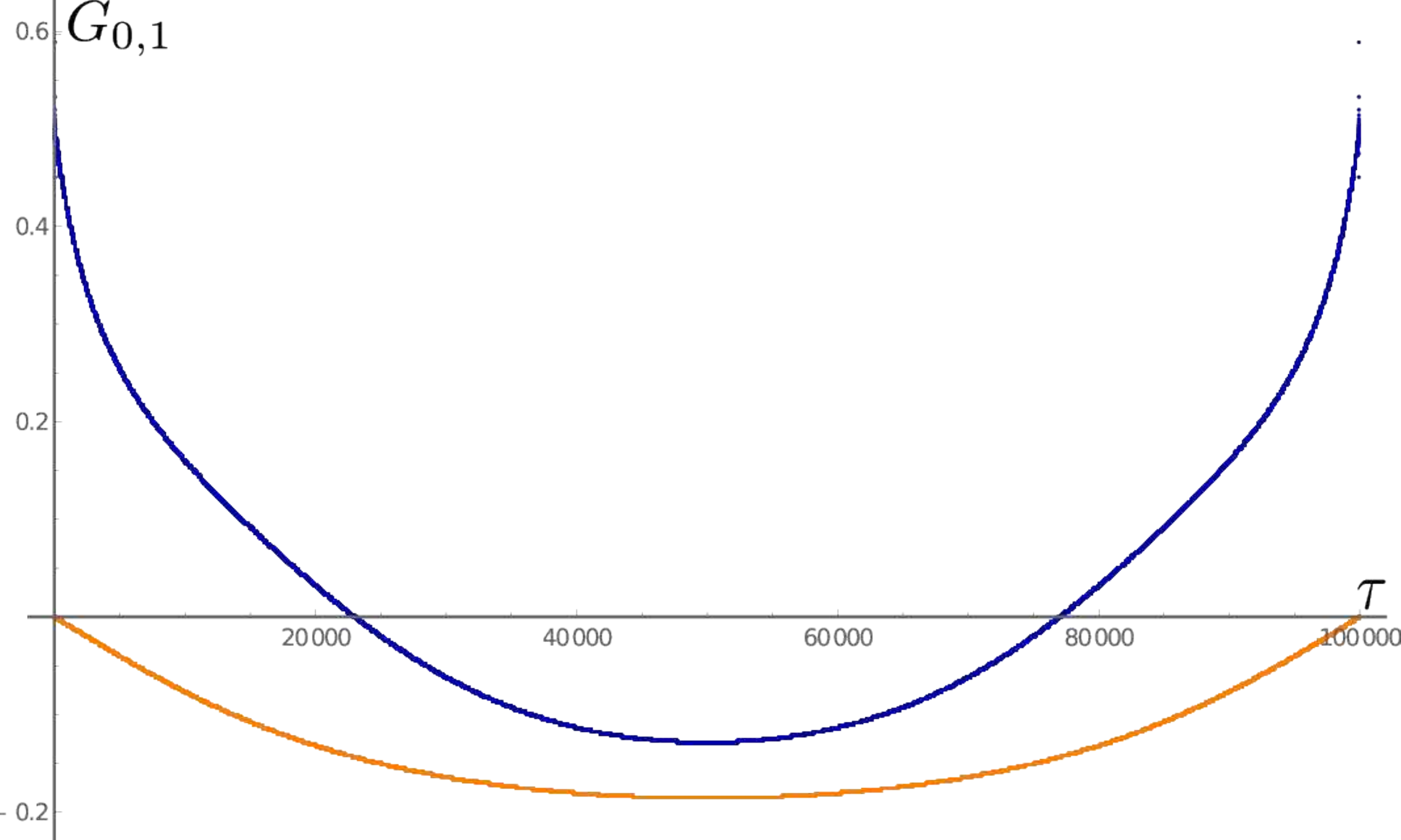}D.
\includegraphics[scale=0.17]{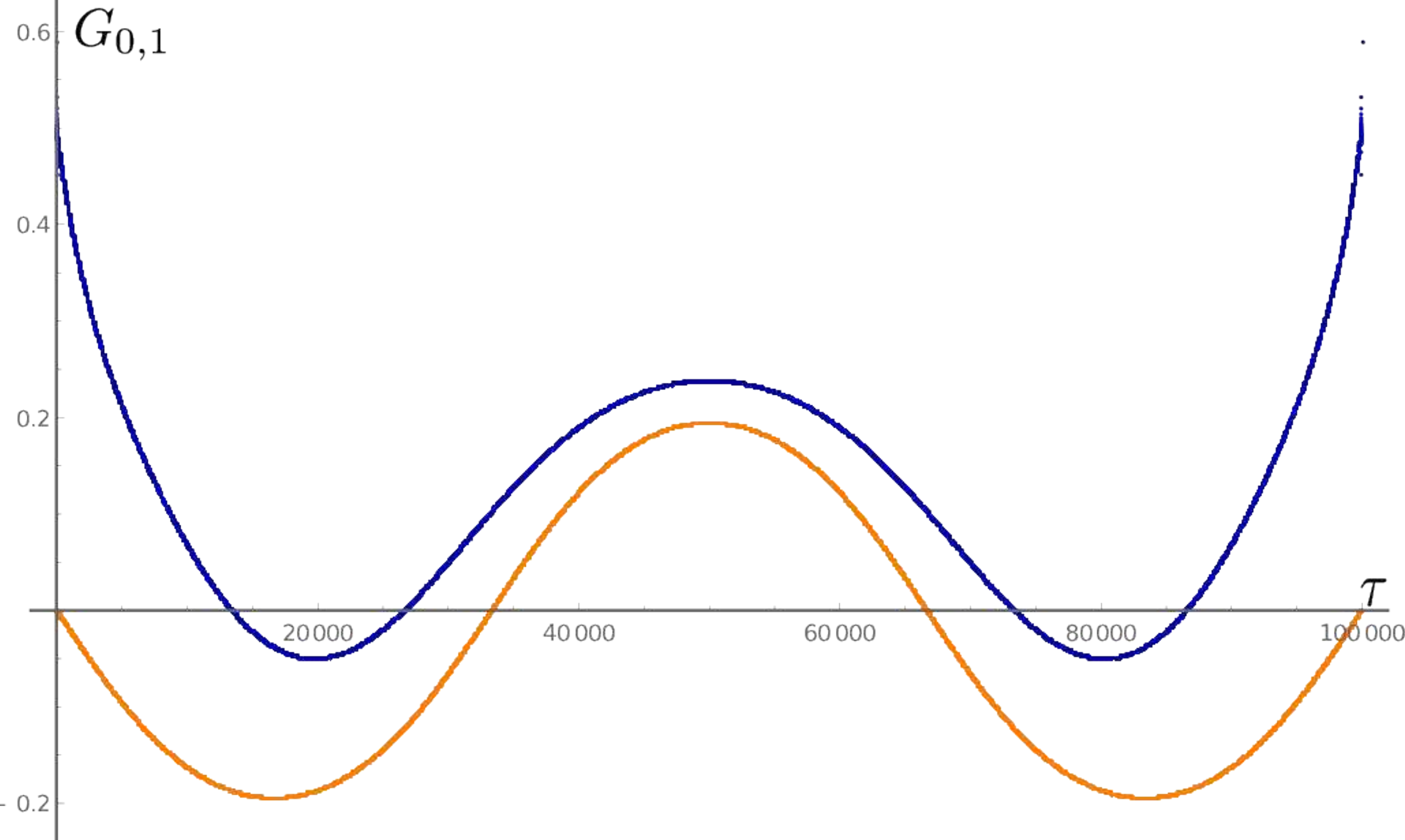}E.
\includegraphics[scale=0.17]{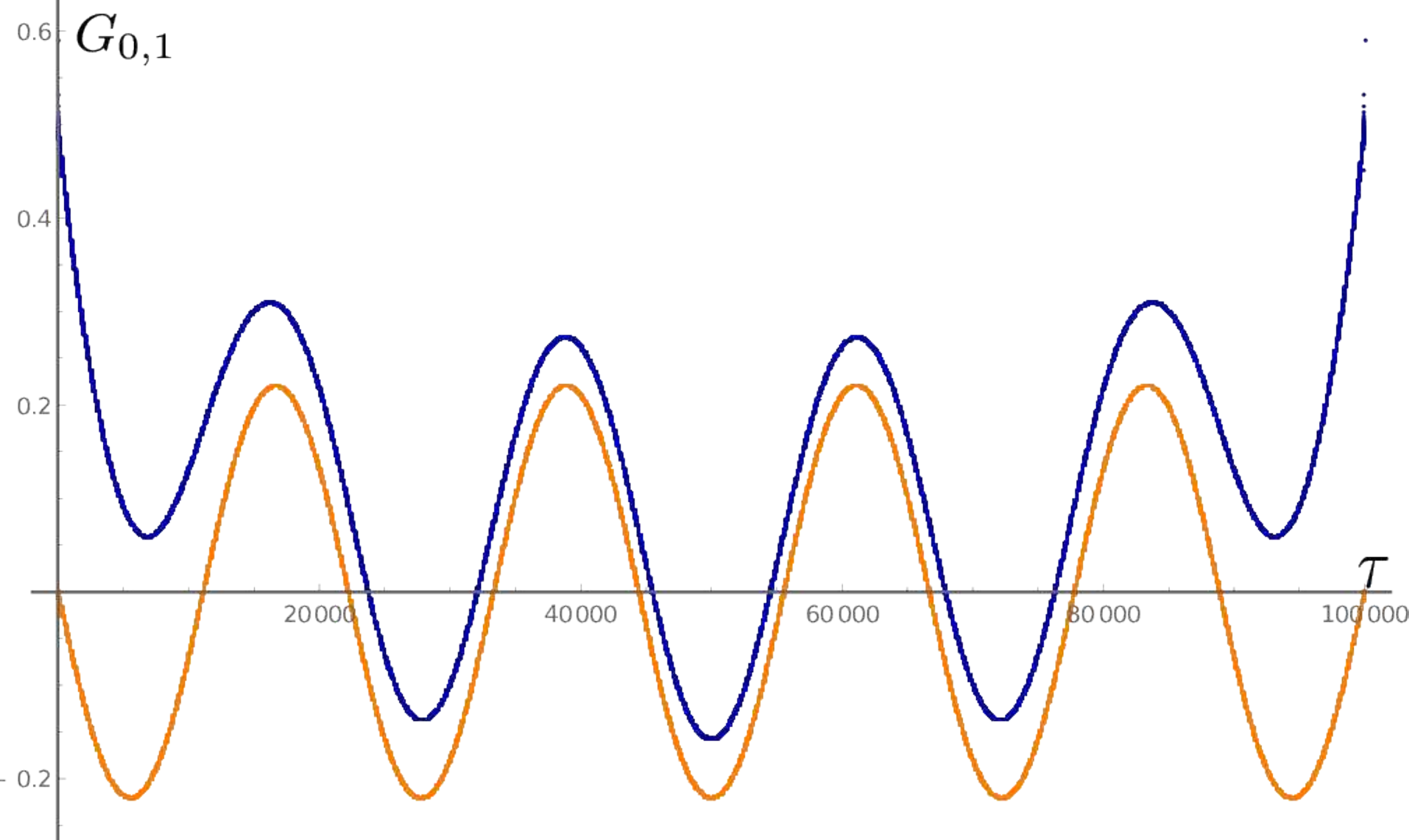}F.\\

\includegraphics[scale=0.17]{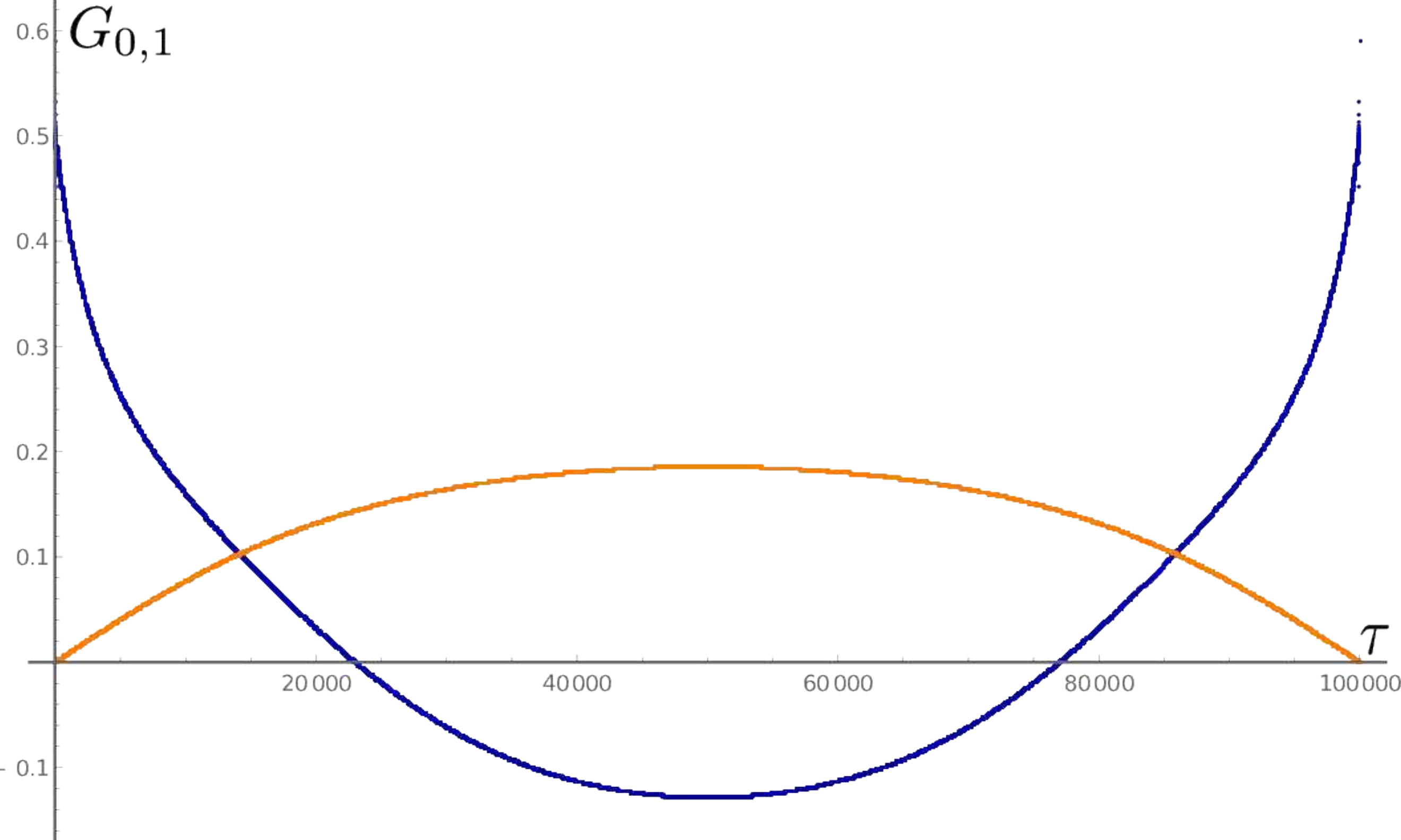}G.
\includegraphics[scale=0.17]{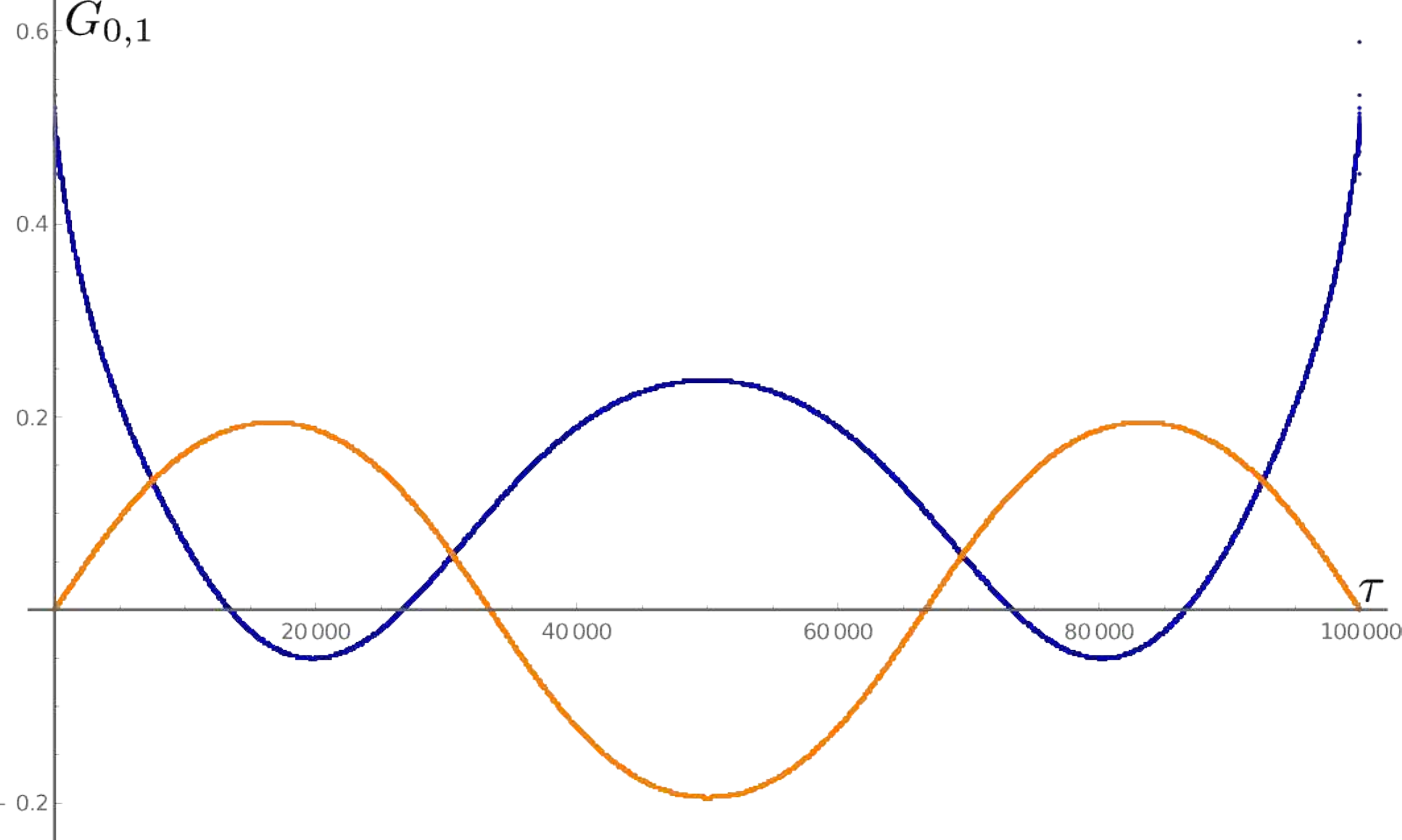}H.
\includegraphics[scale=0.17]{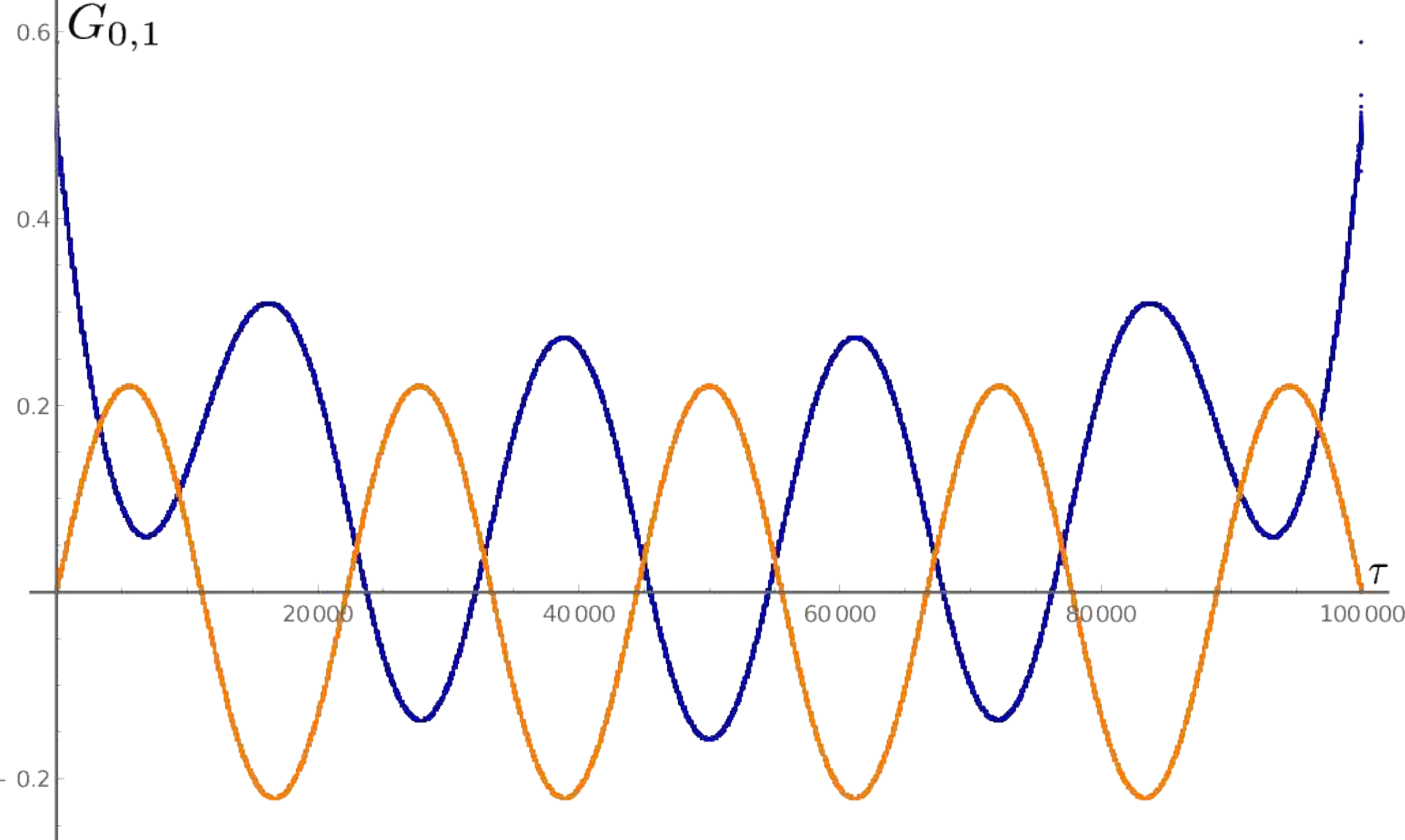}I.\\

\includegraphics[scale=0.17]{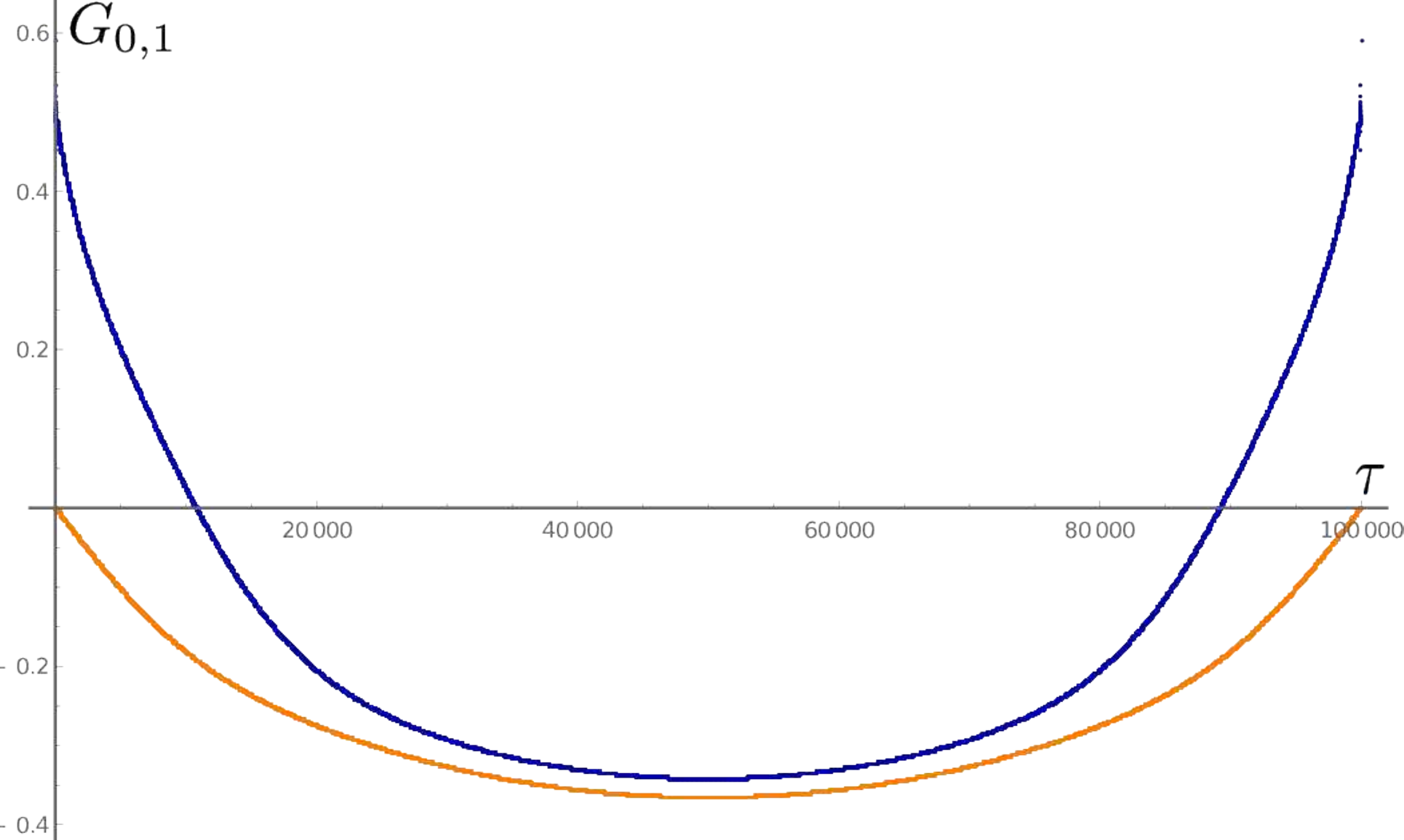}J.
\includegraphics[scale=0.17]{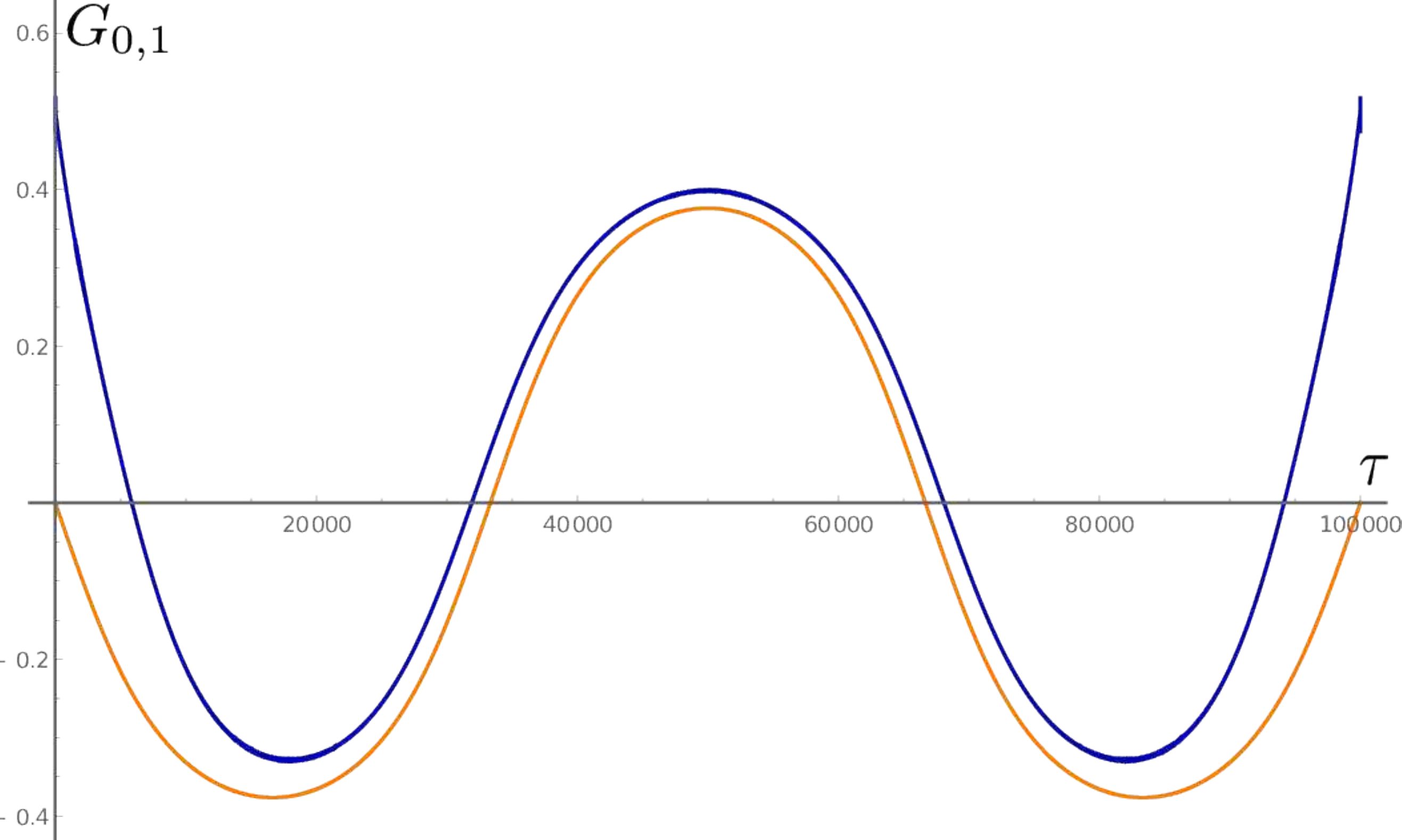}K.
\includegraphics[scale=0.17]{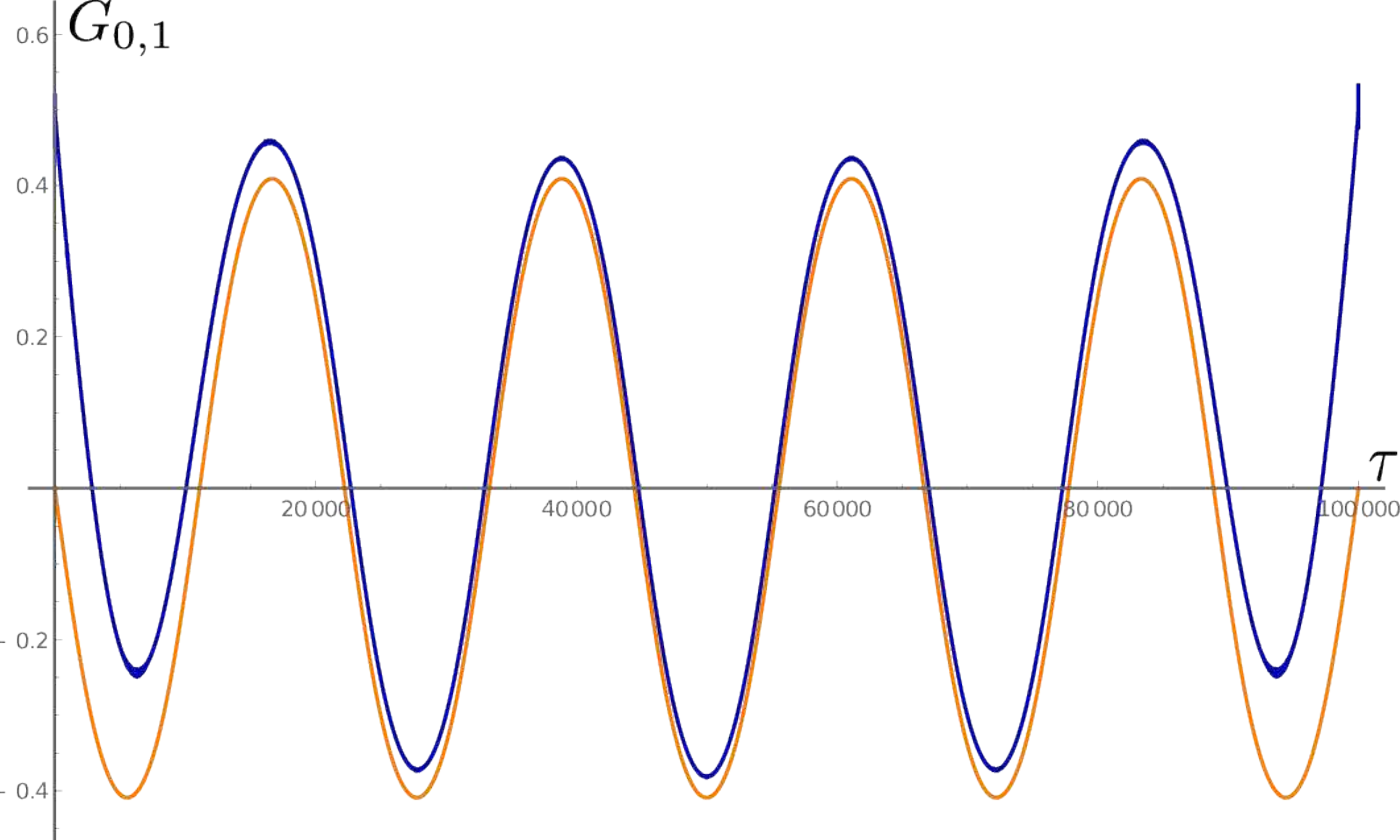}L.\\

$\,\,\,\,$\\
\caption{$G_0$ (blue curve) and $G_1$ (orange curve) as a function of Euclidean time on replica-nondiagonal solutions for $q=4$. The parameters are set at $\beta = 2\pi$ and $J = 10$. 
}
\label{fig:iterations-RS-M>0}
\end{figure}

\subsection{The results}

When studying the saddle points of the replica partition function at finite $M$, the obtained solutions indicate that for every $q=2$ solution, that we choose as initial condition as discussed above, there exists a solution of the $q=4$ model. This generalizes the observation, made in \cite{Cotler16} for the replica-diagonal solutions, to the replica-nondiagonal symmetric case. The solutions shown on Fig.\ref{fig:iterations-RS-M>0} are obtained by iterating from the following $q=2$ solutions:
\begin{itemize}
\item[A.] $M=4$. $G_{0,1}(\omega_0) = -G_{0,1}(\omega_{-1}) = G_{0,1}^{(3)}(\omega_0)$; $G_{0,1}(\omega_n) = G_{0,1}^{(1)}(\omega_n)$ for all other $n$.
\item[B.] $M=4$. $G_{0,1}(\omega_1) = -G_{0,1}(\omega_{-2}) = G_{0,1}^{(3)}(\omega_1)$; $G_0(\omega_n) = G_0^{(1)}(\omega_n)$ for all other $n$.
\item[C.] $M=4$. $G_{0,1}(\omega_4) = -G_{0,1}(\omega_{-5}) = G_{0,1}^{(3)}(\omega_4)$; $G_{0,1}(\omega_n) = G_{0,1}^{(1)}(\omega_n)$ for all other $n$.
\item[D.] $M=2$. $G_{0,1}(\omega_0) = -G_{0,1}(\omega_{-1}) = G_{0,1}^{(3)}(\omega_0)$; $G_{0,1}(\omega_n) = G_{0,1}^{(1)}(\omega_n)$ for all other $n$.
\item[E.] $M=2$. $G_{0,1}(\omega_1) = -G_{0,1}(\omega_{-2}) = G_{0,1}^{(3)}(\omega_1)$; $G_0(\omega_n) = G_0^{(1)}(\omega_n)$ for all other $n$.
\item[F.] $M=2$. $G_{0,1}(\omega_4) = -G_{0,1}(\omega_{-5}) = G_{0,1}^{(3)}(\omega_4)$; $G_{0,1}(\omega_n) = G_{0,1}^{(1)}(\omega_n)$ for all other $n$.
\item[G.] $M=2$. $G_{0,1}(\omega_0) = -G_{0,1}(\omega_{-1}) = G_{0,1}^{(4)}(\omega_0)$; $G_{0,1}(\omega_n) = G_{0,1}^{(1)}(\omega_n)$ for all other $n$.
\item[H.] $M=2$. $G_{0,1}(\omega_1) = -G_{0,1}(\omega_{-2}) = G_{0,1}^{(4)}(\omega_1)$; $G_0(\omega_n) = G_0^{(1)}(\omega_n)$ for all other $n$.
\item[I.] $M=0.5$. $G_{0,1}(\omega_4) = -G_{0,1}(\omega_{-5}) = G_{0,1}^{(4)}(\omega_4)$; $G_{0,1}(\omega_n) = G_{0,1}^{(1)}(\omega_n)$ for all other $n$.
\item[J.] $M=0.5$. $G_{0,1}(\omega_0) = -G_{0,1}(\omega_{-1}) = G_{0,1}^{(3)}(\omega_0)$; $G_{0,1}(\omega_n) = G_{0,1}^{(1)}(\omega_n)$ for all other $n$.
\item[K.] $M=0.5$. $G_{0,1}(\omega_1) = -G_{0,1}(\omega_{-2}) = G_{0,1}^{(3)}(\omega_1)$; $G_0(\omega_n) = G_0^{(1)}(\omega_n)$ for all other $n$.
\item[L.] $M=0.5$. $G_{0,1}(\omega_4) = -G_{0,1}(\omega_{-5}) = G_{0,1}^{(3)}(\omega_4)$; $G_{0,1}(\omega_n) = G_{0,1}^{(1)}(\omega_n)$ for all other $n$.
\end{itemize}

We have also studied this class of solutions in the limit $M \to 0$. For this purpose we add a third stage of iterations, where we keep $q$ fixed, but change the value of $M$ from some finite initial value to zero during iterations. Evaluating $|| \Delta G_{0,1} ||^2 $ and the left hand side of the equations of motion, we observe that the sequence of functions obtained this way fails to converge to any solution of the saddle point equations, other than the standard replica-diagonal saddle. When starting from a replica-nondiagonal solution at finite $M$, we observe that the iterated functions develop discontinuities as $M \to 0$. Thus, from our numerical evidence we conclude that the singular behavior in the $M \to 0$ limit, that we see in the analytic $q=2$ solutions, persists for nondiagonal solutions in the $q=4$ model. Therefore, the main result of our numerical investigation is that at finite replica number we get an infinite number of nontrivial replica-nondiagonal saddle points, whereas in zero replicas limit we do not obtain any replica-nondiagonal solutions of the exact saddle point equations. 

\subsection{On-shell action}

\begin{figure}[t]
\centering
\includegraphics[scale=0.4]{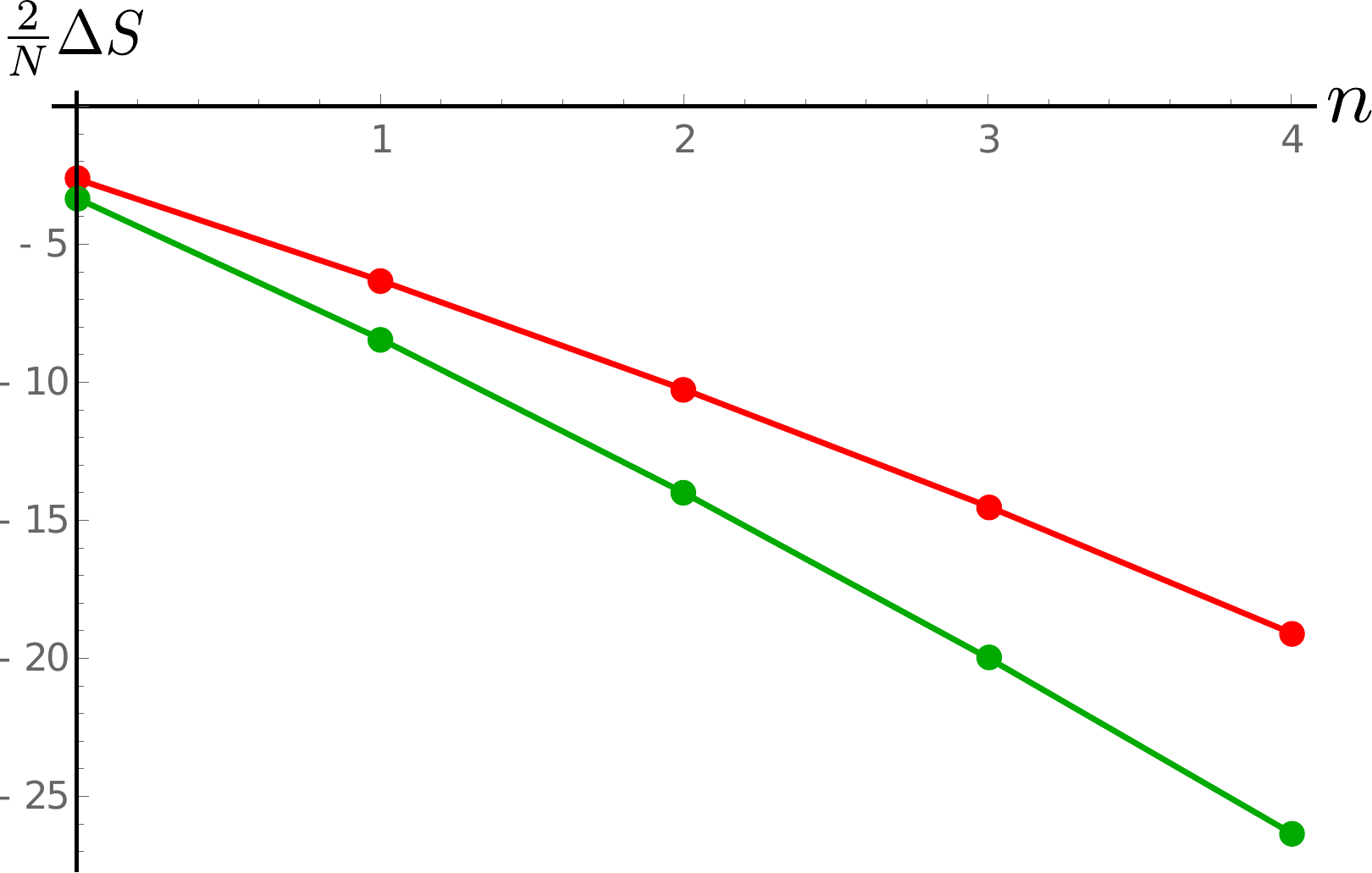}A.
\includegraphics[scale=0.4]{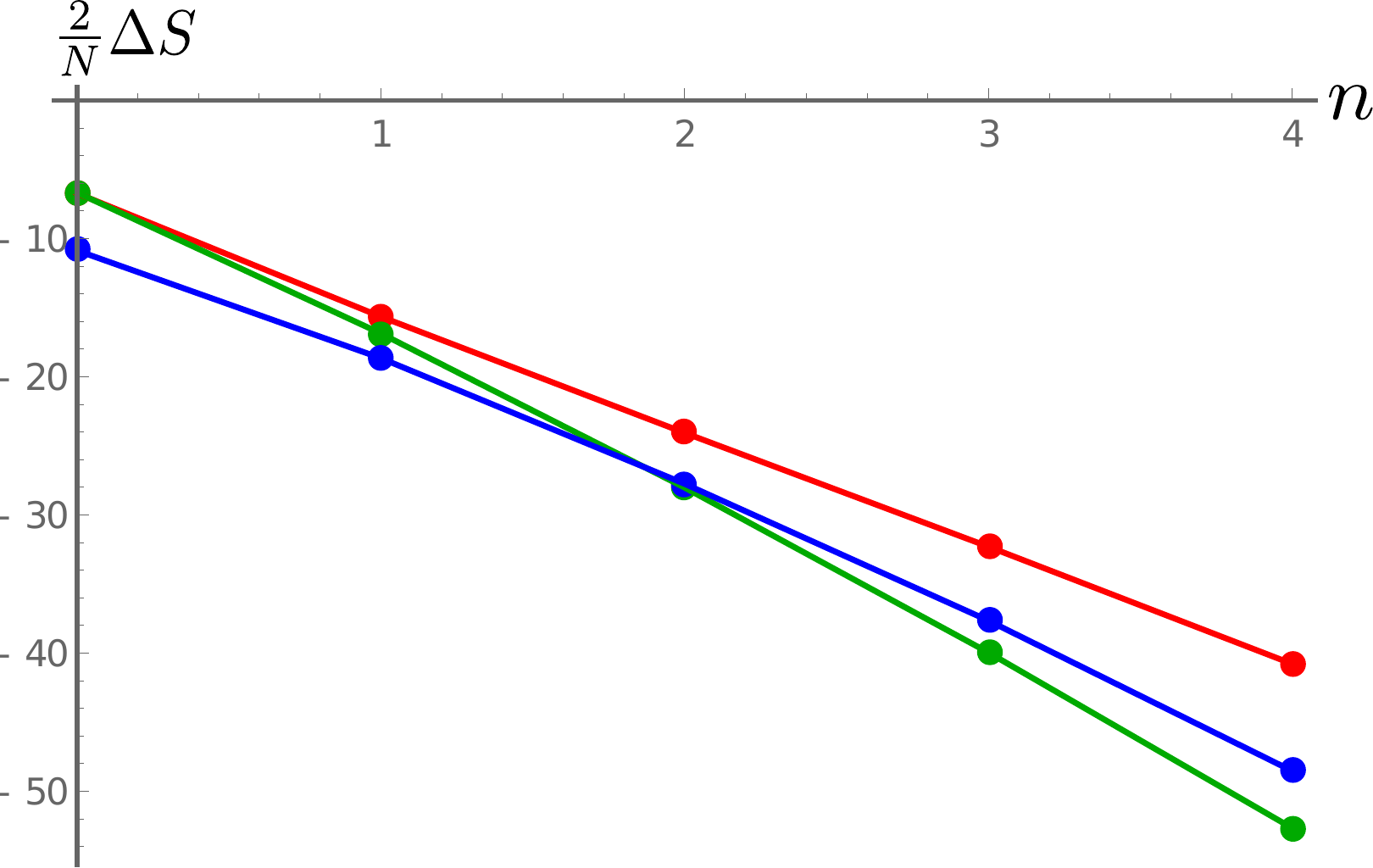}B.
\caption{On-shell action on the solutions as a function of $n$, which labels the Matsubara frequency pair, for which in the initial $q=2$ solution the $G^{(1)}$ was replaced by either $G^{(2)}$ (green), $G^{(3)}$ (red), or $G^{(4)}$ (blue). \textbf{A}: $M=2$. The solutions corresponding to $G^{(3)}$ and $G^{(4)}$ in this case have equal values of the action (shown by the red curve). \textbf{B}. $M=4$. Here $\beta J = 20 \pi$. 
}
\label{fig:action-M}
\end{figure}

We compute the on-shell action on the replica-nondiagonal solutions at finite $M$ defined by the formula (\ref{S_M}). it appears that in the $q=4$ case all the nondiagonal saddles that we have constructed for $M >1$ are subleading, similarly to the $q=2$ case. We study the difference 
\be
\frac{2}{N} \Delta S = \frac{2}{N} (S(\text{standard}) - S(\text{nondiagonal}))\,,\ee
where $S(\text{standard})$ is the value of the on-shell action on the standard replica-diagonal saddle (times $M$), and $S(\text{nondiagonal})$ is the value of the action on a particular replica-nondiagonal solutions. On Fig.\ref{fig:action-M} we plot $\frac{2}{N} \Delta S $ as a function of the label $n$ of a Matsubara mode, which was chosen to be other than $G^{(1)}$ in the $q=2$ trial function of the corresponding numerical solution. 
For the nondiagonal $G^{(2)}$, $G^{(3)}$ and $G^{(4)}$ branches $\Delta S$ decays with $n$ according to what appears to be a power law $n^{\alpha}$ with the exponent $\alpha$ determined by $M$ (as well as coupling). We also confirm the approximately linear decay on the diagonal $G^{(2)}$ branch, which was stated in \cite{Cotler16}. 

\subsection{Remark on the large replica number limit}

We conclude our discussion of exact replica-symmetric nondiagonal saddle points in $q=2$ and $q=4$ models by considering the large replica number limit, $M\to\infty$. Namely, we note that the solutions in this limit become replica-diagonal. In the $q=2$ case this is evident from the analytic solutions (\ref{G03})-(\ref{G14}). From these formulae we see that the nondiagonal terms $G_1^{(3)}$ and $G_1^{(4)}$ vanish in the limit $M\to \infty$, while the diagonal terms $G_0$ have a well defined non-vanishing limit. 
For the $q=4$ case the situation is similar. One can show that Eqs (\ref{SD-RS-1-diag})-(\ref{SD-RS-2}) are reduced in the limit
$M\to\infty$ to the equations for the diagonal case
\bea
 -i \omega G_0(\omega) - G_0(\omega) \Sigma_0(\omega) =1,\,\,\,\\
\Sigma_{0}(\tau, \tau') = J^2 G_{0}(\tau, \tau')^{q-1}.
\eea
So in principle one can treat the nondiagonal terms as the $1/M$-corrections. 

The diagonality of the solutions in the $M \to \infty$ limit leads to the following identity: 
\bea\label{FF}
\lim_{M\to\infty} \lim_{N\to\infty}\frac{1}{MN}\log\frac{\overline{Z^M}}{\overline{Z}^M}=0.
\eea
This can be interpreted as a sort of self-averaging relation for the partition function
\bea
\overline{Z^M}\,\approx\, \overline{Z}^M
\eea
for large $M$ and $N$.

\section{Replica-nondiagonal solutions at strong coupling}
\label{sec:Factorized}

We have found in the above section that the limit of zero replicas is singular in the replica-nondiagonal solutions of the exact saddle point equations, and thus there are no replica-nondiagonal solutions at $M=0$ in the class which we studied. Now we will consider the saddle point equations in the strong coupling (IR or conformal) limit, where we omit the time derivative. By neglecting the UV source term, which is diagonal in replicas, we thus allow for a much bigger set of possible solutions. The aim of the present and the subsequent sections is to construct solutions of the SYK model in the strong coupling limit in $M \to 0$ limit. 

In the strong coupling limit $\beta J \gg 1$ at finite replica number $M$ the saddle point equations (\ref{saddle-point-1}), (\ref{saddle-point-2}) which follow from \eqref{Zreplica} read:
\bea
 \int d\tau' G_{\alpha\beta}(\tau, \tau') \Sigma_{\beta\gamma}(\tau', \tau'') &=&  -\delta_{\alpha\gamma}\delta(\tau-\tau'')\,; \label{saddle-point-1-conformal}\\\nn\\
 \Sigma_{\alpha\beta}(\tau, \tau') &=& J^2 G_{\alpha\beta}(\tau, \tau')^{q-1}\,. \label{saddle-point-2-conformal}
\eea

The saddle point equations (\ref{saddle-point-1-conformal}),(\ref{saddle-point-2-conformal}) in the strong coupling limit are invariant under the transformations, which are induced by separate time reparametrization $\tau \to f_\alpha(\tau)$ in every replica \cite{Kamenev}:
\bea
&& G_{\alpha\beta}(\tau, \tau') = f'_\alpha(\tau)^\Delta f'_\beta(\tau')^\Delta G_{\alpha\beta}(f_\alpha(\tau), f_\beta(\tau'))\,;\label{reparam-replicas-G}\\
&& \Sigma_{\alpha\beta}(\tau, \tau') = f'_\alpha(\tau)^{1-\Delta} f'_\beta(\tau')^{1-\Delta} \Sigma_{\alpha\beta}(f_\alpha(\tau), f_\beta(\tau'))\,. \label{reparam-replicas-S}
\eea
In other words, in the general replica-nondiagonal case the emergent conformal symmetry extends to the group  $\text{diff} (S^1)^{\times M}$. 

\subsection{Separation of variables in the IR limit}

We are going to study the solutions of saddle point equations at strong coupling using the particular ansatz, where the time and replica dependencies are factorized \cite{Kamenev}:
\be
G_{\alpha\beta}(\tau, \tau') = g(\tau, \tau') P_{\alpha\beta}\,. \label{ansatz-Kamenev}
\ee

The main advantage of using the ansatz (\ref{ansatz-Kamenev}) is that we can construct analytic solutions. First, we substitute  $\Sigma_{\alpha\beta}$ from the second equation \eqref{saddle-point-2-conformal} into (\ref{saddle-point-1-conformal}), and we are left with the equation for $G$: 
\bea
J^2 \int d\tau' \sum_\beta G_{\alpha\beta}(\tau, \tau') G_{\beta\gamma}(\tau', \tau'')^{q-1}&=&-  \delta_{\alpha\gamma}\delta(\tau-\tau'')\,. \label{SPG1mm}
\eea
We substitute the factorized ansatz (\ref{ansatz-Kamenev}): 
\be
J^2 \int d\tau' g(\tau, \tau') g(\tau', \tau'')^{q-1} P_{\alpha\beta} P^{ q-1}_{\beta\gamma} = -\delta_{\alpha\gamma} \delta(\tau-\tau'')\,. \label{CG-eq-master}
\ee
The general solution to  \eqref{CG-eq-master} is given by a matrix $P_{\alpha\beta}$ and a function $g(\tau)$ 
that satisfy the matrix equation
\be
\sum_{\beta} P_{\alpha\beta} P_{\beta \gamma}^{q-1} = \delta_{\alpha\gamma} \mathcal{C}\,. 
\label{CG-eq-2}
\ee
and the integral equation
\be
J^2  \int  d\tau' g(\tau, \tau') g(\tau', \tau'')^{q-1} = -\frac1{\mathcal{C}}\,\delta(\tau-\tau'')\,. \label{CG-eq-1}
\ee
here $\mathcal{C}$ is an arbitrary non-zero constant. 
We take the antisymmetric conformal invariant solution of 
\eqref{CG-eq-1}
\be
g(\tau, \tau') = \frac{g_c(\tau,\tau')}{\mC^\Delta}= \frac{b}{\mC^\Delta}\left(\frac{\pi}{\beta}\right)^{2\Delta} \frac{\sgn (\tau-\tau')}{\left|\sin \frac{\pi}{\beta}(\tau-\tau')\right|^{2\Delta}}\,,  \label{CG-g-sol}
\ee
where $b$ is defined by (\ref{b}).  Due to the requirement $G_{\alpha\beta}(\tau, \tau') = -G_{\beta\alpha}(\tau', \tau)$ the matrix $P$ is symmetric $P_{\alpha\beta} = P_{\beta\alpha}$. 
We note that the equation (\ref{CG-eq-master}) has the scaling symmetry under transformations 
\be
g(\tau, \tau') \to \mu\ g(\tau, \tau')\,; \qquad P \to \mu^{-1} P\,.
\ee
Thus we have a scaling degree of freedom which can be fixed arbitrarily. There are two convenient ways two impose the scaling symmetry fixing condition.
\begin{itemize}
\item Normalization constraint. One can fix $\mC = \sum_{\beta} P_{\beta \gamma}^{q} =  1$. The off-diagonal part of the equation \eqref{CG-eq-2} reduces to
\bea
\sum_{\beta} P_{\beta\alpha} P_{\beta \gamma}^{q-1} &=& 0\,, \quad
 \alpha \neq \gamma\,. \label{CG-eq-2''}
\eea
This equation is treated on equal footing with the normalization constraint, and they together determine the matrix $P$.
\item Diagonal constraint. In the present work we instead impose the condition when we fix $P_{\alpha\alpha} = 1$. In this case one first solves the equation for non-diagonal components of $P$, which has the form Eq.(\ref{CG-eq-2''}), and then computes $\mC$ to completely determine the solution according to 
\be
\sum_{\beta} P^q_{\alpha\beta}  =  \mathcal{C}\,.
\label{Constraint}
\ee
This approach is more convenient for study of specific solutions, and we employ it throughout the paper. 
\end{itemize}
The previous considerations in fact mean that ultimately we deal with the normalized replica matrix, such that
\be\label{rescaling}
G_{\alpha\beta}(\tau)=g_c(\tau)\tilde{P}_{\alpha\beta}, \,\,\,\,\,\,\tilde{P}_{\alpha\beta}=\frac{1}{\mC ^{1/q}}P_{\alpha\beta},
\ee 
where $\tilde P_{\alpha\beta}$ is the normalized matrix
\be
\sum_{\beta} \tilde P^q_{\alpha\beta} =1\ee

 We can write the equation (\ref{CG-eq-2}) in the matrix form. To do that, we introduce the Hadamard matrix product: 
\be
(A\circ B)_{ij}=A_{ij} B_{ij} \quad \text{(no sum)}\,.\ee
The degree $(q-1)$ is then understood as the matrix degree with respect to the Hadamard multiplication, and we write the equation (\ref{CG-eq-2}) as follows: 
\be
P\cdot P^{ \circ (q-1)}= \mC I \label{CG-eq-matrix}\,,
\ee
where $P^{\circ r} =\underbrace{P\circ P\circ...P}_{r}$, and $I$ is the identity matrix. 

As a side remark, we note that the equation (\ref{CG-eq-matrix}) with $\mC = 1$ can be interpreted as the saddle point equation for the $0$-dimensional version of the SYK model \cite{Arefeva18}. Thus in principle one can treat the factorized ansatz (\ref{ansatz-Kamenev}) as a sort of dimensional reduction of the SYK model. 

We also note that the factorized ansatz (\ref{ansatz-Kamenev}) breaks the reparametrization symmetry (\ref{reparam-replicas-G})-(\ref{reparam-replicas-S}) symmetry down to a single copy of $\diff(S^1)$, which acts in the same way as in the replica-diagonal case, see eqs. (\ref{repG-RD}),(\ref{repS-RD}). The solution (\ref{CG-g-sol}) spontaneously breaks it down further to $SL(2, \RR)$. 
By acting with the (\ref{reparam-replicas-G})-(\ref{reparam-replicas-S}) transformations on (\ref{ansatz-Kamenev}), we can generate other analytic replica-nondiagonal solutions. The most general form of the solution which can be obtained this way is the follows: 
\be
G_{\alpha\beta}(\tau, \tau') = \mathcal{G}_{\alpha\beta}(\tau, \tau') \mathcal{P}_{\alpha\beta}(\tau, \tau')\,,
\ee
where 
\bea
&& \mathcal{G}_{\alpha\beta}(\tau, \tau') = g(f_\alpha(\tau), f_\beta(\tau'))\,;\\
&& \mathcal{P}_{\alpha\beta}(\tau, \tau') =  f'_\alpha(\tau)^\Delta f'_\beta(\tau')^\Delta P_{\alpha\beta} \quad \text{(no sum)}\,.
\eea
In particular, these transformations can alter time dependence of replica-nondiagonal components and thus lead to physically more interesting solutions. An example is considered in the section \ref{sec:Glass-like}.

\subsection{The approach}
\label{sec:Approach}

To solve the equations (\ref{CG-eq-2}), we use the Parisi ansatz \cite{Mezard91,ParisiBook} for the matrix $P$. For the definition and properties of Parisi matrices, see appendix \ref{sec:Parisi}. The Parisi matrix $P$ of the rank $l$ is characterized by the parameters $a_0, \dots, a_l$. The key properties of the Parisi matrix form that make it especially suitable for solving the equation (\ref{CG-eq-2}), are the following: 
\begin{itemize}
\item[1.] A Parisi matrix satisfies the constraint 
\be
\sum_{\beta} P^q_{\alpha\beta}  =  \mathcal{C}\, \quad \forall \alpha
\label{Constrain}
\ee
 identically, for any $q$ with some $\mC$. This is easy to see because every line and every column of a Parisi matrix contains all of its parameters, so the sum across every line and column is the same. 
\item[2.] The Parisi matrices form an algebra with respect to both direct matrix product and the Hadamard matrix product (the proof is presented in the section \ref{sec:ParisiAlgebra}). This guarantees the consistency of the matrix equation (\ref{CG-eq-matrix}) (see representation (\ref{repr})). 
\end{itemize}

For different possible configurations, we use the terminology analogous to the context of spin glass solutions: 
\begin{itemize}
\item $l=1$, $a_0 \neq 0$, $a_1 = 0$ - replica-diagonal solution
\item $l=1$, $a_0 \neq a_1 \neq 0$ - replica-symmetric solution
\item $l > 1$, $a_0 \neq a_1 \neq \dots \neq a_l$ - $(l-1)$-th step of replica symmetry breaking. At finite replica number the equations are easily solved in the complex domain for any rank $l$ of the Parisi ansatz. 
\end{itemize}

In the framework of the Parisi ansatz we rewrite the equations (\ref{CG-eq-2''}),(\ref{Constraint}) using the formulae (\ref{w0}),(\ref{wj}) with $b_j = a_j^{q-1}$:
\bea 
&& 0 = a_j a_0^{q-1}  + a_0 a^{q-1}_j + \sum_{i < j} (a_i a^{q-1}_j + a_ja^{q-1}_i) (m_{i+1} - m_i) -  m_{j} a^{q}_j +  \sum_{i > j-1}a^{q}_i (m_{i+1} - m_i).\nn\\ \label{jcomp}&& \\
&&\mC =  a^{q}_0 + \sum_{j=1}^{l} a_j^{q} (m_{j+1} - m_j) ; \label{0comp}
\eea

Our aim is to obtain and study solutions in the limit $M \to 0$. We approach the problem of finding the solutions as follows\footnote{Our approach is different from that in the spin glass studies (e.g. \cite{SK,Georges00,Mezard91,Anninos16,Hemmen78}), where it is conventional to derive the expression for the free energy in the $M \to 0$ limit first, then minimize it on the class of configurations restricted by the assumed ansatz. Instead of extremizing the action on the restricted class of configurations, we find extrema of the action, under the assumption that the equations of motion continue to define the saddle points after the zero replica limit.}. We will take the limit $M \to 0$ first, directly in the saddle point equation (\ref{CG-eq-matrix}) and in the on-shell action. Then we will find the solutions and calculate the (regularized) on-shell action on them. 

In general, the strong coupling limit of SYK contains UV divergences, and they will appear in the action, evaluated on an IR solution. Therefore, if one wants to understand the role of solutions in this limit when submerged into complete SYK model, one has to perform the renormalization, by fitting to the numerical solutions of the exact equations\footnote{A perturbative in $(\beta J)^{-1}$ approach in the leading order amounts to accounting for the reparametrization soft modes, as explained in \cite{MScomments,Kitaev17}.}. Because we were not able to find solutions of the exact saddle point equations in zero replicas limit as discussed in section \ref{sec:Numerics}, we expect that for any replica-nondiagonal solution in the IR limit this renormalization will bring the value of the on-shell action to be equal to the standard saddle value. Thus we do not expect that the nondiagonal solutions that we construct in $M=0$ case in the strong coupling limit make any contribution to the physical free energy in the complete SYK model. Nevertheless, we will be interested in calculating the regularized free energy in the strong coupling limit on nondiagonal solutions, and study its difference from the regularized free energy on the diagonal conformal solution in order to analyze the dominance of saddles in the leading order of the strong coupling expansion.

In the next subsection we will consider the solutions of the factorized form (\ref{ansatz-Kamenev}) in the simplest case, when $q=2$ and $M$ is non-zero integer. We employ the strategy, outlined above, in the one-step replica symmetry breaking case and discuss the corresponding solutions in detail in the next section \ref{sec:RSB1}. The rest of this section is focused on technical aspects of regularization and calculation of the on-shell action and of the limit $M \to 0$. 

\subsection{Factorized solutions in the $q=2$ model}
\label{sec:Factorized-q=2}

To provide some more motivation for consideration of solutions of the factorized form, let us consider the simplest example of replica-symmetric solutions in the $q=2$ model at finite $M$. In this case exact solutions are given by formulas (\ref{G03}),(\ref{G13}),(\ref{G04}),(\ref{G14}). We show in this subsection that in the IR limit these solutions have the form (\ref{ansatz-Kamenev}). This fact illustrates the connection between the exact saddle points (\ref{G01})-(\ref{G14}) and factorized solutions of the form (\ref{ansatz-Kamenev}) in the strong coupling limit of the $q=2$ model. 

First of all, let us discuss what we expect in the strong coupling limit. Note that if $q=2$, then in the frequency space the function $g(\omega)$ is piecewise constant, as can be seen from equation (\ref{gco}) by setting $\Delta = \frac12$: 
\be
g_c(\omega) \sim \sgn(\omega)\,.
\ee
That means that for $q=2$ the full solutions of the form (\ref{ansatz-Kamenev}) in the frequency space are also piecewise-constant:
\be
G_{\alpha\beta}(\omega) = C \sgn (\omega) P_{\alpha\beta}\,.\label{Gfactorized-q=2}
\ee
We want to check whether the exact replica-nondiagonal solutions, given by the formulae (\ref{G03}),(\ref{G13}),(\ref{G04}),(\ref{G14}), assume this form in the IR limit. To take the IR limit in these solutions, we set $\omega \to +0$\footnote{Note that this approach to taking the IR limit is not suitable for general $q$. In general one has to perform the strong coupling expansion and select the leading IR asymptotic (we elaborate on this in section \ref{sec:StrongCouplingExp}). However just setting $\omega \to +0$ works in $q=2$.}. We get 
\bea
G_{0}^{(3)}&=&\frac{ i\left(1-\frac2M\right)}{J}\,; \label{G03-IR}\\
G_{1}^{(3)}&=& -i\frac{2}{J M}
\,;\label{G13-IR}\\
G_{0}^{(4)}&=&\frac{-i\left(1-\frac2M\right)}{J}\,; \label{G04-IR}\\
G_{1}^{(4)}&=& i\frac{2}{J M}\,. \label{G14-IR}
\eea
If one takes the IR limit as $\omega \to -0$, then one obtains the same expressions with different overall signs because of the $\sgn$-functions in the solutions (\ref{G03}),(\ref{G13}),(\ref{G04}),(\ref{G14}). 

Thus, these solutions have the form 
\be
G_{\alpha\beta} = \frac{i}{J} \sgn(\omega) P_{\alpha\beta}\,,
\ee
where 
\bea
P_{\alpha\alpha} &=& \pm \left(1-\frac2M\right) \quad \forall \alpha\,;\\
P_{\alpha\beta} &=& \mp \frac{2}{M}\quad \forall \alpha \neq \beta\,,
\eea
is the replica-symmetric Parisi matrix. 

Having confirmed the general form (\ref{Gfactorized-q=2}), we now only have to check that the equation (\ref{CG-eq-2}) for the matrix $P$ is satisfied. The off-diagonal equation (\ref{jcomp}) in this case reads
\bea 
-2 \left(1-\frac2M\right) \frac{2}{M} + \left(\frac{2}{M}\right)^2 (M-2) =0\,.
\eea
It is easy to see that this is true identically for any $M \neq 0$. Now we can find the normalization constant $\mC$ from the equation (\ref{0comp}): 
\bea 
\mC&=&  \left(1-\frac2M\right)^2 +  \left(\frac{2}{M} \right)^2 (M-1)= 1\,.
\eea
Thus we have shown that the solutions (\ref{G03})-(\ref{G14}) reduce to solutions of the form (\ref{ansatz-Kamenev}) in the IR limit. 

\subsection{Treatment of the Pfaffian term}

\subsubsection{Pfaffian factorization}

To obtain the expression of on-shell action for factorized solutions from the partition function (\ref{Zreplica}), we first need to evaluate the Pfaffian term. In the IR limit $\dd_\tau \to 0$ the factorized ansatz for replica-nondiagonal solutions allows for the factorization of the Pfaffian. To prove it, we can write
\be
\Pf (-\hat{\Sigma}_{\alpha\beta}) = \int D\chi \exp \left(-\frac12\int d\tau d\tau' \chi_\alpha(\tau) \Sigma_{\alpha\beta}(\tau, \tau') \chi_\beta(\tau')\right)\,. \label{PfS}
\ee
We can diagonalize the $\Sigma_{\alpha\beta}(\tau, \tau')$ by making the transition to the frequency space. We write 
\bea
\chi(\tau) & =&\frac1\beta \sum_{n \in \ZZ} \tilde{\chi}(\omega_n) \e^{-i \omega_n \tau}\,;\\
\Sigma_{\alpha\beta}(\tau_1, \tau_2) & =&  \frac1\beta\sum_{n \in \ZZ} \tilde{\Sigma}_{\alpha\beta} (\omega_n) \e^{-i \omega (\tau_1-\tau_2)}\,, 
\eea
where $\omega_n$ are Matsubara frequencies defined in (\ref{Matsubara}). Then the quadratic form reads 
\bea
\frac12\int d\tau_1 d\tau_2 \chi_\alpha(\tau_1) \Sigma_{\alpha\beta}(\tau_1, \tau_2) \chi_\beta (\tau_2) = \frac1\beta\sum_{n \in \ZZ_+} \tilde{\chi}_\alpha (-\omega_n) \tilde{\Sigma}_{\alpha\beta}(\omega_n) \tilde{\chi}_\beta(\omega_n)\,.\label{chiSchi}
\eea
Note that $\bar{\tilde{\chi}}_\alpha (\omega_n)=\tilde{\chi}_\alpha (-\omega_n)  $. 

On the factorized solution (\ref{ansatz-Kamenev}), we write
\be
\Sigma_{\alpha\beta}(\tau, \tau') = \Sigma_c(\tau, \tau') \mC^{\Delta-1} P^{q-1}_{\alpha\beta} \Rightarrow \tS_{\alpha\beta}(\omega_n) = \tS_c(\omega_n) \mC^{\Delta-1} P^{q-1}_{\alpha\beta}\,,
\ee
where $\Sigma_c(\tau, \tau') = J^2 g_c(\tau, \tau')^{q-1}$, and $g_c$ is the conformal propagator given by (\ref{Gconf}). Substituting this and (\ref{chiSchi}) into (\ref{PfS}), we evaluate the path integral as 
\bea\label{FPf}
&& \int \prod_{n\in \ZZ_+} \prod_\alpha d \tilde{\chi}_\alpha(\omega_n)
d \bar{\tilde{\chi}}_\alpha(\omega_n) \exp\left(-\frac1\beta \sum_{n\in \ZZ_+} \bar{\tilde{\chi}}_\alpha(\omega_n) \tS_c(\omega_n) \mC^{\Delta-1} P_{\alpha\beta}^{q-1} \tilde{\chi}_\beta(\omega_n)\right)\\ 
&& = \prod_{n\in \ZZ_+} \left[ \left(- \frac1\beta\tS_c (\omega_n)\right)^{M} \det(\mC^{\Delta-1} P_{\alpha\beta}^{q-1})\right] = \prod_{n \in \ZZ_+}  \left(- \frac1\beta\tS_c(\omega_n)\right)^{M} \times \prod_{n \in \ZZ} \det(\mC^{\Delta-1} P_{\alpha\beta}^{q-1})^{1/2}\,.\nn\\ \label{FPf2}
\eea

The factor depending on $\tS$ is the same as in the replica-diagonal case. The second factor is the contribution from replicas, it is an infinite degree of the determinant of a Parisi matrix. The latter is calculated in the $M \to 0$ limit in section (\ref{sec:ParisiDet}), see eq.(\ref{logDet}). To calculate its contribution in the action, we only need to introduce an appropriate regularization by introducing a cutoff at some large $n$ 
in the product.

\subsubsection{Regularization} \label{Regularization}
To finally separate out the contribution of the replica matrix in the Pfaffian, we need to regularize it. In the present work we use a  direct regularization, which is a hard cutoff in the frequency space such that the validity of the strong coupling regime is preserved: 
\be
\label{omega-J}
|\omega_n| \leq J\,.
\ee
This corresponds to the restriction on number of Matsubara modes
\be
\left| n + \frac12 \right| \leq \frac{\beta J}{2\pi}\,.
\ee
and corresponds to  the cutoff  of $\dim{ \mathcal{H}}_f$, where $\mathcal{H}_f$ is a finite dimension subspace of the infinite dimension space  $\mathcal{H}$ of functions on the circle,
\be
 \dim \mathcal{H}_f\equiv d_f\,\simeq \frac{\beta J}{\pi}\,.
\ee
 In this regularization we can write
\be
\prod_{n \in \ZZ} \det(\mC^{\Delta-1} P_{\alpha\beta}^{q-1})^{1/2} = \det(\mC^{\Delta-1} P_{\alpha\beta}^{q-1})^{d_f/2}\,,
\ee
and 
\be
\Tr_{\mathcal{H}_f}\, \mathbf{1}=d_f\,. \label{Lambda}
\ee

The restriction \eqref{omega-J} also supports the validity of the strong coupling, or IR, regime. Indeed,  the contribution of the free propagator (the $\omega$-term) in the denominator of the LHS of \eqref{RD-eq-1} is suppressed as compared to $\Sigma(\omega)$ that  at large $\beta $ can be estimated as 
\be
\Sigma(\omega_n)\sim J^{2\Delta}\omega_n^{1-2\Delta}\ee
and 
$
\omega_n<\Sigma(\omega_n)$
corresponds to the  bound \eqref{omega-J}.

Another  regularization was discussed in \cite{Gurau17}, it  is essentially equivalent to the exponential cutoff in frequencies.

\subsection{
On-shell action  on factorized solutions}
We start with the on-shell action for the partition function (\ref{Zreplica}) at finite $M$ in the strong coupling limit,
\be
\frac{2}{N} S_{M}= - \Tr \log \left(- \hat{\Sigma}_{\alpha\beta}\right) +\int_0^\beta\int_0^\beta d\tau_1 d \tau_2 \sum_{\alpha,\beta} \left(G_{\alpha\beta}(\tau_1, \tau_2) \Sigma_{\alpha\beta}(\tau_1, \tau_2)- \frac{J^2}{q}G_{\alpha\beta}(\tau_1, \tau_2)^q\right)\Big|_{\text{on-shell}}\,.
\ee
We use the saddle point equations (\ref{saddle-point-1-conformal}),(\ref{saddle-point-2-conformal}) and substitute the factorized ansatz (\ref{ansatz-Kamenev}). For the action we get  
\bea
\frac{2}{N} S_M &=& -\Tr \log[- \hat{\Sigma}_{\alpha\beta}]  +\left(1-\frac1q \right)J^2 \int_0^\beta\int_0^\beta d\tau_1 d \tau_2 \sum_{\alpha,\beta} G_{\alpha\beta}(\tau_1, \tau_2)^{q}\nn.
\eea
Substituting for the solution 
\bea G_{\alpha\beta}(\tau, \tau') &=& g_c (\tau, \tau') \frac{1}{\mC^\Delta} P_{\alpha\beta}\,,\\
\Sigma_{\alpha\beta}(\tau, \tau') &=& \Sigma_c(\tau, \tau') \mC^{\Delta-1} P_{\alpha\beta}^{q-1} = J^2 g_c(\tau, \tau')^{q-1} \mC^{\Delta-1} P_{\alpha\beta}^{q-1}\,,\eea
we get
\bea
\frac{2}{N} S_M= -\log\Det[- \mC^{\Delta-1} P^{  (q-1)}_{\alpha\beta}\hat{\Sigma}_c] +\frac{q-1}{q \mC}J^2\sum_{\alpha,\beta} P^{ q}_{\alpha\beta} \int_0^\beta\int_0^\beta d\tau_1 d \tau_2 g_c (\tau_1, \tau_2)^{q}\nn
\eea
Using the factorization property of the $\det$ and the identity 
\be\sum_{\alpha,\beta} P^{ \circ q}_{\alpha\beta}=\mC M\,,\ee
we obtain
\bea
\frac{2}{N} S_M&=&- M\log\Det[- \hat{\Sigma}_c] +d_f\left(1-\frac{1}{q}\right)J^2 M -d_f \log\det[\mC^{\Delta-1} P^{ \circ (q-1)}]\nn \\
&=& M(\fs_{1,c}+\fs_{2,c} +\fs_{3}) \label{Son-shell}
\eea 
where $\fs_{1,c}$ and $\fs_{2,c}$ are defined in \eqref{s1c} and \eqref{s2c} respectively. Note that we use the same regularization in the polynomial term (which amounts to (\ref{Lambda}), as discussed above for the Pfaffian term. The new term unique to replica-nondiagonal factorized solutions is introduced:
\bea\label{fs3}
\fs_{3}=-\frac{d_f}{M}\log\det[\mC^{\frac1q-1} P^{ \circ (q-1)}] \,.
\eea
Note that it is proportional to the UV cutoff parameter $d_f$, which is defined in (\ref{Lambda}). 

\subsection{The $M \to 0$ limit}
\subsubsection{Replica symmetric case}
\label{sec:RS}
Let us first consider the simplest example of a Parisi matrix to illustrate our approach for taking the zero replicas limit. We consider the Parisi matrix $P$ of the first level (as explained in appendix \ref{sec:Parisi}), or the replica symmetric ansatz. Let us take $P_{\alpha\alpha}=a,\,\alpha =1,2,...,M$ and $P_{\alpha\beta}=A$ if $\alpha\neq \beta$.
Here $a$ and $A$ are two generically complex-valued numbers. We have
\bea
P=(P_{\alpha\beta})&=&
\left(
\begin{array}{cccc}
  a& A&A &.\,\,\,.\,\,\,.   \\
  A& a  &A&.\,\,\,.\,\,\,.  \\
  A&   A&a&.\,\,\,.\,\,\,. \\
  .&.&.&.\,\,\,.\,\,\,. 
\end{array}
\right)\,,\,\,\,\,\,\,\,\,\,\,
  P^{ \circ (q-1)}= P_{\alpha\beta}^{q-1}=\left(
\begin{array}{cccc}
  a^{q-1}& A^{q-1}&A^{q-1} &.\,\,\,.\,\,\,.   \\
  A^{q-1}& a^{q-1}  &A^{q-1}&.\,\,\,.\,\,\,.   \\
  A^{q-1}&   A^{q-1}&a^{q-1}&.\,\,\,.\,\,\,. \\
  .&.&.&.\,\,\,.\,\,\,.  
\end{array}
\right) \label{P-RS}
\eea
Note that from \eqref{detP}
\bea
\det(P^{ \circ (q-1)})=(a^{q-1}-A^{q-1})^{M-1}\Big(a^{q-1}+(M-1)A^{q-1}\Big)\,.\eea
The limit $M \to 0$ in this case is taken straightforwardly: 
\bea
\label{M0det}
\lim_{M\to 0}\frac1M\log\det(P^{ \circ (q-1)})=\log (a^{q-1}-A^{q-1})+\frac{A^{q-1}}{a^{q-1}-A^{q-1}}\eea
Equations \eqref{CG-eq-2''} and \eqref{Constraint} read
\bea 
aA^{q-1}+Aa^{q-1} +A^q(M-2)=0, \label{Eq1-M}
\eea
\bea 
\mC&=& a^q+A^q(M-1) \label{Eq2-M}
\eea
and in the limit $M\to 0$ take the form
\bea \label{Eq1}
 aA^{q-1}+Aa^{q-1} -2A^q&=&0,
\\\label{Eq2}
a^q-A^q &=& \mC\,.
\eea
\begin{figure}[t]
\centering
\includegraphics[scale=0.3]{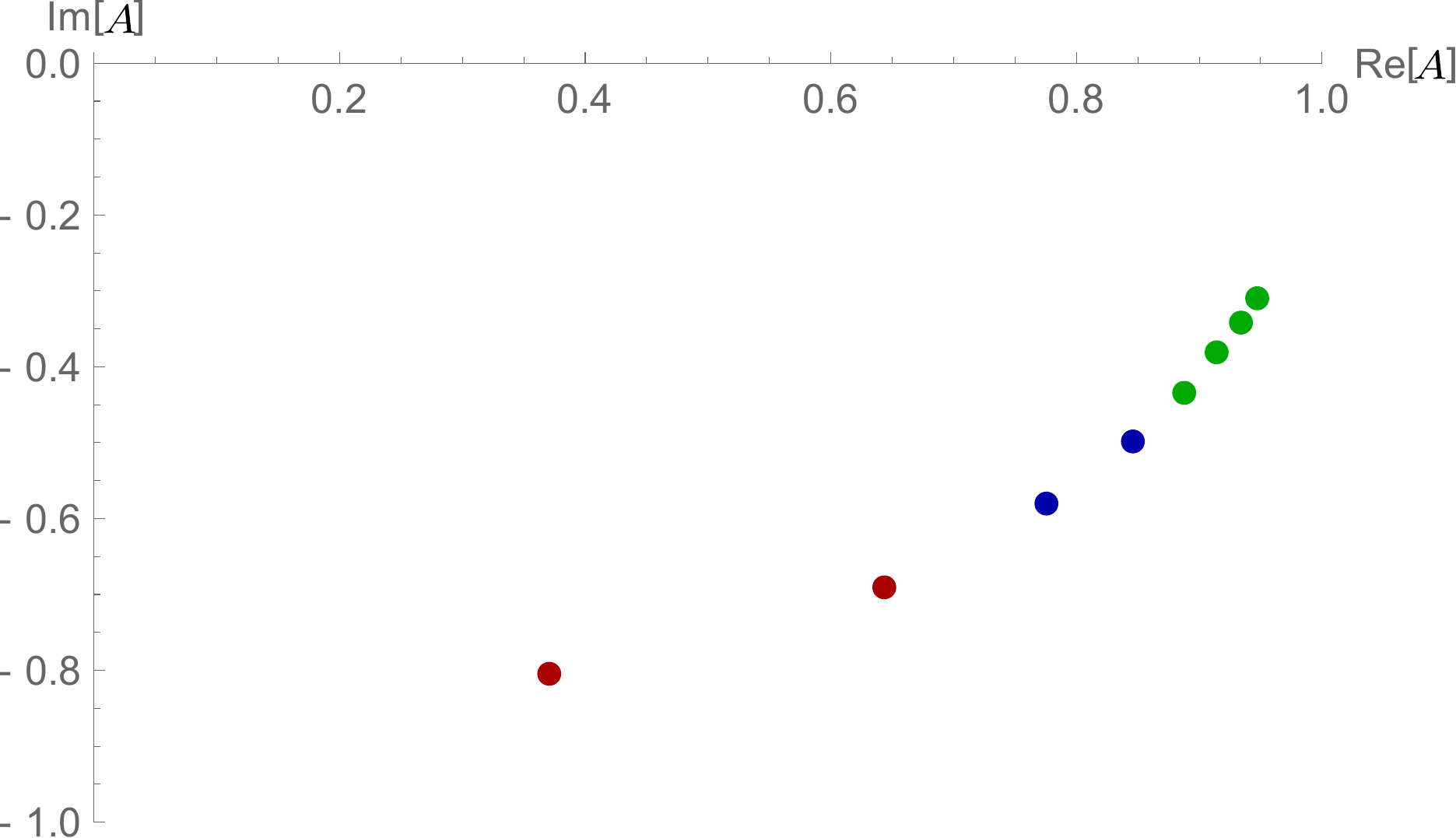}A.
\includegraphics[scale=0.3]{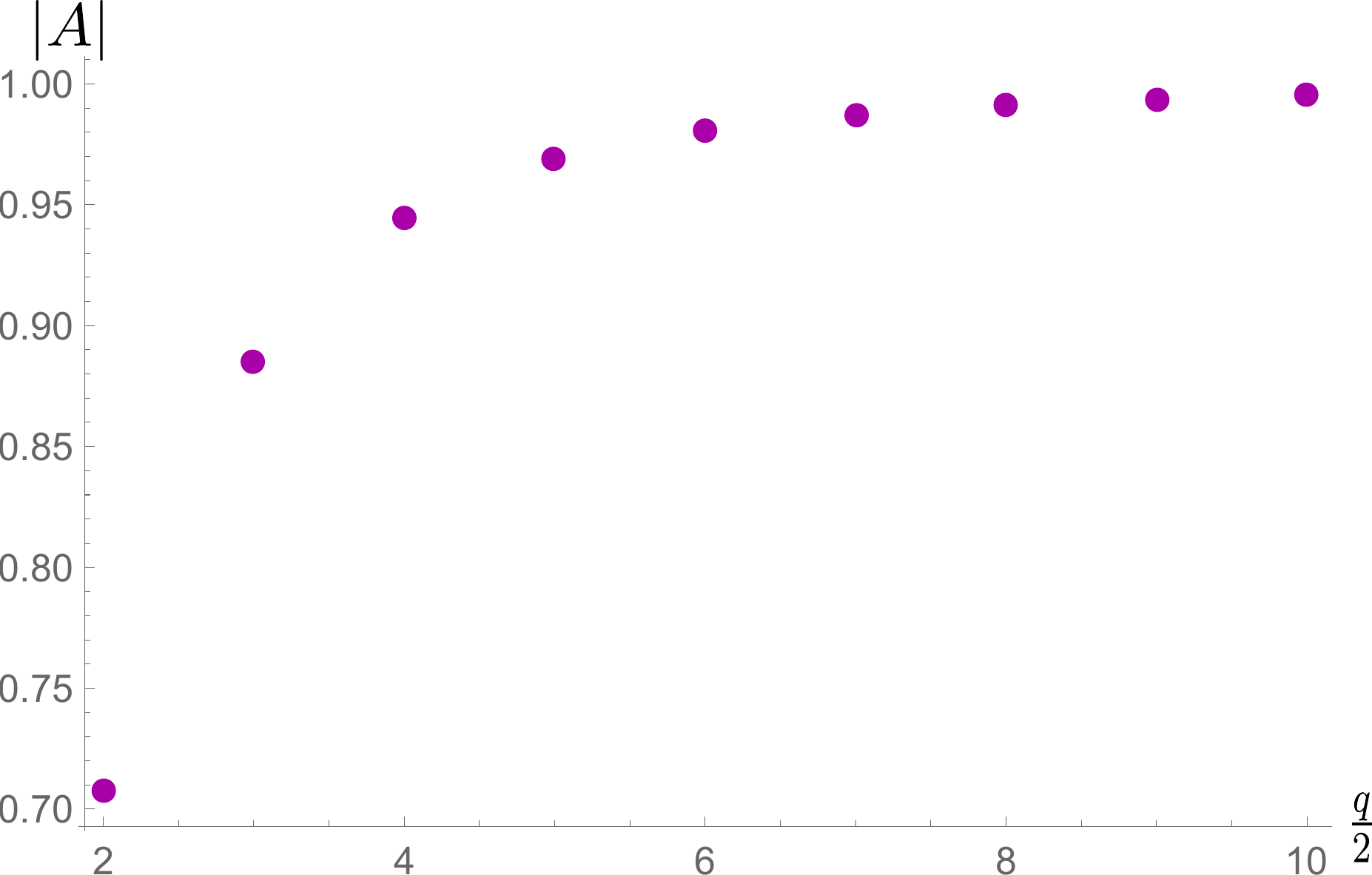}B.
\caption{{\bf A.} Locations of the dominant root of \eqref{Eq1a} for $q=4,6,...20$ (14,16,18,20 - green).
{\bf B} Absolute value of the dominant root of \eqref{Eq1a} for $q=4,6,...20$.}
\label{fig:XM0}
\end{figure}
Fixing the scaling freedom by setting $a=1$,we get the final equations for the $M=0$ case: 
\bea 
\label{Eq1a}
 A^{q-1}+A -2A^q&=&0\,,
\\ \label{Eq2a}
1-A^q &=& \mC \,.
\eea
Now we compute the limit of $\fs_3$ at $M\to 0$. We have
\bea
\label{Egreen-q}
-\lim_{M\to 0}\frac1M\log\det(\mC^{\frac{1}{q}-1}P^{ \circ (q-1)})
=\frac{q-1}{q}  \log (1-A^q) -    \log (1-A^{q-1})-\frac{A^{q-1}}{1-A^{q-1}}  \eea
For $q=2(n-1)$ the equation (\ref{Eq1a}) has $n$ pairs of conjugated roots $(b^{(i)}, \bar{b}^{(i)})$. Each such pair gives the contribution to the regularized free energy, determined by the real part of $\fs_3$, see discussion below in sec. \ref{sec:DeltaF} for more details. Taking into account that the cutoff is $d_f \simeq \beta J$, we can compute the normalized difference between the regularized free energies of the replica-diagonal and nondiagonal solutions:
\bea
\label{Egreen-q-b}
f\equiv\frac{2}{NJ}\Delta F& =&\frac{2}{NJ}(F_{RND}-F_{RD})\\&=&-\lim_{M\to 0} \Re\frac1M\log\det(C^{-1+\frac{1}{q}}P^{ \circ (q-1)})]]\\& =&
\left(\frac{q-1}{q}  \log |1-b^q| -    \log |1-b^{q-1}|-\Re\frac{b^{q-1}}{1-b^{q-1}} \right)\,.
\eea
In the $q=4$ case, $\log|1-b^4|=0$ since  $|1-b^4|=1$.  Then $|1-b^3|^2=1/2$ and therefore $\log |1-b^3|=- \frac{1}{2}\log 2.$. Finally, $\Re \frac{b^3}{1-b^3}=\frac{3}{8}$, and we have 
 \be 
f_4^{(1)}= -\frac38 +\frac{\log 2}{2}=-0.028\,, \label{f_4}
\ee
It is negative, so the value of the regularized free energy on  this solution is below the replica diagonal case.

We present pairs of nontrivial solutions $A=(b_q^{(i)},\bar b_q^{(i)}),$  $i=1,...\frac{q}2-1$ to \eqref{Eq1} for values $q=4,6,8,10$ and the corresponding values of $f_q$ in the table \ref{Table}. 
\begin{table}[t]
\begin{center}
\begin{tabular}{|l | l  | l| }
\hline 
q &  $b_q$ &  ${f_q}$\\ 
\hline 
4 & $ b_4^{(1)} =\left(-1 \pm i \sqrt{7}\right)/4$ & $f_4^{(1)}=-0.028$   \\ 
\hline 
\multirow{2}{*}{6} & $b_6^{(1)}=0.621 \pm 0.502 i;$ & $f_6^{(1)}=0.0652; $\\
 &$b_6^{(2)}=0.371 \pm 0.803 i $  & $f_6^{(2)}=-0.301$ \\ 
\hline 
\multirow{3}{*}{8} & $b_8^{(1)}=-0.757 \pm 0.388 i; $ & $f_8^{(1)}=0.104;$ \\
& $b_8^{(2)}=-0.137 \pm 0.869i;$ & $f_8^{(2)}=-0.080;$ \\
 & $b_8^{(3)}=0.644 \pm 0.690 i $& $f_8^{(3)}=-0.556$\\ 
\hline 
\multirow{4}{*}{10 }& $  b_{10}^{(1)}=-0.824 \pm 0.314 i; $ & $f_{10}^{(1)}=0.125;$\\
& $b_{10}^{(2)}=-0.415 \pm 0.79 i;$ & $f_{10}^{(2)}=0.0132;$\\
& $ b_{10}^{(3)}=0.215 \pm 0.898 i;$ & $f_{10}^{(3)}=-0.244;$ \\
& $b_{10}^{(4)}=0.775 \pm 0.582 i$ & $f_{10}^{(4)}=-0.773$\\ 
\hline 
\end{tabular} 
\end{center}\caption{Complex roots of equation (\ref{Eq1a}) and corresponding values of $f_q$ for different values of $q$. }
\label{Table}
\end{table}
\begin{figure}[t]
\centering
\includegraphics[scale=0.2]{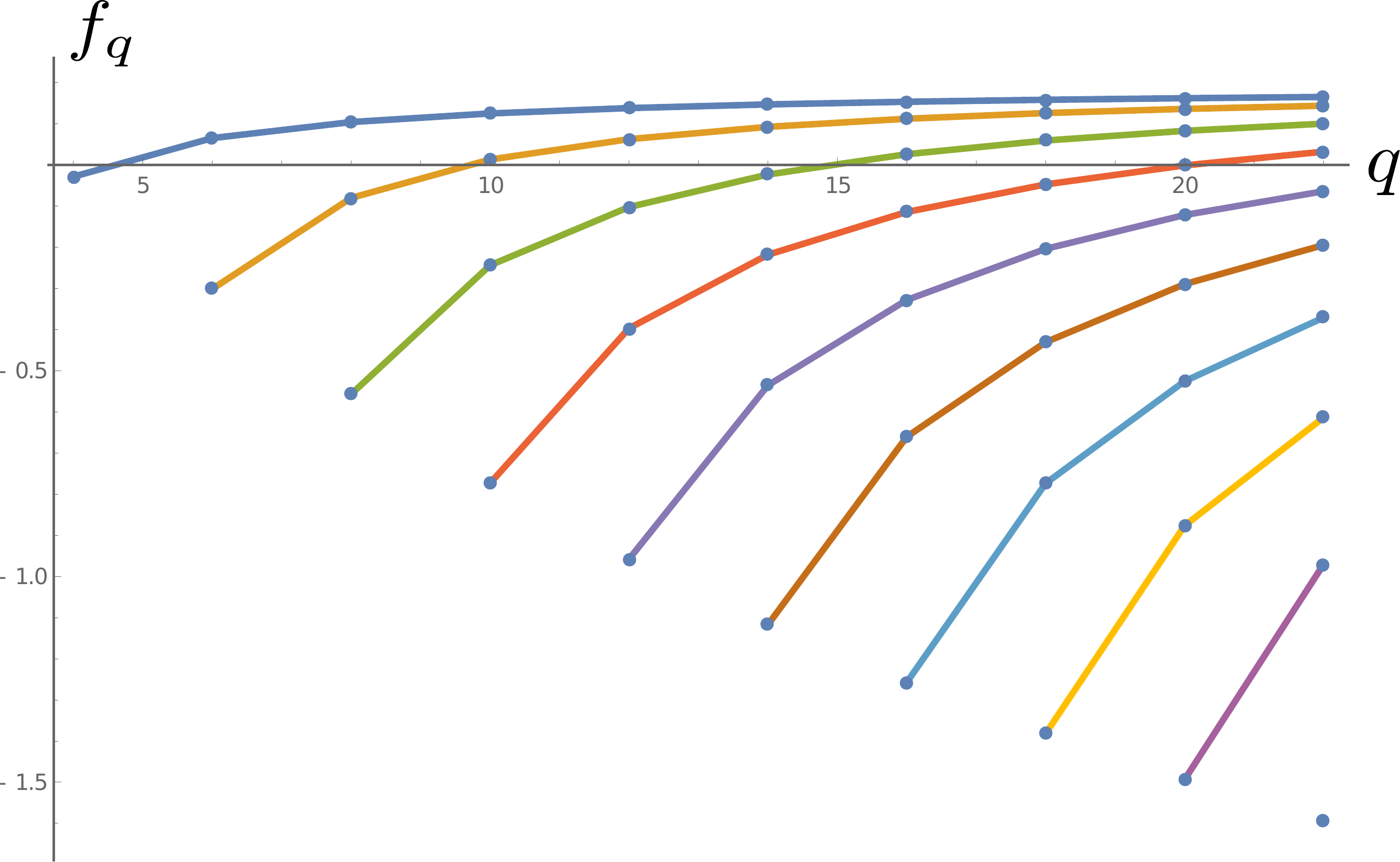} {\bf A.}
\includegraphics[scale=0.3]{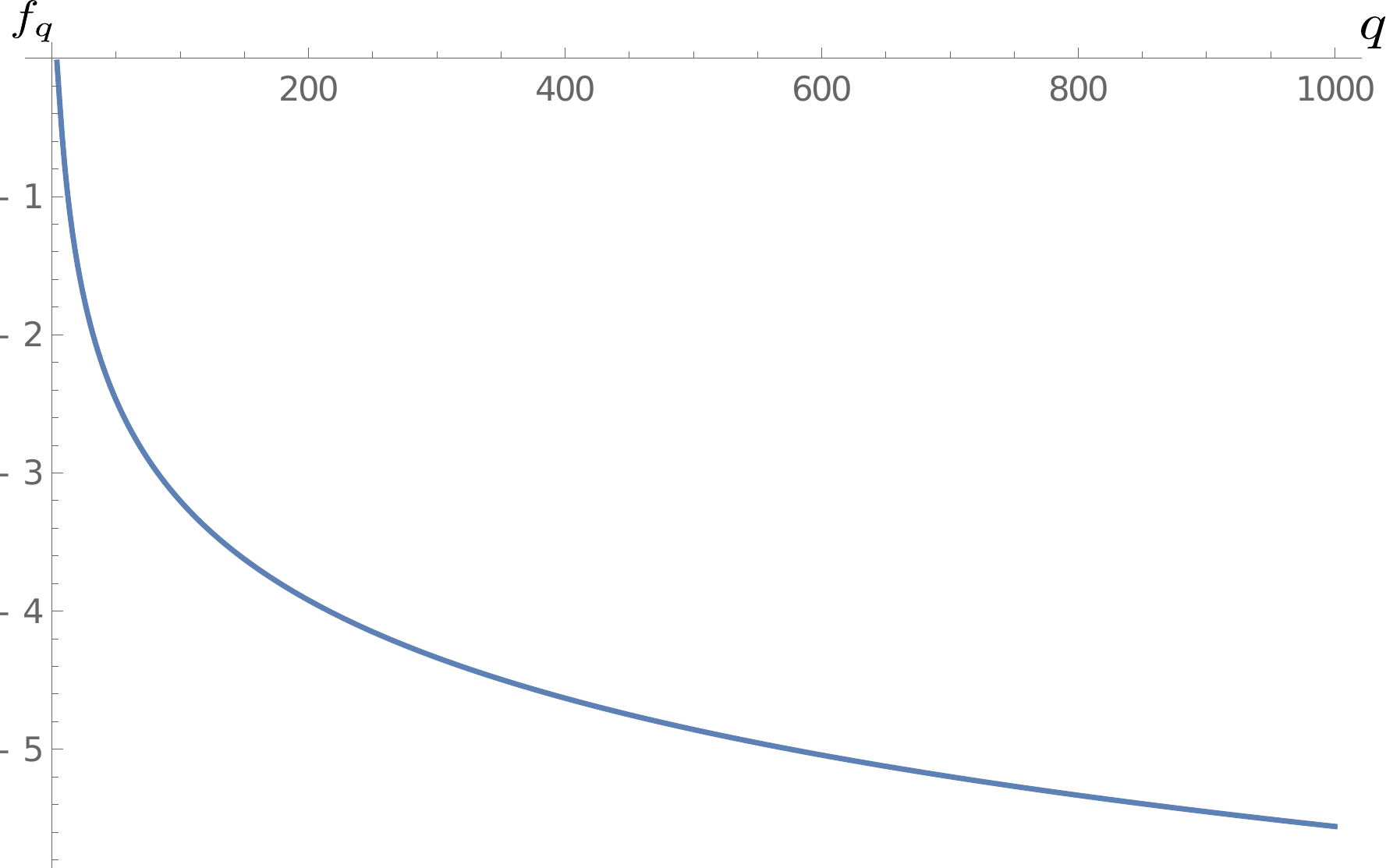}{\bf B.} 
\caption{{\bf A.} Regularized free energy on replica-symmetric saddles at various values of $q$. 
 {\bf B.} Free energy of the lowest replica-symmetric saddle as a function of $q$.  }
\label{fig:FRS-q}
\end{figure}
Comparing the regularized free energies for solutions for every $q$, the last pair of roots in every column have the smallest regularized free energy and are thus dominant. The position of these dominant roots on the complex plane are shown on the Fig.\ref{fig:XM0}A, and the absolute value as a function of $q$ is plotted on Fig.\ref{fig:XM0}B. The values of $f_q$ for other replica-symmetric solutions are plotted on Fig.\ref{fig:FRS-q}A, and the dependence of the $f_q$ for the lowest replica-symmetric saddle on $q$ is presented on Fig.\ref{fig:FRS-q}B.
%

\subsubsection{General Parisi ansatz at $M \to 0$}
To take the limit $M \to 0$ for more general structure of the Parisi matrix, we use the standard trick of \cite{Mezard91}: we introduce the Parisi function of a continuous variable instead of the Parisi matrix parameters, thus performing the analytic continuation from integer values of $M$ to arbitrary positive real number: $a_j\to a(u)$
\bea
\sum_{i=1}^{n} \rho _i  (m_{i+1} - m_i)&\to&\int _1^n\,\rho(v)\,dv\,;\\
\sum_{i=1}^{j} \rho _j  (m_{i+1} - m_i)&\to&\int _1^u\,\rho(v)\,dv\,;\\
\sum_{i=j+1}^{n} \rho _j  (m_{i+1} - m_i)&\to&\int _u^n\,\rho(v)\,dv\,.\eea
and \eqref{jcomp} and \eqref{0comp} become
\bea 
& 0 =&   a_0^{q-1} a(u)  + a_0 a^{q-1}(u)-u a^{q}(u)+
a^{q-1}(u)\int _1^u\,a(v)\,dv+a(u)\int _1^u\,a^{q-1}(v)\,dv\nn\\
& +&\int _u^M\,a^{q}(v)\,dv\,.\\
& \mC =&  a^{q}_0 +\int _1^M\,a^q(v)\,dv\,;\eea
Taking the limit $M\to 0$, we write
\bea 
& 0 =&   a_0^{q-1} a(u)  + a_0 a^{q-1}(u)-u a^{q}(u)-
a^{q-1}(u)\int ^1_u\,a(v)\,dv-a(u)\int ^1_u\,a^{q-1}(v)\,dv\nn\\
& -&\int ^u_0\,a^{q}(v)\,dv\,.\\
& \mC =&  a^{q}_0 -\int _0^1\,a^q(v)\,dv\,;
\eea
Defining the average 
\be
\int _0^1\,a^p(v)\,dv\equiv \langle a^p \rangle\ee
we arrive at the following equations: 
\bea 
& 0 =&    a(u)[a_0^{q-1}-\langle a^{q-1}\rangle ]  + a^{q-1}(u) [a_0-\langle a \rangle] 
 -\int ^u_0[a(v)-a(u)][a^{q-1}(v)-a^{q-1}(u)]\ dv\,,\nn\\\label{24n} &&\\
 & \mC =&  a^{q}_0 -\langle a^q \rangle\label{14n}\,.
\eea
where $u \in [0,\ 1]$. Finally, we fix the scaling freedom by imposing $P_{\alpha\alpha}=a_0 = 1$, or, equivalently, dividing (\ref{24n}) and (\ref{14n}) by $a_0^q$. We arrive at
\bea 
& 0 =&    a(u)[1-\langle a^{q-1}\rangle ]  + a^{q-1}(u) [1-\langle a \rangle] 
 -\int ^u_0[a(v)-a(u)][a^{q-1}(v)-a^{q-1}(u)]\ dv\,,\nn\\\label{24m} &&\\
 & \mC =&  1 -\langle a^q \rangle\label{14m}\,.
\eea
Thus, these integral equations are the final form of the matrix equation (\ref{CG-eq-matrix}) in the zero replicas limit. 

\subsubsection{A comment on the $q=2$ case}

For $q=2$ the equations (\ref{14m}),(\ref{24m}) have the form
\bea 
& 0 =&   2 a(u)[1-\langle a\rangle] 
 -\int ^u_0\,[a(v)-a(u)]^2\,dv \,.\label{q=2eq2}\\
 & \mC =&  1 - \langle a^2 \rangle \,;\label{q=2eq1}
\eea
Taking the derivative on the first equation, we get
\be
2 a'(u)[1-\langle a\rangle] 
 -[a(u)-a(u)]^2+2a'(u) \int ^u_0\,[a(v)-a(u)]\,dv=0\,,
\ee
or
\be
2 a'(u)[1-\langle a\rangle] 
+2a'(u) \int ^u_0\,[a(v)-a(u)]\,dv=0\,.
\ee
If $a'(u)\neq 0$, we get
\be
1-\langle a\rangle
+ \int ^u_0\,[a(v)-a(u)]\,dv=0\,.
\ee
Taking derivative once again, we arrive at
\be
 -a'(u)\int ^u_0\,\,dv=0
\ee
This shows that there is no solution, except for $a'(u)=0$, i.e. $a=A=\text{const}$. 
From the equation (\ref{q=2eq2}) it follows that only two possibilities can be realized: 
\begin{itemize}
\item $A = 0$ - this leads to the regular replica-diagonal solution;
\item $A = 1$ - is not a solution of the full SYK saddle point equations, because it according to (\ref{q=2eq1}) it results in $\mC = 0$.
\end{itemize}
Therefore the replica-diagonal solution is the only valid solution of the $q=2$ SYK model in the zero replicas limit. We note that the proof works only for smooth functions $a(u)$. For the discontinuous function $a(u)$, we have checked that this also is the case on the step function ansatz, and expect this to be true for any function. 

\subsubsection{Contribution to the regularized free energy}
\label{sec:DeltaF}

The free energy is expressed from (\ref{F}) using (\ref{ReplicaTrick}):
\be
-\beta F = \lim_{M \to 0} \frac{\log \overline{Z^M}}{M}\,.
\ee
To calculate it in the large-$N$ approximation, we have to find the saddle point configuration of $\overline{Z^M}\sim \text{exp}(-S_M)$ in the space of replica bilocal fields $G_{\alpha\beta}$ and $\Sigma_{\alpha\beta}$ with the minimal value of the real part of the on-shell action. On the factorized solutions, we can have one or several saddle points with the same value of real part of the action. 

We are specifically interested in the free energy of factorized replica-nondiagonal Parisi saddle points in the conformal limit. The regularized on-shell action in this case, as dictated by (\ref{Son-shell}), separates into the contribution identical to the replica-diagonal on-shell action in the conformal limit plus a contribution from the Parisi replica matrix:
\be
\lim_{M \to 0} \fs_3 = -\lim_{M \to 0} \frac1M d_f\ \tr \log Q\,, \label{S_R}
\ee
where we use the cutoff consistency assumption that $ d_f = \beta J/\pi$, and the matrix $Q$ on the replica space is defined as 
\be
Q = \mC^{\Delta-1} P^{\circ(q-1)}\,. \label{Qdef}
\ee
The matrix Q has the Parisi form and is expanded as (\ref{repr}): 
\be
Q = \sum_{i =1}^M q_i (\mI_{m_{i+1}} - \mI_{m_i}) + q_0 \mI_1\,,
\ee
where the Parisi parameters are defined as follows: 
\be
q_i =\mC^{\Delta-1} a_i^{q-1}\,; \quad i = 0,\dots,M\,.
\ee
In the continuum representation the matrix $Q$ is defined by the constant $q_0$ and the Parisi function $q(u)$, which are defined as (with $a_0=1$ taken into account)
\bea
&& q_0 =\mC^{\Delta-1}\,; \\
&& q(u) =\mC^{\Delta-1} a(u)^{q-1}\,.
\eea
To compute (\ref{S_R}) in the replica limit $M \to 0$, we use the formula for the Parisi matrix tracelog in the continuum representation \cite{Mezard91}: 
\be\label{TrLogP}
\lim_{M \to 0} \frac1M \tr\ \log Q = \log(q_0-\langle q \rangle)+\frac{q(0)}{q_0- \langle q \rangle}-\int_0^1 \frac{dv}{v^2}\log \frac{q_0-\langle q \rangle - [q](v) }{q_0- \langle q \rangle}\,,
\ee
where
\be
[q](u)=\int_{0}^u dv (q(u)-q(v))\,.
\ee
Substituting (\ref{Qdef}) into (\ref{TrLogP}), we obtain the expression: 
\bea
&& \lim_{M \to 0} \frac1M \tr\ \log Q = \log(q_0-\langle q \rangle)+\frac{q(0)}{q_0- \langle q \rangle}-\int_0^1 \frac{dv}{v^2}\log \frac{q_0-\langle q \rangle - [q](v) }{q_0- \langle q \rangle} \\&&=\frac{1-q}{q}\log \mC + \log (1- \langle a^{q-1} \rangle )+ \frac{a^{q-1}(0)}{1- \langle a^{q-1} \rangle} - \int_0^1 \frac{d v}{v^2} \log \frac{1 - \langle a^{q-1} \rangle - [a^{q-1} ] (v)}{1 - \langle a^{q-1} \rangle}\,. \nn\\\label{TrLogQ}
\eea
With $\mC$ substituted using the equation (\ref{14m}), this formula establishes the resulting general expression for the contribution of the replica structure to the on-shell action on a particular solution for the Parisi function $a(u)$.

To be able to compute the free energy, what is left is to describe which saddle points actually contribute in the path integral. We find them by solving the equations of motion for the Parisi matrix $P$. The subtlety here is that we can have multiple saddle points with equal absolute value of the integrands. Each saddle point gives a contribution to the free energy which reads 
\be
F_{\text{saddle}}(k) = \frac{1}{\beta} \lim_{M \to 0} \frac1M S_{M}(k)\,, \label{SM->F}
\ee
where $k$ labels the saddle point. 
If we have a family of $n$ saddle points with free energy values $F_{\text{saddle}}(k)$  such that
\be
\Re\ F_{\text{saddle}}  (i) = \Re\ F_{\text{saddle}}(j)\quad \forall i,j = 1,\dots,n\,, \label{Fsingle}
\ee
we have to sum over them to obtain the full expression for the free energy: 
\bea
F = -\frac{1}{\beta} \log \sum_{k=1}^n \e^{-\beta F_{\text{saddle}}(k)}
=
 \Re\ F_{\text{saddle}}  - \frac{1}{\beta}\log \sum_{k=1}^n \e^{-i \beta\  \Im\ F_{\text{saddle}}(k)}\,. \label{Fmulti}
\eea
Note that in the case of integer $M$ the l.h.s. of the equation (\ref{CG-eq-matrix}) is a polynomial with real coefficients. Therefore is $P_1$ is a complex-valued solution of the equation, then the matrix $P_2 = P_1^*$ is also a solution, as we saw in the replica-symmetric particular case in sec. \ref{sec:RS}. Therefore for every complex saddle point we will also have the complex conjugated saddle point contributing in the sum in (\ref{Fmulti}). Focusing on the case of two complex-conjugated saddle points in the sum, the formula for the free energy is written as 
\be
F = \Re\ F_{\text{saddle}}  - \frac{1}{\beta}\log \left[2 \cos \left( \beta\  \Im\ F_{\text{saddle}}\right)\right]\,. \label{FmultiReal}
\ee
Recall that the saddle point value of the free energy is determined by the on-shell action $S_M$ in the $M \to 0$ limit, see Eq.(\ref{SM->F}). The $S_M$ is expressed by the formula (\ref{Son-shell}). First of all, let us note that the on-shell action is proportional to $N$. The first term in the formula (\ref{FmultiReal}) is a real part of that action, and is also therefore an extensive contribution to the free energy. However, the second term is not proportional to $N$. When taking the large $N$ limit in a special way, so that the singularities in the log are avoided, the second term can be neglected. Therefore, we are left with the real part of the on-shell action (\ref{Son-shell}) defining the free energy at large $N$. 

The $\fs_3$ piece is responsible for non-trivial contributions to the free energy of the replica-nondiagonal factorized solutions. Hence on specific solutions we are most interested in the quantity
\be
\Delta F = F_{RND}-F_{RD}=\lim_{M\to 0} \frac{N}{2\beta}\, \Re\,\fs_3 \,, \label{deltaF}
\ee
and, in particular, in the sign  of $\Delta F$ describing the shift of the replica non-diagonal solutions in respect to the diagonal one.  The negativity of $\Delta F $ for given pair of solutions means that  the replica non-diagonal solution
is the dominant one. This is the quantity in the rigid strong coupling limit where the replica matrix $P$ in the factorized ansatz introduces discrepancy with the replica-diagonal result. 

\section{One-step RSB solution}
\label{sec:RSB1}

Having established the formalism and derived general expressions for equations of motion and on-shell action in the previous section, we are now ready to study solutions more specifically. In this section we focus on the SYK model with $q=4$.  We study the solutions of the equations (\ref{14m}),(\ref{24m}). We restrict ourselves to the solutions for $a(u)$, which can be described by the one-step replica symmetry breaking ansatz, in analogy with the one-step solutions in the spin glass systems \cite{Mezard91,ParisiBook}, such as the Sachdev-Ye model \cite{Georges00}:
\be\label{onestep}
a(u)=A_0+A_1 \theta (u-\mu )\,.\ee
In this formula $\mu$ is a free parameter, to which we will refer as the breakpoint (again, in analogy with \cite{Georges00}). 
The moments of $a(u)$ and integrals which contribute to (\ref{14m}),(\ref{24m}) are evaluated on the ansatz (\ref{onestep}) as follows:
\bea
&& \langle a \rangle = A_0 + A_1 (1-\mu)\,; \\
&& \langle a^3 \rangle = (A_0 + A_1)^3 - A_1 (3 A_0^2 + 3 A_0 A_1 + A_1^2) \mu\,; \\
&& \langle a^4 \rangle = \left(A_0+A_1\right){}^4-A_1 \left(2 A_0+A_1\right) \left(2 A_0^2+2 A_1
   A_0+A_1^2\right) \mu\,;\\
&& \int_{0}^u dv (a(v) - a(u)) (a(v)^3 - a(u)^3) = A_1^2 (3 A_0^2 + 3 A_0 A_1 + A_1^2) \mu\ \theta(u - \mu)\,.
\eea
The equation \eqref{14m} reduces to 
\be\label{q4C}
1 - \left(A_0+A_1\right)^4+A_1 \left(2 A_0+A_1\right) \left(2 A_0^2+2 A_1
   A_0+A_1^2\right) \mu = \mC\,.
   \ee
Meanwhile,  the equation of motion \eqref{24m} decays into two equations which we get  separating the coefficient in front of the step function: 
\bea
&& A_0 \left(A_0^2+1+A_1^3 (\mu -1)+3 A_0 A_1^2 (\mu -1)+4 A_0^2
   A_1 (\mu -1)-2 A_0^3\right)=0\,; \label{ee1}\\
&&A_1 \Big(A_1^2 \left(1+A_0 (3 \mu -7)\right)+3 A_0 A_1 
\left(1+A_0 (\mu -3)\right)+3 A_0^2+1+A_1^3 (\mu -2)-4 A_0^3\Big)=0\,.\nn\\\label{ee2}
   \eea
We take $\mu$ as the free parameter and solve these equations for $A_0$ and $A_1$. Equations (\ref{ee1})-(\ref{ee2}) are algebraic equations of the 4-th order, so we have 16 solutions. We want to pick solutions that describe saddle points of the model, i.e. ones that also solve the equation (\ref{CG-eq-master}) in the $M \to 0$ limit.

We compute the contribution to the regularized free energy (\ref{deltaF}) on these solutions 
\bea
\Delta F&=&
\lim_{M\to 0} \frac{N}{2\beta}\, \Re\,\fs_3 = -\lim_{M \to 0} \frac{N}{2\beta}\,\frac1M d_f\Re[ \tr \log Q]\,
=-\lim_{M \to 0} \frac{J\,N}{2}\,\frac1M \Re[ \tr \log Q]\,,\nn\\\label{a}
\eea
For the tracelog we use the formula (\ref{TrLogQ}), which on the one-step ansatz assumes the following form: 
\bea
&&\lim_{M \to 0} \frac{1}{M} \tr \log Q =\\
&-&\frac{3}{4}
   \log \left(1+A_1 \left(2 A_0+A_1\right) \left(2 A_0^2+2 A_1 A_0+A_1^2\right) (\mu
   -1)-A_0^4\right)\nn\\&+&\log \left(1+A_1 \left(3 A_0^2+3 A_1 A_0+A_1^2\right) (\mu
   -1)-A_0^3\right) \nn\\&+& \frac{A_0^3}{1+A_1 \left(3 A_0^2+3 A_1 A_0+A_1^2\right) \mu -\left(A_0+A_1\right){}^3} \nn\\&+& \left(1-\frac{1}{\mu }\right) \log \left(1-\frac{A_1 \mu \left(3 A_0^2+3
   A_1 A_0+A_1^2\right) }{1+A_1 \left(3 A_0^2+3
   A_1 A_0+A_1^2\right) \mu -\left(A_0+A_1\right){}^3}\right)\,.\label{freeen}
   \eea
Let us note some observations:
\begin{itemize}
\item  All of the complex saddle points, which give the same real part of the action, are organized in mutually conjugated pairs, and we didn't find any instances of multiple pairs of saddle points giving the same real part of the on-shell action, so the formula (\ref{deltaF}) is applicable. 
\item While the general formula (\ref{FmultiReal}) does not necessarily guarantees that the regularized free energy is real-valued, we found this to be the case for all solutions we studied, except for one which we mention below. 
\end{itemize}

\begin{figure}[t]
\centering
\includegraphics[scale=0.30]{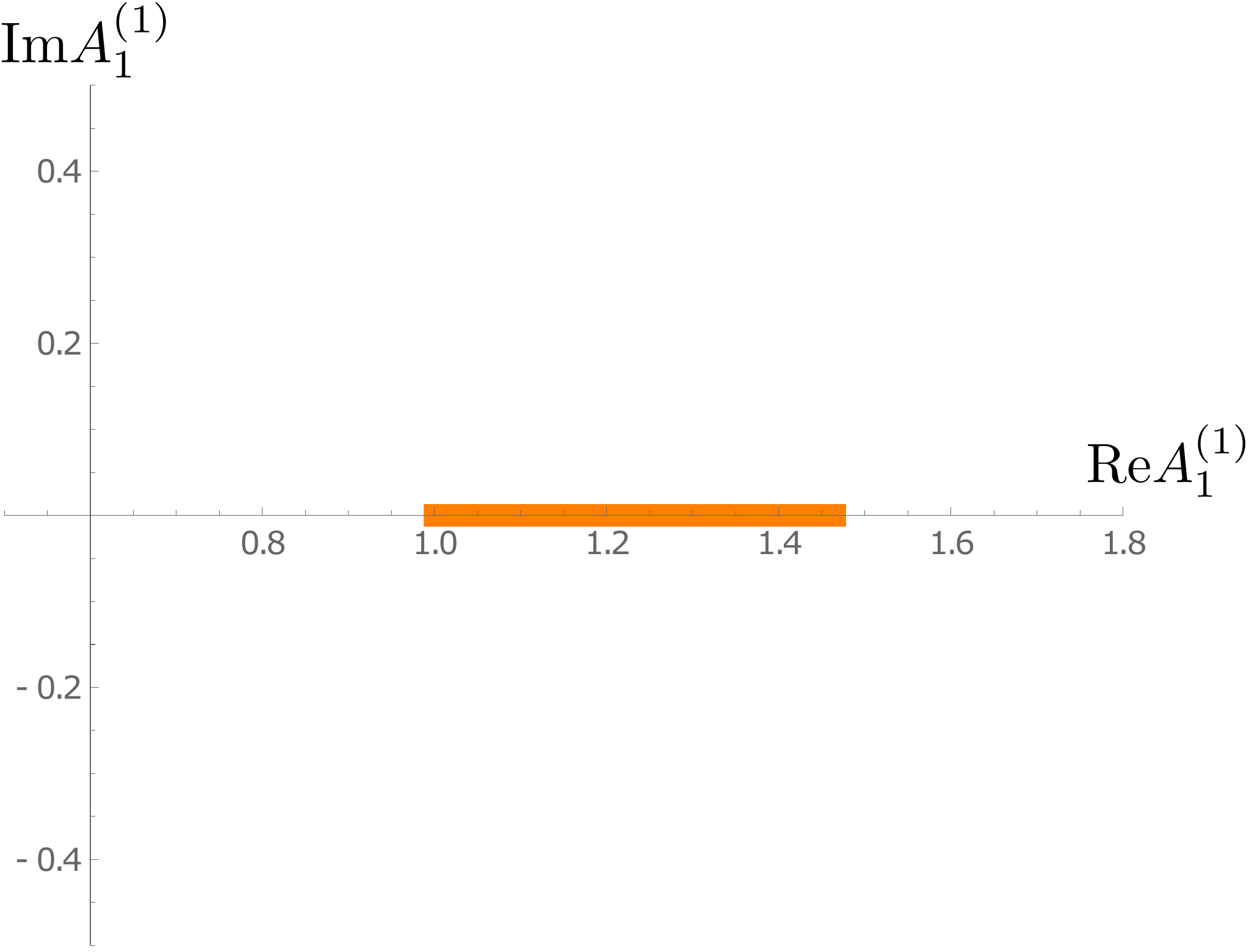} $\,\,\,${\bf A.}$\,\,\,\,\,\,\,$
\includegraphics[scale=0.33]{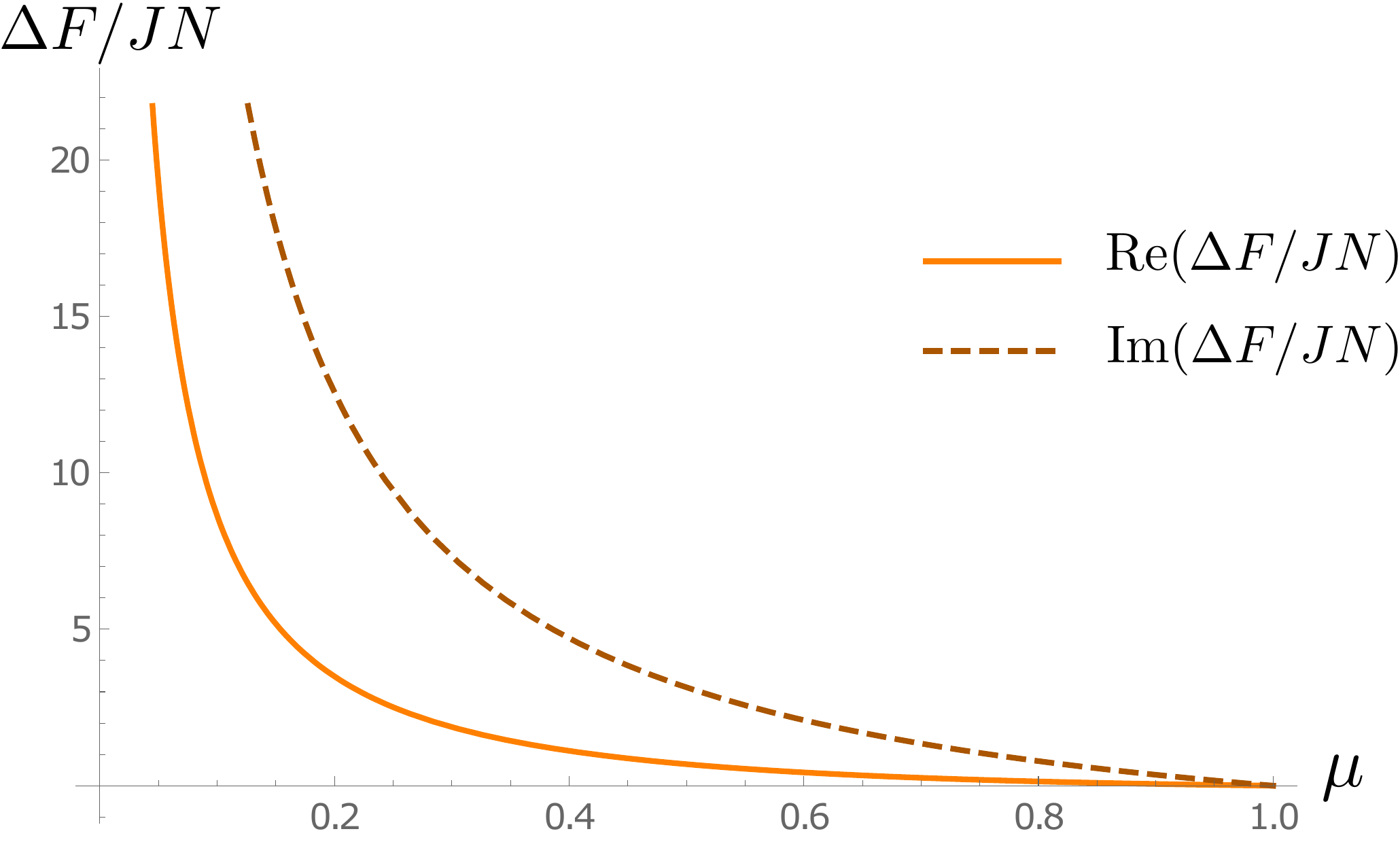}$\,\,\,${\bf B.} 
\caption{{\bf A.} Trajectory of the saddle point \eqref{Areal} parametrized by $\mu$ on the complex plane. {\bf B.} The value of the free energy on solution \eqref{Areal}, the solid line shows the real part and dashed line the imaginary part.}
\label{fig:0A1r}
\end{figure}

The constant solutions correspond to one step function  with $A_1 = 0$,
 see \eqref{onestep}. We have two kinds of solutions in this class: \\
\textbf{1.} Replica-diagonal (paramagnetic) solution: $A_0 = 0$, $A_1 = 0$. \\
\textbf{2.} Replica-symmetric complex-valued solutions. This case is reduced to considerations performed in sec. \ref{sec:RS}, see table \ref{Table}. 

\paragraph{Replica symmetry breaking solutions.} There are also two kinds of solutions with $A_1 \neq 0$:\\
\textbf{1.} $A_0 = 0$. In this case $A_1$ is a solution of the equation which follows from (\ref{ee2}): 
\be
A_1^2 a_0+a_0^3+A_1^3 (\mu -2)=0\label{A1}\ee
This equation has one real and two complex mutually conjugated solutions:
\bea \label{Areal}
 A_1^{(1)} &=& \frac{\sqrt[3]{2} }{3 (2-\mu) \fK} +\frac{\fK}{3 \sqrt[3]{2} (2-\mu)}+\frac{1}{3 (2-\mu )}\,;\\\label{Acompl}
 A_1^{(2,3)} &=& -\frac{\left(1\mp i \sqrt{3}\right) }{3\, 2^{2/3} (2-\mu) \fK} -\frac{\left(1 \pm i
   \sqrt{3}\right)\fK }{6 \sqrt[3]{2} (2-\mu)}+\frac{1}{3 (2-\mu)}\,,
\eea
where
\bea
\fK&=&\sqrt[3]{-\fk-\sqrt{ \fk^2-4 }},\,\,\,\,\,\,\,\fk=-27  \mu ^2+108 \mu -110 \,.\eea

\begin{figure}[t]
\centering
\includegraphics[height=4cm,width=7.7cm]{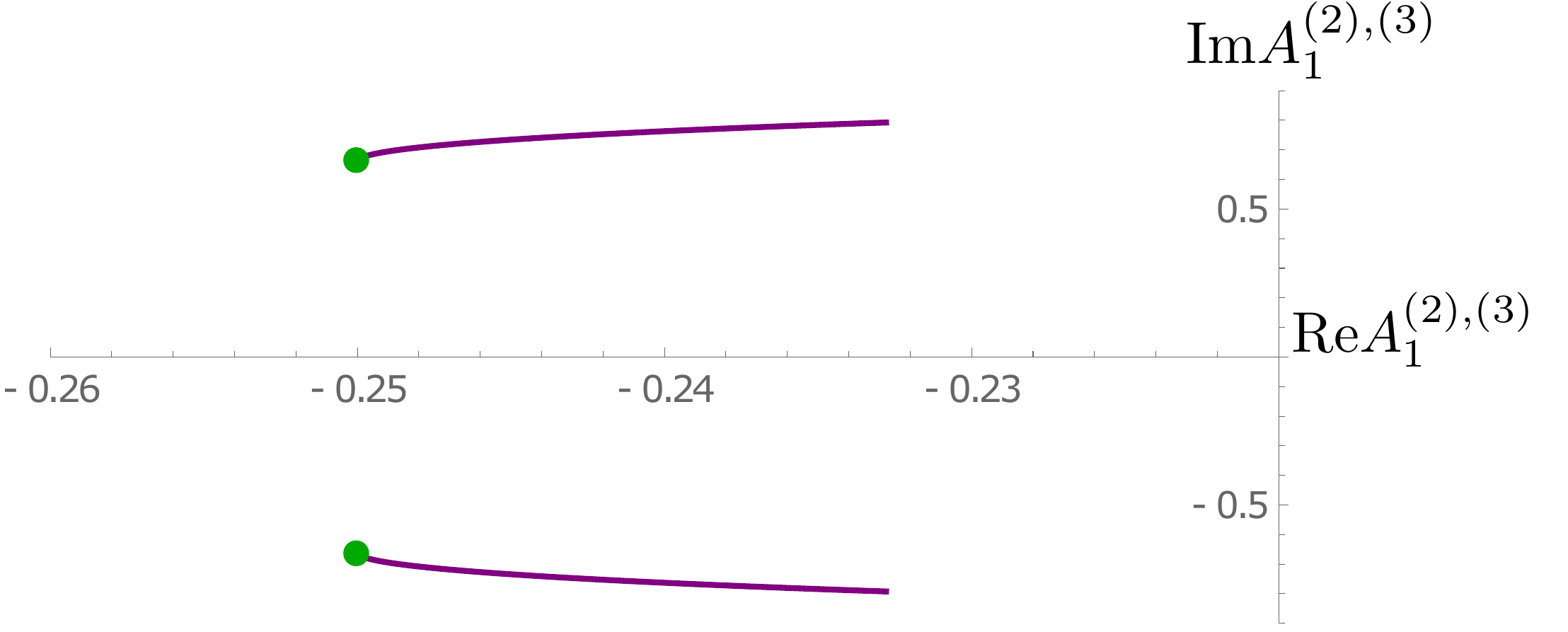}{\bf A.} $\,\,\,$
\includegraphics[height=4cm,width=6cm]{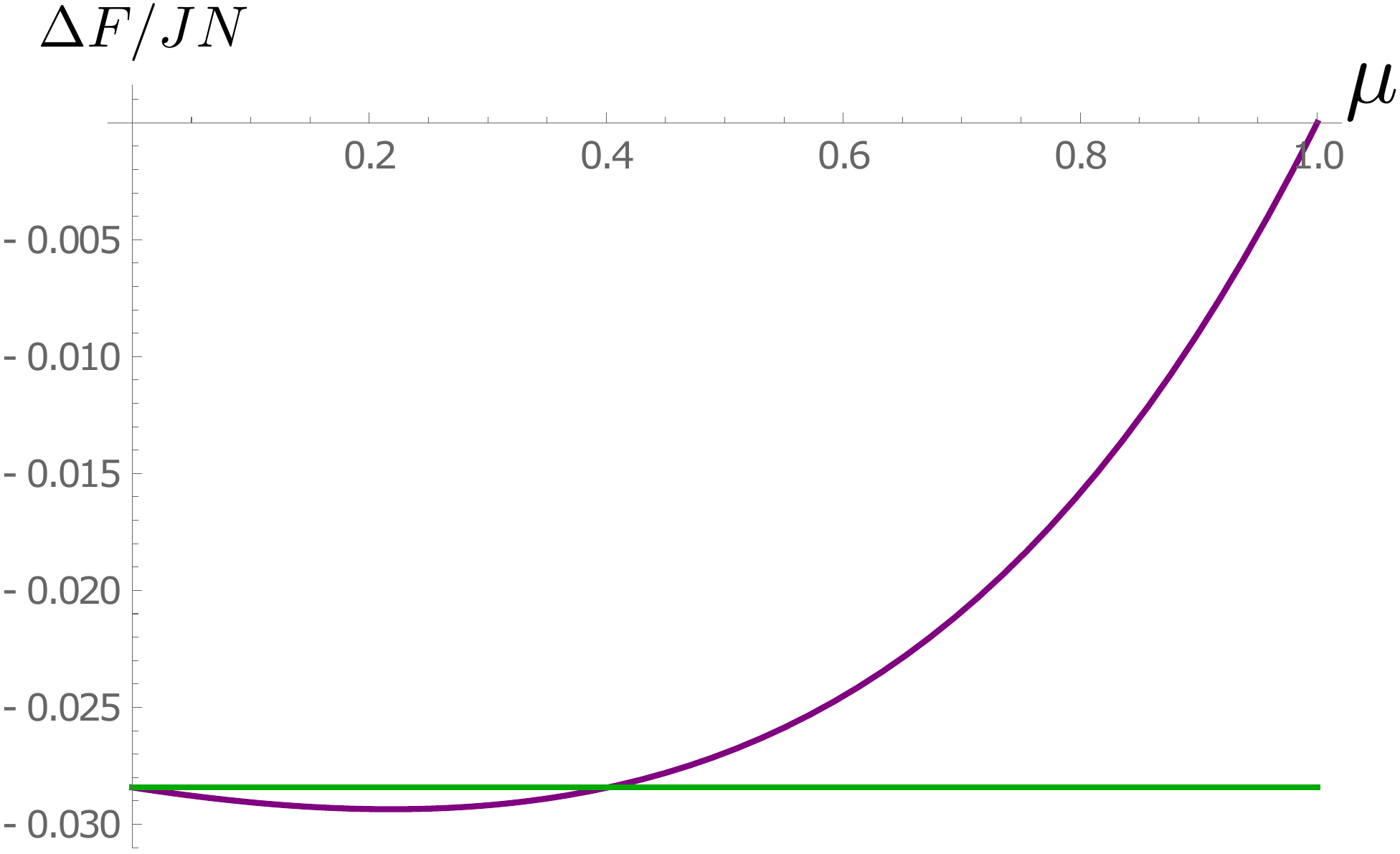}{\bf B.} 
\caption{{\bf A.} Trajectories of saddle points \eqref{Acompl} parametrized by $\mu$ on the complex plane. The green points show the location of the replica symmetric solutions. {\bf B.} The value of the regularized free energy on solution \eqref{Acompl} as a function of $\mu$. The green line is the value of the free energy for replica-symmetric solutions (\ref{f_4}).}
\label{fig:0A1c}
\end{figure}

The regularized free energy for the real solution labeled by $(1)$ is complex valued.
The location of solution \eqref{Areal} on the complex plane and its contribution to the free energy is presented on Fig.\ref{fig:0A1r}. Note that the free energy diverges at  the real solution  near the limiting case $\mu=0$.  

The solutions (\ref{Acompl}) are mutually complex conjugated. Their trajectories on the complex plane parametrized by $\mu$ and their contribution to the regularized free energy is presented on Fig.\ref{fig:0A1c}. We see that at $\mu=0$ the positions of these solutions coincide with the positions of pure constant complex solutions (see table \ref{Table} at $q=4$), shown by green points. This is not surprising since  for $\mu=0$ our step function solution in fact corresponds to the constant solution.

\begin{figure}[t]
\centering
\includegraphics[scale=0.35]{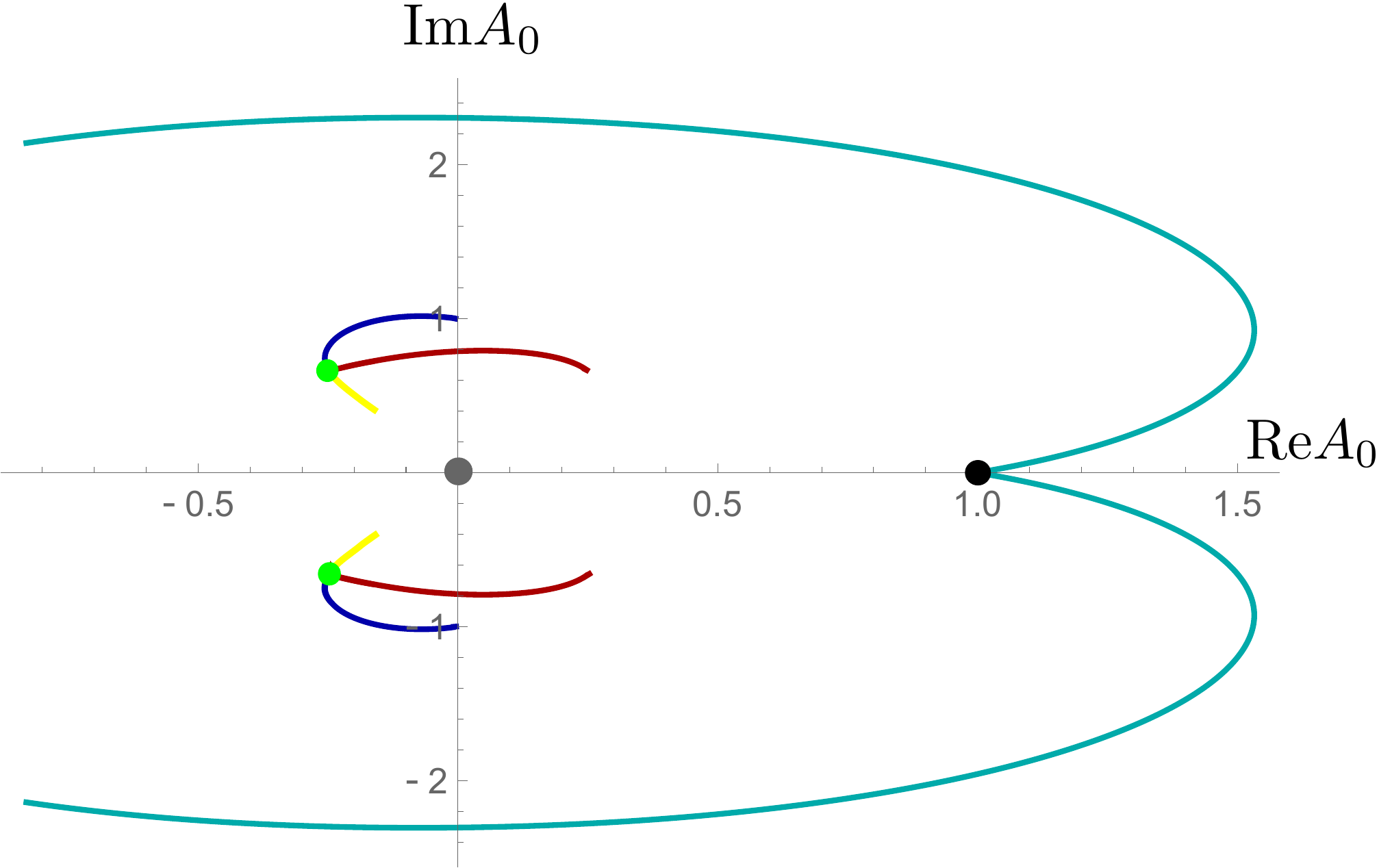}A. 
\includegraphics[scale=0.3]{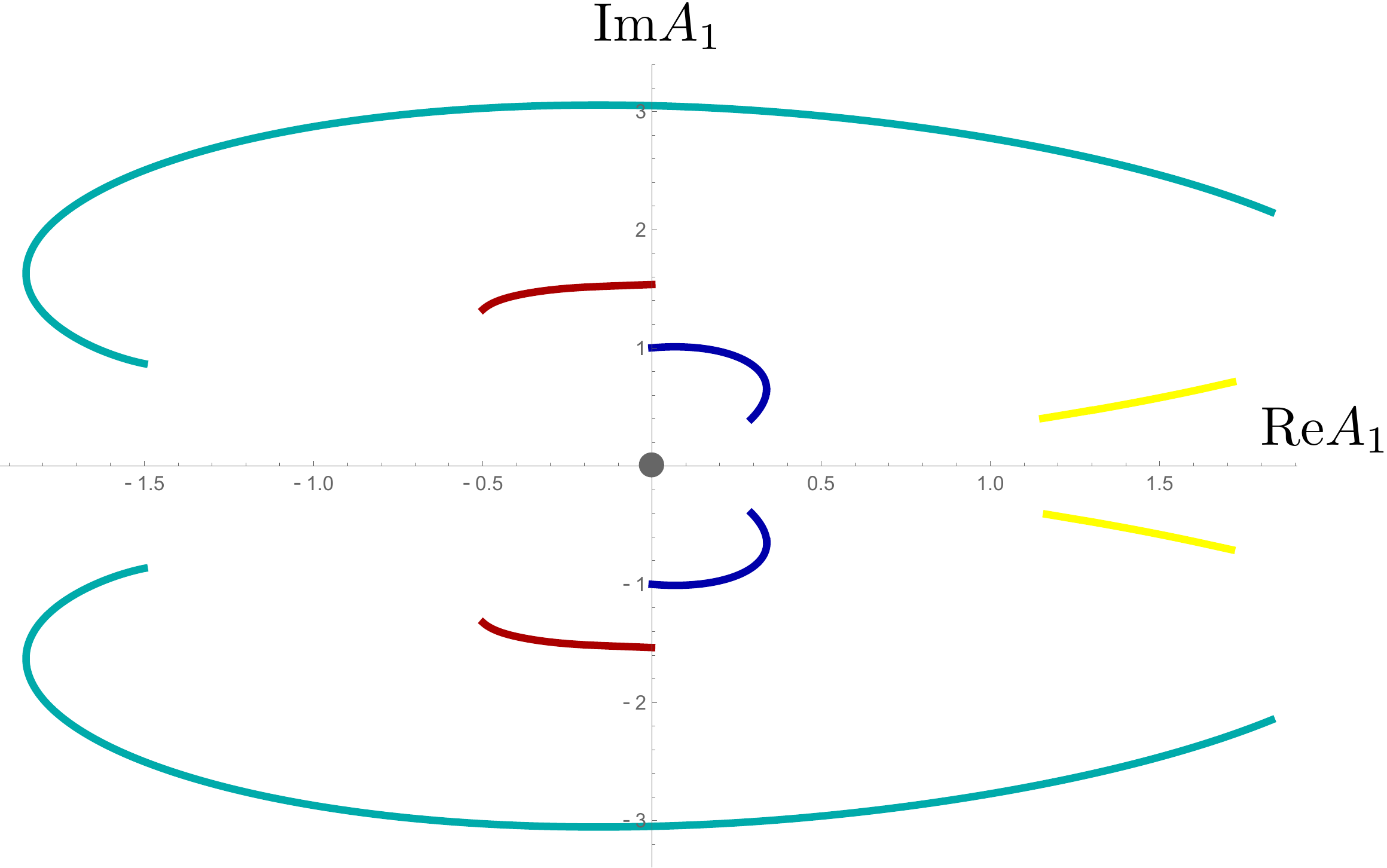}B.
\caption{Trajectories of saddle points parametrized by $\mu$ on the complex plane. The black point is the corresponding replica-symmetric solution with $a_0 = A_0$. The green points are the complex replica-symmetric solutions.}
\label{fig:A0-A1}
\end{figure}

\begin{figure}[t]
\centering
\includegraphics[scale=0.7]{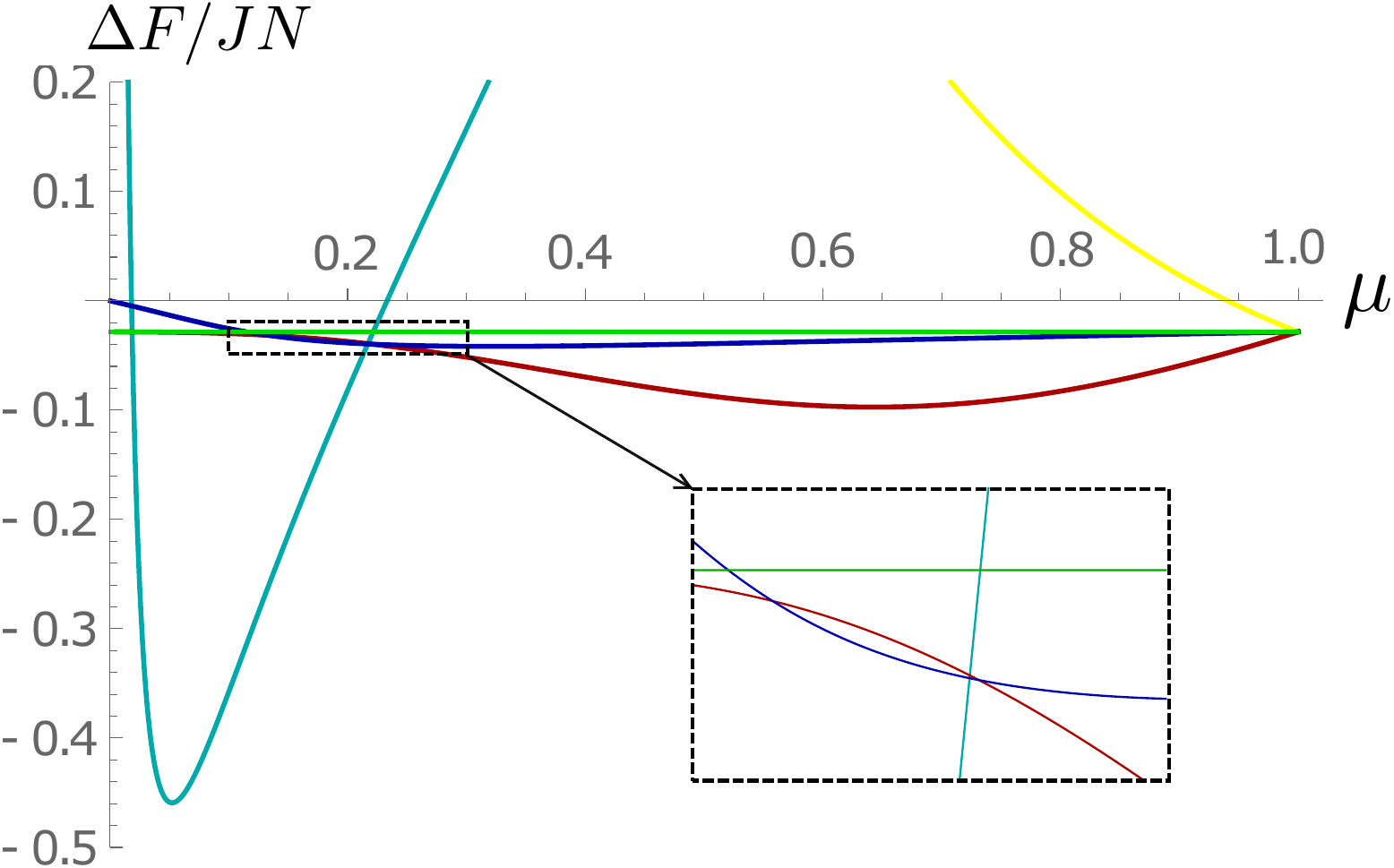}
\caption{The plot of the replica dependent part of the free energy on numeric solutions. Green line is the free energy on complex replica symmetric solution. All colors here correspond to the colors in the Fig.\ref{fig:A0-A1}.}
\label{fig:FreeEn}
\end{figure}

\textbf{2.} Second group of RSB solutions is characterized by non-zero both $A_0$ and $A_1$. In this case the degree of the algebraic equations (\ref{ee1}),(\ref{ee2}) is too high to obtain explicit analytic solutions, but they can be solved numerically. We plot these solutions on the complex plane on Fig.\ref{fig:A0-A1}. Among these RSB solutions there are some that turn into the real replica-symmetric solution with $A_0 = a_0$ at either $\mu = 0$ or $\mu = 1$, whereas others reduce to the complex replica-symmetric solutions. We have calculated the contribution to the regularized free energy of the remaining solutions. The dependence of $\Delta F$ on the breakpoint parameter $\mu$ is presented on Fig.\ref{fig:FreeEn}. We see that there are three local minima below the replica-diagonal value. 

\section{Some implications of factorized solutions}
\label{sec:Implications}

\subsection{Spin-glass-like metastable states}
\label{sec:Glass-like}

As we discussed in the section \ref{sec:Factorized}, we can use the extended reparametrization symmetry (\ref{reparam-replicas-G}),(\ref{reparam-replicas-S}) to generate more physically interesting solutions from a factorized solution. For example, we can make the time dependence of replica-offdiagonal components of the field $G_{\alpha\beta}$ differ from the time dependence of the diagonal components if we act with the set of diffeomorphisms $V_\alpha \in \text{diff}(S^1)$ which is arranged in such a way that 
\be
\forall \alpha, \beta:\ \alpha \neq \beta\,,\;  V_\alpha = V_\beta \circ U_\beta,\ \text{where}\ U_\beta \in SL(2, \RR)\,.
\ee
In this case all diagonal components will have the same time dependence, whereas the time dependence of the off-diagonal components will be different, because of the extra $SL(2, \RR)$ transformation which acts only on one of the times. 

\paragraph{Example.}
Let us fix $M =2$. We will work in terms of dimensionless times on the thermal circle
\be
\theta_{1,2} = \frac{2\pi}{\beta} \tau_{1,2}\,.
\ee
Consider the following pair of diffeomorphisms, where the first transformation is identity and the second one is a fraction linear transformation on the circle: 
\bea
&& \theta \to \varphi_1 (\theta) = \theta\,;\\
&& \theta \to \varphi_2(\theta)\,:\ \e^{i\varphi_2(\theta)} = \frac{a \e^{i \theta} + b}{c \e^{i \theta} + d}\,,\ a d-b c = 1\,.
\eea
In this case the diagonal components $G_{11}$ and $G_{22}$ are invariant under these transformations and are given by 
\be
G_{11} = g(\tau_1, \tau_2) P_{11}\,;\quad  G_{22} = g(\tau_1, \tau_2) P_{22}\,.
\ee
However, the off-diagonal part transforms non-trivially: \be
G_{12} \to \const (d + c\ \e^{i \theta_2})^{-2 \Delta} \left| \e^{i \theta_1} - \frac{a \e^{i \theta_2} + b}{c \e^{i \theta_2} + d}\right|^{-2\Delta} P_{12}\,.
\ee
In order to analyze the dependence on real time, we set $\theta_1 = 0$ and $\e^{i \theta_2} = e^{\frac{2\pi}{\beta} t_2}$. Then, we obtain
\be
G_{12}(0, t_2) \sim \left(d - b - (a-c) \e^{\frac{2\pi}{\beta}t_2}\right)^{-2\Delta}\,.  \label{glasslike-offdiag}
\ee
This is the decay law of the off-diagonal component. The exponent does not dominate until the characteristic time scale of 
\be
t_2^* = \frac{\beta}{2\pi} \log \frac{d-b-1}{a-c}\,.
\ee
Since the $SL(2,\RR)$ is non-compact, by fine tuning the parameters of the M\"obius transformation, we can make this time scale as long as we like, so that
\be
 \log \frac{d-b-1}{a-c} \gg 0\,.
\ee
 For comparison, the diagonal components behave like 
\be
G_{11}(0, t_2) \sim (1 - \e^{\frac{2\pi}{\beta} t_2})^{-2\Delta}\,, \label{glasslike-diag}
\ee
and it starts decaying right away. This means that at short times $t_2 < t_2^*$ the transformed solution is frozen, behaving like a kind of spin glass. However, after $t_2^*$ the thermal fluctuations destroy the spin-glass-like configuration with the same speed as a regular replica-diagonal configuration. 

\subsection{On relation to thermofield double and AdS$_2$ gravity}
\label{sec:Holography}
\subsubsection{Thermofield double from two replicas}

The emergent (extended) conformal invariance suggests that the factorized solutions might have a holographic interpretation in terms of nearly AdS$_2$ gravity, like the replica-diagonal solution does \cite{Maldacena16,Jevicki16,Harlow18}. The purpose of this section is to explore this correspondence on a toy example of factorized solution for $M = 2$. 

In this case we need to solve the equation (\ref{CG-eq-2}) for the $q=4$ and $M=2$. Assuming the replica-symmetric form for $P$ like in (\ref{P-RS}), we arrive at the equations (\ref{Eq1-M})-(\ref{Eq2-M}) with $M=2$ and $q=4$: 
\bea 
aA^{3}+Aa^{3}&=&0\,; \label{Eq1-M=2}\\
a^4+A^4 &=&\mC\,. \label{Eq2-M=2}
\eea
Fixing the scaling with $a=1$, we arrive at the solution with 
\bea
&& P_{11} = P_{22} \equiv a = 1\,; \qquad P_{12} = P_{21} \equiv A =  i\,; \label{M=2sol}\\
&& \mC = 2\,.
\eea
We now apply a pair of reparametrizations to the solution using (\ref{reparam-replicas-G}):
\be
f_1(\tau) = \tau\,; \qquad
f_2(\tau) = \tau + \frac{\beta}{2}\,. \label{TFDtrans}
\ee
Each of these transformations belongs to $SL(2, \RR)$, so diagonal elements of $G$ are unchanged, whereas the off-diagonal elements transform non-trivially. The transformed solution is given by 
\bea
&& G_{11} (\tau_1, \tau_2) = G_{22} (\tau_1, \tau_2) =  \frac{b}{2^{\Delta}}\left(\frac{\pi}{\beta J}\right)^{2\Delta} \frac{\sgn (\tau_1-\tau_2)}{\left|\sin \frac{\pi}{\beta}(\tau_1-\tau_2)\right|^{2\Delta}}\,; \label{G11} \\
&& G_{12} (\tau_1, \tau_2) = G_{21} (\tau_1, \tau_2) =  \frac{b}{2^{\Delta}}\left(\frac{\pi}{\beta J}\right)^{2\Delta} \frac{i\ \sgn (\tau_1-\tau_2)}{\left|\cos \frac{\pi}{\beta}(\tau_1-\tau_2)\right|^{2\Delta}}\,. \label{G12}
\eea
When performing the analytic continuation to the Lorentzian signature, it is evident that (\ref{G11}),(\ref{G12}) are, correspondingly, one-sided and two-sided correlators of a Majorana fermion in the thermofield double state \cite{Maldacena01,Caputa15,Harlow18}. The transformations (\ref{TFDtrans}) make it explicit that the two-sided (off-diagonal) correlator can be obtained from a one-sided (diagonal) correlator by moving one of the endpoints halfway along the thermal circle \cite{Caputa15}. Thus, a replica-nondiagonal large $N$ saddle point of the SYK model with $M=2$ replicas in the conformal limit can describe the purification of a single replica of SYK model at finite temperature in the form of the thermofield double state. 

\subsubsection{Comparison with semiclassical holographic computation}

We can compare the factorized solution (\ref{G11}),(\ref{G12}) with what we can get from a holographic computation of correlators in the AdS$_2$ spacetime. We consider the Lorentzian AdS$_2$ spacetime, which is described in embedding space of signature $(-\ -\ +)$ as a hyperboloid defined by the equation \cite{Maldacena16,Harlow18}
\be
-T_1^2 - T_2^2 + X^2 = -1\,.
\ee
We are interested in the Schwarzschild coordinate patch in the AdS$_2$. Let us denote by symbols $L$ and $R$ two boundaries of the AdS$_2$. In this case there are two corresponding choices of the Schwarzshild patch, which are described by the parametrizations \cite{Harlow18}:
\bea
&& T_1^L = \frac{r}{R}\,; \quad T_1^R = -\frac{r}{R}\,; \label{T-1}\\
&& T_2^{L,R} = \sqrt{\frac{r^2}{R^2}-1} \sinh R t\,; \label{T-2}\\
&& X^{L,R} = \sqrt{\frac{r^2}{R^2}-1} \sinh R t\,. \label{X}
\eea
Here $r, t \in \RR$. The horizon radius is related to the temperature as $R = \frac{\pi}{\beta}$. The induced metric in both cases is the same, 
\be
ds^2 = -(r^2 - R^2) dt^2 + \frac{dr^2}{r^2 - R^2}\,.
\ee
Suppose we want to calculate a correlator of heavy (scalar) operator $\mathcal{O}$ of the dimension $\Delta_{\mO}$ inserted on the same boundary. In the limit $\Delta_\mO \gg 1$, the correlator is determined\footnote{In this discussion we do not go into the details about analytic continuations and $i\epsilon$-prescription for correlators. We note that one can restore the analytic structure of the geodesic correlators by modifying the prescription to include the specific phase factors \cite{Arefeva16}. } by the length $\mL$ of the geodesic anchored on the boundary endpoints:  
\be
G_{\Delta_{\mO}}(t_1, t_2) \sim \e^{- \Delta_\mO \mL(t_1, t_2)}
\ee
Thus we need the boundary-to-boundary geodesics. The one-sided geodesics give complex-valued lengths because the interval is timelike, but we can circumvent this by making analytic continuation to the Euclidean signature $\tau = i t$ in the parametrization and $T_2 \to i T_2$ in the embedding space. To find the geodesic length between the points $1$ and $2$, we can use the relation
\be
\cosh \mL(1, 2) = \langle \vec{Y}(1), \vec{Y}(2) \rangle\,, \label{coshL}
\ee
where $\vec{Y}$ denotes a point in the embedding space, and angular brackets denote the scalar product in the embedding space. 
We choose the endpoints on the boundary, i. e. so that $r_1 = r_2 = r_0 \to \infty$. Taking the asymptotic with respect to $r_0$ and subtracting the divergent part, we obtain for one-sided correlator
\be
G_{\Delta_{\mO}}^{LL}(\tau_1, \tau_2) \sim \left(\frac{1}{2 \sin \frac{\pi}{\beta} (\tau_1 - \tau_2)}\right)^{\Delta_\mO}\,,
\ee
and for the two-sided correlator 
\be
G_{\Delta_{\mO}}^{LR}(\tau_1, \tau_2) \sim \left(\frac{1}{2 \cos \frac{\pi}{\beta} (\tau_1 - \tau_2)}\right)^{\Delta_\mO}\,.
\ee
We see that the time dependence of these semiclassical correlators is captured properly by the $2$-replica result (\ref{G11}),(\ref{G12}). However, it is clear that the structure of the Parisi matrix $P$ will not be captured by the bulk theory on the leading semiclassical level. 

In conclusion to this remark we would like to note that in the holographic derivation we performed the analytic continuation to the Euclidean signature while trating it purely formally. It seems plausible that the actual bulk spacetime which would have to correspond to our analytically continued Lorentzian AdS$_2$ would be similar to the double cone construction, described by authors of \cite{Saad18}.

\subsection{Comments on solutions beyond the strong coupling limit}

\subsubsection{(In)applicability of separation of variables}

Now let us make some comments about possible continuation of the factorized solutions (\ref{ansatz-Kamenev}) beyond the strong coupling regime. The non-conformal saddle point equation under the assumption of separability of variables (\ref{ansatz-Kamenev}) reads
\be
P_{\alpha \gamma} (\dd_\tau g)(\tau, \tau'') - J^2 \int d\tau' g(\tau, \tau') g(\tau', \tau'')^{q-1} P_{\alpha\beta} P_{\beta \gamma}^{q-1} = -\delta_{\alpha\gamma} \delta(\tau - \tau'')\,. \label{SPG1-nonconf}
\ee
Taking the diagonal and off-diagonal part, one can write 
\bea
&&P_{\alpha \alpha} (\dd_\tau g)(\tau, \tau'') - J^2 \int d\tau' g(\tau, \tau') g(\tau', \tau'')^{q-1} \mC_{\alpha} = - \delta(\tau - \tau'')\,; \label{SPG1-nonconf-diag}\\
&& P_{\alpha \gamma} (\dd_\tau g)(\tau, \tau'') - J^2 \int d\tau' g(\tau, \tau') g(\tau', \tau'')^{q-1} P_{\alpha\beta} P_{\beta \gamma}^{q-1} = 0\,, \quad \alpha \neq \gamma\,.\label{SPG1-nonconf-offdiag}
\eea
On the Parisi ansatz we have $P_{\alpha \alpha} = a_0$, so $\mC_{\alpha} = \mC$ remains true beyond the conformal limit. 

However, the ansatz in the form (\ref{ansatz-Kamenev}) is inconsistent with the full saddle point equation (\ref{SPG1-nonconf}). Let us assume without loss of generality\footnote{The Parisi form assumption here is mainly for streamlining the notations. The statement is true for a generic matrix $P$ with properties $P_{\alpha \alpha} = a_0$ and $\mC_{\alpha} = \mC$.} that $P$ is a Parisi matrix, which can be represented using (\ref{repr}):
\be
P = \sum_i a_i (\mI_{m_{i+1}}-\mI_{m_i}) + a_0 I_M\,,
\ee
and that we fixed the scaling freedom by setting $\mC_\alpha = \mC = 1$. In this case the equation (\ref{SPG1-nonconf-diag}) can be written in operator form as 
\be
a_0 \dd_\tau \hat{g} - J^2 \hat{g} * \hat{g}^{\circ(q-1)} = \bf{1}\,, \label{factorized-to-exact-1}
\ee
where $\bf{1}$ denotes the delta-function. To rewrite the the off-diagonal equation (\ref{SPG1-nonconf-offdiag}) in the convenient form, we introduce the Parisi matrix $Q$:
\be
Q := P \cdot P^{\circ(q-1)} = \sum_i w_i (\mI_{m_{i+1}}-\mI_{m_i}) + w_0 I_M\,.
\ee
Now we can expand the entire (\ref{SPG1-nonconf-offdiag}) in terms of the Parisi algebra generators and rewrite it in terms of the components: 
\be
a_i \dd_\tau \hat{g} - J^2 w_i \hat{g} * \hat{g}^{\circ(q-1)} =0\,. \label{factorized-to-exact-2}
\ee
We can extract the derivative term from (\ref{factorized-to-exact-1}) and substitute it into (\ref{factorized-to-exact-2}), which yields the equation
\be
J^2 \hat{g} * \hat{g}^{\circ(q-1)} = - \left(1-\frac{w_i}{a_i} a_0\right)^{-1} {\bf 1}\,.
\ee
This equation says that the function $g(\tau, \tau')$ is a conformal correlator, up to a constant factor. However, this contradicts the diagonal equation (\ref{factorized-to-exact-1}). This means that there are no exact saddles which have the factorized form, and we have to break an assumption e.g. about the time independence of $P$. To find the exact UV completion of a factorized solution, we have to modify the factorized ansatz. One can show that this modification must be replica-nondiagonal.

\subsubsection{Strong coupling expansion}
\label{sec:StrongCouplingExp}

One would like to look for solutions of the equations (\ref{saddle-point-1}),(\ref{saddle-point-2}), which we will write in the operator form as 
\bea
\kappa \dd_\tau \hat{G}_{\alpha\gamma} - \sum_\beta \hat{G}_{\alpha\beta} * \hat{\Sigma}_{\beta\gamma} &=& \bf{1}\times \delta_{\alpha\gamma} \,;\\
\hat{\Sigma}_{\alpha\beta} &=& J^2 \hat{G}_{\alpha\beta}^{\circ(q-1)}\,. \label{SPG1-op}
\eea
Based on the above considerations, one has to look for the solution in the form 
\be
G_{\alpha\beta}(\tau, \tau') = g(\tau, \tau') P_{\alpha\beta} + \Phi_{\alpha\beta}(\tau, \tau')\,. \label{ansatz-exact}
\ee
Perturbatively, the solution can be constructed using the strong coupling expansion in $\kappa$: 
\be
\Phi_{\alpha\beta} = \kappa \varphi_{\alpha\beta}^{(1)} + \kappa^2 \varphi_{\alpha\beta}^{(2)} + \dots \,.
\ee
In the position space, the $\Sigma$ is then expanded as follows: 
\be
\Sigma_{\alpha\beta} = g^{q-1} P_{\alpha\beta}^{q-1} + \kappa (q-1) J^2 g^{q-2} P_{\alpha\beta}^{q-2} \varphi_{\alpha\beta}^{(1)} + \dots\,.
\ee
Next we substitute the expansions into (\ref{SPG1-op}) and equate the powers in $\kappa$ to $0$. The $\kappa^0$ equation gives us the saddle point equations for the factorized configurations in the IR limit (\ref{CG-eq-1}),(\ref{CG-eq-2}). They are solved by $g$ given by (\ref{CG-g-sol}) and a Parisi matrix $P$. 

The $\kappa^1$ equation reads 
\be
\frac{1}{J^2} P_{\alpha\gamma} \dd_\tau \hat{g} = \sum_\beta \left[ \hat{\varphi}_{\alpha\beta}^{(1)}* \hat{g}^{\circ(q-1)} P_{\beta\gamma}^{q-1} + (q-1) P_{\alpha\beta} \hat{g} * (\hat{g}^{\circ(q-2)} \hat{\varphi}_{\beta\gamma}^{(1)}) P_{\beta\gamma}^{q-2}\right]\,. \label{kappa1}
\ee
This is again a linear inhomogeneous integral equation for $\hat{\varphi}_{\beta\gamma}^{(1)}$. It is not very tractable analytically in general, but one can make some remarks: 
\begin{itemize}
	\item If $P = I$, then the equation (\ref{kappa1}) and equations for higher degrees of $\kappa$ ensure that $\Phi_{\alpha\beta} = 0$ for $\alpha \neq \beta$, as can be expected from general intuition and known numerical solutions of exact saddle point equations in the replica-diagonal case (see e.g. \cite{MScomments}).  
	\item The replica dependence in the equation (\ref{kappa1}) would mean that if one was to take the replica limit $M \to 0$ like we do in the strong coupling limit in section \ref{sec:Factorized}, this will mean that one would arrive at the integral equation for a function $\varphi(\tau, \tau'; u)$ of three variables, with mixed integrals in time and replica variables. 
\end{itemize}

\section{Discussion}
\label{sec:Discus}

In this paper we have found and discussed replica-nondiagonal saddle points of the replica partition function (\ref{Zreplica}) of the SYK model. The obtained results are schematically presented on Table \ref{table:Sol}. Let us summarize and comment on these results in more details.
\begin{table}[h!]
\centering
\begin{tabular}{|c|c|l|c|l|}
\hline
Replica number $M$        & \multicolumn{2}{c|}{Exact}                                                & \multicolumn{2}{c|}{Strong coupling limit}                       \\ \hline
\multicolumn{1}{|l|}{}    & q=2                                   & \multicolumn{1}{c|}{q=4}          & q=2                                   & \multicolumn{1}{c|}{q=4} \\ \hline
$M>1$, $M \in \ZZ$ & \multicolumn{1}{l|}{$S_{RND}>S_{RD}$} & $S_{RND}>S_{RD}$                  & \multicolumn{1}{l|}{$S_{RND}<S_{RD}$} & $S_{RND}<S_{RD}$         \\ \hline
$0<M<1$                   & \multicolumn{1}{l|}{$S_{RND}<S_{RD}$} & $S_{RND}<S_{RD}$                  & \multicolumn{1}{l|}{}                 &                          \\ \hline
$M \to 0$                 & no solutions                          & \multicolumn{1}{c|}{no solutions} & no solutions                          & $F_{RND}<F_{RD}$         \\ \hline
\end{tabular}
\caption{A summary of replica-nondiagonal solutions considered in the present work. $S_{RND}$ denotes the on-shell action on the dominant nondiagonal solution (among obtained ones), $F_{RND}$ denotes the lowest regularized free energy on a nondiagonal solution.}
\label{table:Sol}
\end{table}

First, we have studied the exact nondiagonal saddles, analytically in the $q=2$ model and numerically in the $q=4$ model.
\begin{itemize}
\item[\textbf{1.}] For $q=2$ we found exact analytic replica-nondiagonal solutions of the saddle point equations, given by the formulae (\ref{G03})-(\ref{G14}), for arbitrary $M > 0$. As we have discussed, these solutions are singular in the limit $M \to 0$. It is easy to check that the on-shell action in the zero replica limit is also singular, which means that there is no well-defined limit of zero replicas on this class of solutions. Another significant property of these solutions is the non-analyticity in coupling constant at $J=0$. This confirms the nonperturbative nature of the nondiagonal solutions, discussed in the section \ref{sec:weak-coupling}. 

\item[\textbf{2.}] Our numerical study shows that for every exact $q=2$ nondiagonal solution there is a numerical real-valued nondiagonal solution in the $q=4$ model. It appears that the lack of the zero replicas limit is also present in the interacting case. The solutions are plotted on the Fig.\ref{fig:iterations-RS-M>0}. Note that in the IR region the diagonal and nondiagonal part exhibit similar behavior up to a numerical factor, and the absolute value of that factor approaches to $1$ as $M$ decreases. We also argue that these solutions probably can be analytically constructed in terms of the $1/M$ expansion, which we will investigate in the future work.

\item[\textbf{3.}] We have shown that the solutions, that we have constructed in the $q=2$ case, are suppressed by the diagonal saddle in the replica path integral for $M>1$ (however, they can dominate in the replica partition function in the $0<M<1$ case). In the $q=4$ SYK the replica-nondiagonal saddles are also subleading for $M>1$. The fact that nondiagonal saddle points are suppressed at integer $M$ is in agreement with the statement that the SYK partition function is self-averaging at large $N$ \cite{Bagrets16,Cotler16,Saad18,Gross16,Garcia-garcia16}, as verified by exact diagonalization numerics. 
\end{itemize}

Second, we have studied the analytic nondiagonal solutions in the strong coupling limit $\beta J \gg 1$. We focused on the class of solutions with the time dependence given by the conformal propagator, and the replica dependence given by a Parisi matrix. The key findings are the following. 

\begin{itemize}

\item[\textbf{4.}] We were able to obtain solutions in the $M \to 0$ limit, using the Parisi ansatz analytic continuation. We have obtained the nondiagonal replica-symmetric solutions, and the solutions with one-step replica symmetry breaking. Meanwhile, the findings of the previous part suggest that these solutions have no well-defined UV completion at $M = 0$, and therefore seem to have no influence on the thermodynamics of SYK at finite coupling. Nevertheless, we calculated the regularized free energy on the solutions. Among them there are some that have global minimum regularized free energy value smaller than the replica-diagonal free energy. We have also checked that at finite $M$ the replica-symmetric solution dominates over the diagonal saddle in the strong coupling limit for $q=4$ and $q=2$. 

\item[\textbf{5.}] We have illustrated on an $M =2$ example that the extended reparametrization symmetry in the IR limit has interesting consequences, since it can be used to obtain solutions that have spin-glass-like dynamics for a finite amount of time. Another solution that can be obtained using reparametrizations is related by analytic continuation to the purification of the replica-diagonal SYK by the thermofield double. 
\end{itemize}

An important technical question that we left unattended in this work is the problem of stability of the replica-nondiagonal solutions that we have obtained in the interacting $q=4$ model. The numerical solutions of the exact saddle point equations appeared to be stable to small enough fluctuations in the initial condition. As far as the factorized solutions in the strong coupling regime are concerned, the following comments can be made. We can say that the replica-diagonal solution is stable with respect to the replica-diagonal fluctuations of the field $G(\tau_1, \tau_2)$, because the quadratic part of the action is determined by the ladder kernel and it has non-negative eigenvalues \cite{MScomments}. The numerical investigations of \cite{Fu16,Gur-Ari18} confirm that there is no instability to general fluctuations of the replica-diagonal saddle point. Since the time dependence of the factorized solutions is the same, we can expect that they also would be stable with respect to the replica-diagonal fluctuations. However, we also can expect that some of the solutions discussed in the section \ref{sec:RSB1} can be unstable to general fluctuations, because they are defined by pairs of complex saddles, and one would need to take into account fluctuations around both saddles for every solution. It could also be instructive to consider other solutions with several steps or continuous replica symmetry breaking to verify these observations. We leave these questions for the future study. 

It is also important to clarify the role of the replica-nondiagonal solutions obtained in the present paper in regards to the previous results about replica-nondiagonality in SYK obtained in the literature. As was mentioned in the introduction, the papers \cite{Sachdev92,Georges00,Fu16,Ye18,Gur-Ari18,Caracciolo18} studied the question of the spin glass phase in SY and SYK models, and they argue against the existence of such a phase in fermionic models (but they do not rule out replica-nondiagonal solutions conclusively in general). The exact replica-nondiagonal solutions, constructed in the present work in sections \ref{sec:q=2} and \ref{sec:Numerics} are time dependent (so do not generally describe glassy physics), and, more importantly, are always subleading in the replica path integral at $M > 1 $. Therefore, they do not introduce new phases, but remain a small nonperturbative effect within the $1/N$ expansion of the partition function and some annealed correlators, see Appendix \ref{sec:Correlators}. 

The second class of solutions, that we study in this paper, is restricted to the IR limit of the SYK model. We treat the IR limit of SYK as essentially a low-energy effective field theory with a UV cutoff, and we found nontrivial phase structure in this EFT by studying the regularized free energy. The fact that we have not found any replica-nondiagonal saddles, contributing to the free energy after the zero replicas limit beyond the IR limit, hints that  the phase structure of an EFT does not generally match the phase structure of the UV completion (see also discussion in the end of sec.\ref{sec:Approach}). It is worth noting that the effect of destruction of conformal replica-nondiagonal saddle points by the UV corrections was encountered by authors of \cite{Saad18} in the two-replica case. They explicitly show that the leading UV-correction to the replica-nondiagonal conformal saddle point of the spectral form factor introduces an instability to the contribution of this solution to the path integral. We can expect that at least some of our solutions in the $M \to 0$ limit suffer from the same effect beyond the strong coupling limit.

Let us note that the solutions that we have constructed can be interesting in the following aspects. 
\begin{itemize}

\item[(i)] We have shown that the annealed quantites, which would require a finite number of replicas, do have nontrivial repica-nondiagonal saddle points, but they are subleading in the replica partition function. In the case of $M=2$ and complex-valued $\beta$ it is shown that non-trivial saddles can become dominant and are crucial for the quantum dynamics of black holes in the work \cite{Cotler16,Saad18}. Therefore, an interesting problem is to study the counterparts of our saddles in the spectral form-factor, and to see if they can be responsible for the long-time behavior of this quantity. 

\item[(ii)] The factorized solutions, which we discussed, also have the emergent conformal symmetry in the strong coupling regime, which suggests the applicability of the holographic description in some sense as well. At finite replica number this would suggest that we need some kind of asymptotically AdS$_2$ space with multiple boundaries. However, we are not aware of such solutions except for the AdS$_2$ itself, which we connected in the section \ref{sec:Holography} to the $M=2$ solution. It would be interesting to find holographic duals to other replica structures. 
\end{itemize}

\section*{Acknowledgments}

We are grateful to Ksenia Bulycheva, Juan Maldacena, Andrey Mikhailov, Douglas Stanford and  Zhenbin Yang for useful discussions and remarks. We are also particularly grateful to Douglas Stanford for sharing the code for numerical solution of saddle point equations in the replica-diagonal case.
Our Mathematica code for finding nondiagonal numerical solutions is availible on request. Parts from this work were presented at "Quarks 2018" conference in Valday, Russia in the talk \cite{ValdayAKTV} and at "Strings 2018" conference at OIST, Okinawa in a poster presentation. The work is supported by the Russian Science Foundation (project 17-71-20154).

\appendix

\section{Parisi matrices}
\label{sec:Parisi}

\subsection{Definition} 

\begin{figure}[t!]
 \centering \includegraphics[scale=0.3]{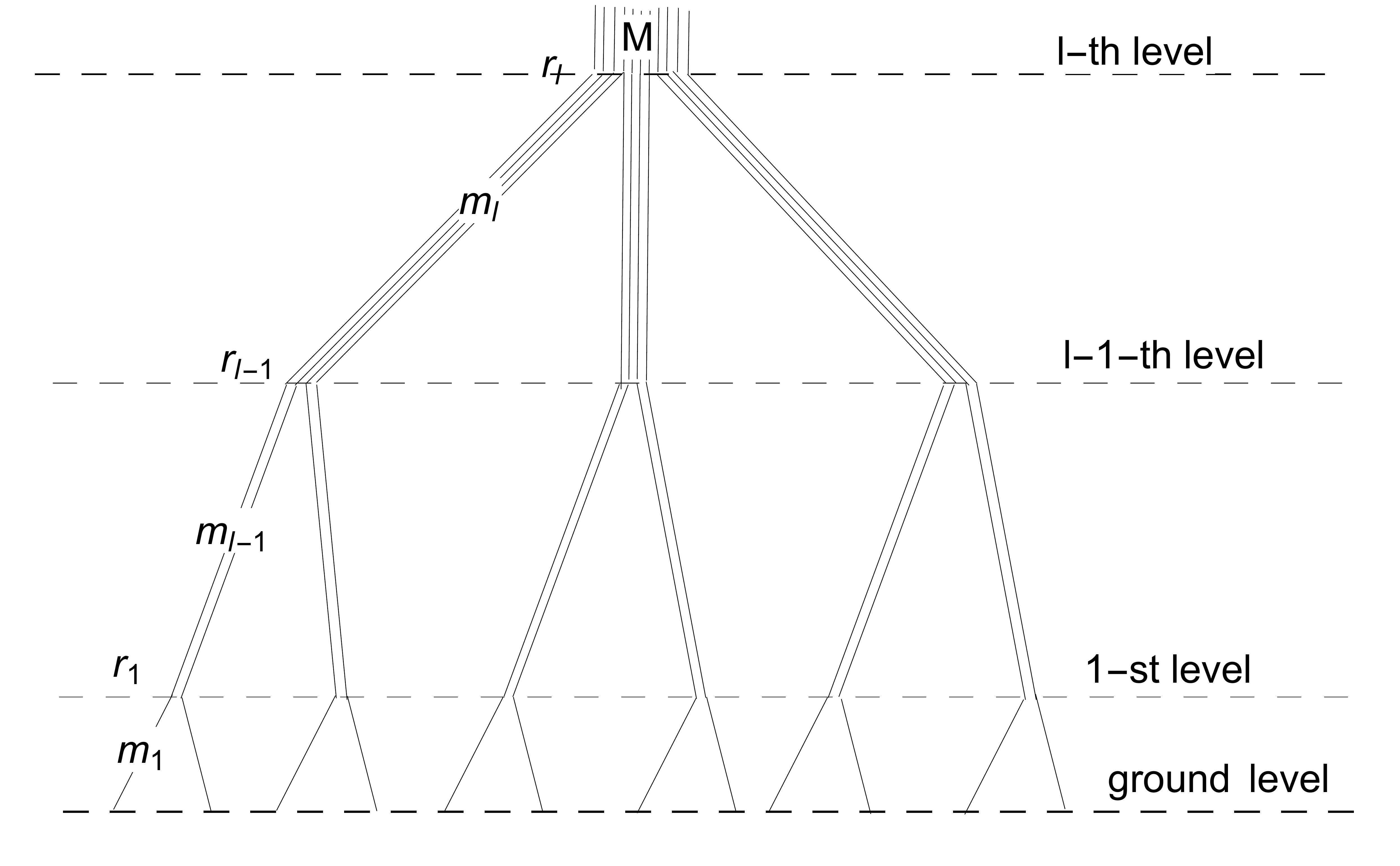}
 \caption{Illustratuion of the tree structure of a Parisi matrix with given ramifications $\{ r_1, r_2, r_3... r_l\}$. The dimensions $\{m_1, m_2, r_3... m_l,M\}$ are represented by thickness of the branches}
  \label{Fig:TT}
\end{figure}

 A symmetric $M\times M$ matrix  $ Q $ is called the Parisi matrix if it is defined in the following way. 
The Parisi matrix is associated with a tree\be
\fT=\fT( r_1, r_2, r_3... r_l)\ee
characterized by the set $\{ r_1, r_2, r_3... r_l\}$ satisfying 
\be
r_1\cdot r_2...r_l=M\ee
Note that here the order of $r_i$ is important. From this set of numbers one constructs a set of number
\be
\fD_\fT=\fD_\fT( m_1, m_2, r_3... m_l,M),\ee
where $m_i, i=1,...l$   are
\be
m_i=
\prod _{j=1}^{i-1}r_j.\label{mi}\ee
As follows from \eqref{mi} the number $r_i, m_i$ and $m_{i+1}$
are related as 
\be
m_{i+1}=m_i\cdot r_i,\,\,\,\,\,i=1,...l-1,\,\,\,\,\,M=m_l\cdot r_l.\ee
The numbers $\{r_l, ...r_1\}$  characterize the ramifications in the given tree and $\{m_l, ...m_1\}$, $m_1=1$ characterize the thickness  of  branches  corresponding to the same tree, see Fig.\ref{Fig:TT}. $l$ is a number of branch points\footnote{One can repeat the the above description of the tree in other words: each tree is characterized by dividing the elements of M into $ r_l $ groups with
$ m_l $ elements in each, 
i.e. \be
M=r_l\cdot m_l.
\ee Then we divide $m_l$ elements on $r_{l-1}$ groups with $m_{l-1}$ elements in each, i.e. 
\be m_l=m_{l-1}r_{l-1}
\ee and so on. On the last step we left with $r_1$ elementary elements, i.e  dim $r_1=1$.}. We call the number $l$ the \textit{rank} of the Parisi matrix. 

{\bf Definition.}  {\it The Parisi matrix $Q$ associated with the given tree $\fT$ with a set $\fD_\fT$ is defined as follows}
\bea
Q_{a,a}&=&q_0,\nn\\
Q_{a,b}&=&Q_{b,a}=q_i,\,\,\, \text{if}\,\,\, 
\left\{
\begin{array}{ccc}
 \Big[\frac{a-1}{m_i}\Big]&\neq&\Big[\frac{b-1}{m_i}\Big]    \\
  &  \text{and} &   \\
  \Big[\frac{a-1}{m_{i+1}}\Big]&=&\Big[\frac{b-1}{m_{i+1}}\Big]    
 \end{array}\right\}\\
 a,b=1,...,M\label{DP}
\eea
\subsection{Representation in terms of block matrices $\mathcal{I}_{m_i}$}
It is convenient to represent the Parisi matrix \eqref{DP} using the family of the block matrices $\mathcal{I}_{m_i}$ composed on 1's and 0's
\be
\mathcal{I}_{m_i}= I_{M/m_i}\otimes\fJ _{m_i}\label{II}\ee
Here $\fJ_l$ are Hadamard identity matrices of dimension $m_i$: 
\be
(\fJ_i)_{kj}= 1\,, \quad k,j = 1,\dots,m_i\,.
\ee
and $I_p$ are the usual unit matrix of dimension $p$, $\otimes $ is the tensor product. The matrix $Q$ defined by \eqref{DP} can be represented as 
\be
Q=\sum_{i=1,\,\,m_i\in \fT}^{l}\,q_i(\mathcal{I}_{m_{i+1}}-\mathcal{I}_{m_{i}})+q_0\mathcal{I}_{1}
\label{repr}\ee 
Note that in according with \eqref{II}
\be
\mathcal{I}_{m_1}\equiv\mathcal{I}_{1}=\fJ _{1}\otimes I_{M}=I_{M}\label{IM}\ee
Notice that representation \eqref{repr} is equivalent to the following representation
\bea
Q&=&q_0\mathcal{I}_{1}+q_1(\mathcal{I}_{m_{2}}-\mathcal{I}_{m_{1}})+
q_2(\mathcal{I}_{m_{3}}-\mathcal{I}_{m_2})+...+q_l(\mathcal{I}_{m_{l+1}}-\mathcal{I}_{m_{l}})
\nn\\&=&(q_0-q_1){I}_{M}+(q_1-q_2)\mathcal{I}_{m_{2}}+
(q_2-q_3)\mathcal{I}_{m_{3}}+...+(q_{l-1}-q_l)\mathcal{I}_{m_{l}}+q_l\fJ_M\,,
\label{reprI}
\eea
where we use \eqref{IM} and 
\be
\mathcal{I}_{m_{l+1}}=\fJ_M\,. \label{AIM}\ee
Rewriting \eqref{reprI} we get
\bea
Q&=&(q_0-q_1)\mathcal{I}_{1}+\sum _{i=1}^{l-1}(q_i-q_{i+1})\mathcal{I}_{m_{i+1}}+q_l\fJ_M\nn\\
&=&\sum _{i=0}^{l-1}(q_i-q_{i+1})\mathcal{I}_{m_{i+1}}+q_l\fJ_M\label{reprII}
\eea
\subsection{Algebra of Parisi matrices} 
\label{sec:ParisiAlgebra}
The block matrices $\mathcal{I}_{m_{i}}$ satisfy the following relations
\bea
\mathcal{I}_{m_{i}}\mathcal{I}_{m_{j}}=\mathcal{I}_{m_{j}}\mathcal{I}_{m_{i}}=m_i \mathcal{I}_{m_{j}}, \,\,\,\mbox{for} \,\,\,\,i\leq j\label{prod}\eea
From this statement follow the 

\begin{Lemma} The space of Parisi matrices with fixed tree $\fT\{ r_1, r_2, r_3... r_l\}$ is an algebra under both regular and Hadamard matrix products.
\end{Lemma}

\textbf{Proof.} The closeness under the Hadamard product is obvious. For the direct matrix product this follows from representation   
\eqref{repr} and the following properties \eqref{prod} of $\mathcal{I}_{m_i}$ matrices corresponding to the same tree. Indeed
we define the two Parisi matrices of the rank $l$ ($a_{l+1}=0$):
\bea
&&A = \sum_{i=1,\,\,m_i\in \fT}^{l}\,a_i(\mathcal{I}_{m_{i+1}}-\mathcal{I}_{m_{i}})+q_0\mathcal{I}_{1}\,;\\
&& B= \sum_{i=1,\,\,m_i\in \fT}^{l}\,b_i(\mathcal{I}_{m_{i+1}}-\mathcal{I}_{m_{i}})+b_0\mathcal{I}_{1}\,.
\eea
We want to calculate their product, which we denote as 
\be
W = A \cdot B\,.
\ee
We proceed as follows: 
\bea
&& W = a_0 b_0 \mI_1 + \sum_i (b_0 a_i + a_0 b_i) (\mathcal{I}_{m_{i+1}}-\mathcal{I}_{m_{i}}) + \text{four terms}\,.
\eea
The four terms come from the multiplication of two brackets between each other. Let us evaluate carefully each term using the relation (\ref{prod}). The first term gives
\bea
&& \mathrm{T} _1:= \sum_i \sum_j a_i b_j \mI_{m_{i+1}} \mI_{m_{j+1}} = \sum_j \sum_{i < j} a_i b_j m_{i+1} \mI_{m_{j+1}} + \sum_j a_j b_j m_{j+1} \mI_{m_{j+1}} + \sum_j \sum_{i > j} a_i b_j m_{j+1} \mI_{m_{i+1}} \nn
\\
&& = \sum_j \sum_{i < j} (a_i b_j + a_j b_i) m_{i+1} \mI_{m_{j+1}} + \sum_j a_j b_j m_{j+1} \mI_{m_{j+1}}\,.
\eea
The second term yields
\bea
&& \mathrm{T} _2:= - \sum_i \sum_j a_i b_j \mI_{m_i} \mI_{m_{j+1}} = - \sum_j \sum_{i < j+1} a_i b_j m_i \mI_{m_{j+1}} -\sum_j  a_{j+1} b_j m_{j+1} \mI_{m_{j+1}} - \sum_j \sum_{i > j+1} a_i b_j m_{j+1} \mI_{m_{i}} \nn\\
&& \!\!\!\!\!\!= - \sum_j \sum_{i < j} a_i b_j m_i \mI_{m_{j+1}} - \sum_{j} a_j b_j m_j \mI_{m_{j+1}} -  \sum_j \sum_{i < j} a_j b_i m_{i+1} \mI_{m_{j}}
\,.
\eea
We relabeled some summation indices and canceled two terms here. The third term can be obtained from $T_2$ by making the replacement $a \leftrightarrow b$:
\be
\mathrm{T} _3:= - \sum_i \sum_j a_i b_j \mI_{m_i+1} \mI_{m_{j}} =  - \sum_j \sum_{i < j} b_i a_j m_i \mI_{m_{j+1}} - \sum_{j=i} a_j b_j m_j \mI_{m_{j+1}} -  \sum_j \sum_{i < j} b_j a_i m_{i+1} \mI_{m_{j}} \,.
 \ee
The fourth term yields: 
\bea
 \mathrm{T} _4:&=& \sum_i \sum_j a_i b_j \mI_{m_i} \mI_{m_{j}} = \sum_j \sum_{i < j} a_i b_j m_i \mI_{m_{j}} + \sum_{j = i} a_j b_j m_j \mI_{m_j} + \sum_j \sum_{i < j} a_j b_i m_i \mI_{m_j}\\
&=&  \sum_j \sum_{i < j}( a_i b_j + a_jb_i) m_i \mI_{m_{j}} + \sum_{j = i} a_j b_j m_j \mI_{m_j} 
\eea
where we again relabeled the indices in the last term. Therefore, we have presented the four terms as a linear combination of the $\mI$-matrices, more specifically
\be
\text{four terms} = \sum_{j}( U_j \mI_{m_{j+1}} + V_j \mI_{m_j} )\,.
\ee
As it will become clear below, this is already proves the lemma. 
Let us write down the $U$ and $V$ explicitly:
\bea
&& U_j = \sum_{i < j} (a_i b_j + b_i a_j) (m_{i+1} - m_i) + a_j b_j (m_{j+1} - 2 m_j)\,.\\
&& V_j = -\sum_{i < j} (a_i b_j + b_i a_j) (m_{i+1} - m_i) + a_j b_j m_j\,.
\eea
As we see, there is a difference between the two:
\be
D_j : = U_j + V_j = a_j b_j (m_{j+1} - m_j)\,,
\ee
and therefore in $W$ there is a rogue term 
\be
\sum_j D_j \mI_{m_{j+1}}
\ee
However, we can deal with it using a trivial formula: 
\be
\mI_{m_k} = \mI_1 + \sum_{j < k} (\mI_{m_{j+1}} - \mI_{m_{j}})\,. \label{trivial}
\ee
Thus, we can write the term as follows: 
\be
\sum_{i} D_i \mI_{m_{i+1}} = \sum_i D_i \mI_1 + \sum_i \sum_{j < i+1} D_i (\mI_{m_{j+1}} - \mI_{m_{j}}) = \sum_i D_i \mI_1 + \sum_j \sum_{i > j-1} D_i (\mI_{m_{j+1}} - \mI_{m_{j}})\,.
\ee
Therefore, the lemma is proved. We obtain that $W$ is a Parisi matrix 
\be \label{product}
W = \sum_j w_j (\mI_{m_{j+1}} - \mI_{m_j}) + w_0 \mI_1\,,
\ee
where its parameters are given by
\bea 
w_0 &=& a_0 b_0 + \sum_j a_j b_j (m_{j+1} - m_j) ; \label{w0}\\
 w_j &=& b_0 a_j + a_0 b_j + \sum_{i < j} (a_i b_j + b_i a_j) (m_{i+1} - m_i) - a_j b_j m_{j} +  \sum_{i > j-1} a_i b_i (m_{i+1} - m_i)\nn\\
& =& b_0 a_j + a_0 b_j + \sum_{i < j} (a_i b_j + b_i a_j) (m_{i+1} - m_i) - a_j b_j m_{j} +  \sum_{i > j} a_i b_i (m_{i+1} - m_i)\nn\\
&+&a_j b_j(m_{j+1} - m_j)\label{wj}\,.
\eea
\qed\\
These formulae can be used to directly solve the equation (\ref{CG-eq-matrix}), if one takes $b_j = a_j^{q-1}$. 

\subsection{Determinant of the Parisi matrix}
\label{sec:ParisiDet} 

The eigenvalues of the Parisi matrix $Q$ are given by the formulae \cite{Khr}
\bea
\lambda_0&=&q_0 -q_1\\
\lambda_i&=&q_0 -q_1+\sum _{j=1}^i(q_j-q_{j+1})m_j,\,\,\,\,\,\,i=1,2,...,l-1 \\
\lambda_l&=&q_0 -q_1+\sum _{j=1}^{l-1}(q_j-q_{j+1})m_j+q_l m_{l}.\eea

Throughout the paper we compute $\log \det Q$ for different Parisi matrices both at finite and zero $M$. We illustrate the calculation for $l=1$ and present the result for general $l$. 

In the case of $l = 1$, the determinant for arbitrary $M$ reads: 
\be
\det Q = (q_0 - q_1)^{M-1} (q_0 + (M-1) q_1) \,.
\label{detP}\ee
For the logarithm we write 
\bea
&& \log \det Q = (M-1) \log (q_0 - q_1) + \log (q_0 - q_1) +\log \left(1 + \frac{M q_1}{q_0 - q_1} \right) \nn\\&& =  M \log (q_0 - q_1) + \log \left(1 + \frac{M q_1}{q_0 - q_1} \right)\,. \label{TrLogQ-finM}
\eea
Taking the limit $M \to 0$, we arrive at the expression 
\be
\lim_{M\to 0} \frac{1}{M} \log \det Q = \log(q_0-q_1) + \frac{q_1}{q_0-q_1}\,.
\ee
The cases of higher $l$ can be calculated analogously. The main idea is to extract the part from the lowest degree bracket which is the same as the next degree bracket, which will always be of the odd degree and carry a singular contribution. 

For arbitrary $l$ we obtain the expression
\bea
\lim_{M\to 0} \frac{1}{M} \log \det Q &=& \frac{1}{2}\log(q_0-q_1)+\sum_{i=2}^{l-1} \frac{1}{2^{i}}\log(q_0+\sum_{m=0}^{i-1}2^m q_{m+2}-2^i q_{i+1}) \label{logDet}\\
&+&\frac{1}{2^{k-1}}\log (q_0+\sum_{m=0}^{l-1} 2^m q_{m+2}-2^{l}q_l) +\frac{1}{2}\frac{q_l}{q_0+\sum_{m=0}^{l-1}2^m q_{m+2}-2^{l}q_l}\,.\nn
\eea

\section{The action on replica-nondiagonal solutions at finite $M$}
\label{sec:appB}

We start with the on-shell action for the path integral (\ref{Zreplica}) at finite $M$:
\be
\frac{2}{N} S_{M}= - \Tr \log \left(\delta_{\alpha\beta} \dd_\tau- \hat{\Sigma}_{\alpha\beta}\right) +\int_0^\beta\int_0^\beta d\tau_1 d \tau_2 \sum_{\alpha,\beta} \left(G_{\alpha\beta}(\tau_1, \tau_2) \Sigma_{\alpha\beta}(\tau_1, \tau_2)- \frac{J^2}{q}G_{\alpha\beta}(\tau_1, \tau_2)^q\right)\Big|_{\text{on-shell}}\,. \label{S_M}
\ee
We use the saddle point equations (\ref{saddle-point-1}),(\ref{saddle-point-2}) and rewrite it as
\bea
\frac{2}{N} S_M &=& -\Tr \log[\delta_{\alpha\beta}\dd_\tau- \hat{\Sigma}_{\alpha\beta}]  +\left(1-\frac1q \right)J^2 \int_0^\beta\int_0^\beta d\tau_1 d \tau_2 \sum_{\alpha,\beta} G_{\alpha\beta}(\tau_1, \tau_2)^{q} \Big|_{\text{on-shell}}\,.
\eea
We renormalize the logarithmic term by subtracting the free part $-M \Tr\log(\dd_\tau)$ \cite{Kitaev17,MScomments} and denote it as: 
\be
 \fs_1 = - \Tr \log \det\left(\delta_{\alpha\beta} + \frac{\hat{\Sigma}_{\alpha\beta}}{-\dd_\tau}\right)\Big|_{\text{on-shell}} \,,\label{s1}
 \ee
where the determinant is taken over the replica indices. We denote the polynomial term as 
\be
\fs_2 = \left(1-\frac1q \right)J^2 \int_0^\beta\int_0^\beta d\tau_1 d \tau_2 \sum_{\alpha,\beta} G_{\alpha\beta}(\tau_1, \tau_2)^{q}\Big|_{\text{on-shell}}\,. \label{s2}
\ee

\paragraph{Replica-symmetric ansatz.} In this case we have two dynamical variables $G_{0,1}$ (and corresponding $\Sigma_{0,1}(\tau)=J^2 G_{0,1}^{q-1}(\tau)$). To evauate the Pfaffian term in the frequency space we use the formula for the determinant of a Parisi matrix (\ref{TrLogQ-finM}) for every frequency: 
\be
\log \det\left(\delta_{\alpha\beta} + \frac{\Sigma_{\alpha\beta}(\omega)}{i \omega}\right) = M \log \left(1 + \frac{\Sigma_0(\omega)}{i \omega} - \frac{\Sigma_1(\omega)}{i \omega}\right) + \log \left(1+M \frac{\Sigma_1(\omega)}{i\omega + \Sigma_0(\omega) - \Sigma_1(\omega)}\right)\,. \label{logdet-dd-s}
\ee
At zero temperature, $\fs_1$ reads
\be
\fs_1= -\frac{\mathcal{V}}{2\pi} \int_{-\infty}^\infty d\omega \left[M \log \left(1 + \frac{\Sigma_0(\omega)}{i \omega} - \frac{\Sigma_1(\omega)}{i \omega}\right) + \log \left(1+M \frac{\Sigma_1(\omega)}{i\omega + \Sigma_0(\omega) - \Sigma_1(\omega)}\right)\right]\,. \label{fs1-RS}
\ee
At finite temperature we have instead the sum over Matsubara frequencies (\ref{Matsubara}):
\be
\fs_1 = -\sum_{n=-\infty}^{\infty} \left[M \log \left(1 + \frac{\Sigma_0(\omega_n)}{i \omega_n} - \frac{\Sigma_1(\omega_n)}{i \omega_n}\right) + \log \left(1+M \frac{\Sigma_1(\omega_n)}{i\omega_n + \Sigma_0(\omega_n) - \Sigma_1(\omega_n)}\right)\right]\,.
\ee
For the polynomial term, we have the expression 
\be
\fs_2 = \left(1-\frac1q\right) J^2 \int d\tau_1 d\tau_2 \left( M G_0(\tau_1, \tau_2)^q + M(M-1) G_1 (\tau_1, \tau_2)^q \right)\,. \label{fs_2-RS}
\ee

\section{Observables and disorder: annealed vs quenched}
\label{sec:Correlators}

Having found replica-nondiagonal solutions in $q=2$ and $q=4$ SYK models, it is appropriate to discuss which observables will be affected by these saddle points. For this purpose, it is useful to review the two types of correlation functions, which one can consider in a disordered model.

First, let us remind that the Lagrangian of the SYK model is given by \cite{Polchinski16,MScomments,Kitaev17}
\be
L[\psi, \bj] =- \frac12 \sum_i \psi_i \frac{d}{d \tau} \psi_i -  \frac{i^{q/2}}{q!} \sum_{i_1, i_2,\dots,i_q} j_{i_1 i_2\dots\i_q} \psi_{i_1} \psi_{i_2} \dots \psi_{i_q}\,, \label{Lsyk} 
\ee
which is used to construct the generating functional of SYK correlation functions for a fixed realization of disorder ${\bf j} = \{ j_{i_1 i_2\dots\i_q} \}$: 
\be
Z_{\bj}(\beta; \eta) = \int D\psi \exp\left[-\int_0^\beta d\tau L[\psi, \bj] + \int_0^\beta \eta_i (\tau) \psi_i (\tau) d\tau \right]\,. \label{Zreplica-in}
\ee
Here we have introduced the fermionic sources $\eta_i(\tau)$. The correlation functions in a fixed realization of disorder are defined as usual: 
\be
\langle \psi_{i_1}(\tau_1)\dots \psi_{i_k} (\tau_k) \rangle_{\bj}= \frac{1}{Z_{\bj}(\beta; 0)} \frac{1}{k!} \left.\frac{\delta}{\delta \eta_{i_1}(\tau_1)} \dots \frac{\delta}{\delta \eta_{i_k}(\tau_k)} Z_{\bj}(\beta; \eta)\right|_{\eta = 0}\,. \label{correlators}
\ee
As a final preliminary remark, the average over the disorder of a function $f({\bf j})$ is performed by taking the integral
\be
\overline{f({\bf j})} = \int d {\bf j}\ P({\bf j}) f({\bf j})\,, \label{averageF}
\ee
where the Gaussian distribution $P({\bf j})$ is given by (\ref{Gaussian}). 

\subsection{Quenched quantities}

In the quenched quantities the disorder averaging is performed on the last step. The quenched correlators are defined as 
\be
\overline{\langle \psi_{i_1}(\tau_1)\dots \psi_{i_k} (\tau_k) \rangle}= \overline{\frac{1}{Z_{\bj}(\beta; 0)} \frac{1}{k!} \left.\frac{\delta}{\delta \eta_{i_1}(\tau_1)} \dots \frac{\delta}{\delta \eta_{i_k}(\tau_k)} Z_{\bj}(\beta; \eta)\right|_{\eta = 0}}\,. \label{quenched}
\ee
Because the nominator and denominator are averaged together, these quantities require replica trick, and specifically they require taking the limit $M \to 0$. Below we explicitly derive the representation of the quenched correlators in terms of the bilocal replica field path integral. 

\subsubsection{Replica-diagonal case}

We start with the singlet two-point function of fermions in a single copy of SYK\footnote{We also emphasize the $N$ dependence for purposes of the further discussion}: 
\be
\mG(\tau_1, \tau_2; N) := \frac1N \sum_{i=1}^N \overline{\langle \psi_i (\tau_1) \psi_i (\tau_2) \rangle} = \frac{1}{2!N} \sum_{i=1}^N \overline{\frac{1}{Z_{\bj}(\beta; 0)} \left.\frac{\delta}{\delta \eta_{i}(\tau_1)} \frac{\delta}{\delta \eta_{i}(\tau_2)} Z_{\bj}(\beta; \eta)\right|_{\eta = 0}}\,. \label{quenched2point}
\ee
Now let us derive the expression for $\mG$ in terms of the disorder-averaged bilocal replica field path integral. Let us define the bilocal field correlator in the disorder-averaged SYK with $M$ replicas: 
\bea
\fG (\tau_1, \tau_2; N,M)&=&\label{repfG} \int \prod_{\alpha,\beta = 1}^M D G_{\alpha\beta} D\Sigma_{\alpha\beta}\ \e^{-N S[G_{\alpha\beta}, \Sigma_{\alpha\beta}]} G_{MM}(\tau_1, \tau_2)\,.
\eea
In terms of this quantity, the quenched correlator is written as following: 
\be
\mG(\tau_1, \tau_2; N) = \lim_{M \to 0} \fG (\tau_1, \tau_2; N,M)\,.\label{quenched-diag}
\ee

Let us sketch the proof. On the first step we take the derivatives in (\ref{quenched2point}) and write the nominator partition function explicitly as the path integral. For the normalizing denominator, we use the following formal identity \cite{SpinGlassBook}:
\be
Z_{\bj}(\beta; 0)^{-1} = \lim_{M \to 0} Z_{\bj}(\beta; 0)^{M-1}\,.
\ee
We get (the summation over color indices is implicit)
\be
\mG(\tau_1, \tau_2; N) = \frac{1}{N} \overline{\lim_{M \to 0} Z_{\bj}(\beta; 0)^{M-1}\int D\psi^M\ \e^{-S_{\text{SYK}}[\bj, \psi^M]} \psi_i^M(\tau_1) \psi_i^M (\tau_2) }\,,
\ee
where we have assigned a replica index $M$ to fermion fields participating in the last path integral. We can rewrite this as the path integral over $M$ replicas with insertions supported on the $M$-th replica: 
\be
\mG(\tau_1, \tau_2; N) = \frac{1}{N} \overline{\lim_{M \to 0}\int \prod_{\alpha=1}^M D\psi^\alpha\ \e^{-\sum_{\alpha=1}^M S_{\text{SYK}}[\bj, \psi^\alpha]} \psi_i^M(\tau_1) \psi_i^M (\tau_2) }\,,
\ee
The averaging over the disorder can be performed in this replica integral in the same way as in the partition function (\ref{Zreplica}). After that, we can rewrite this in terms of the bilocal fields with the fermions integrated out: \footnote{This identity holds up to some numerical constants which come from the measure \cite{Kitaev17}, which we omit}:
\be
\overline{\int \prod_{\alpha'=1}^M D\psi^{\alpha'}\,\e^{-\sum_{\alpha'=1}^M S_{\text{SYK}}[\bj, \psi^{\alpha'}]}\frac1N\sum _i \psi_i^M(\tau_1) \psi_i^M (\tau_2) } = \int \prod_{\alpha,\beta=1}^M D G_{\alpha\beta} D\Sigma_{\alpha\beta}\ \e^{-N S[G_{\alpha\beta}, \Sigma_{\alpha\beta}]} G_{MM}(\tau_1, \tau_2)\,. 
\ee
Thus, we obtain the formula (\ref{quenched-diag}):
\be
\mG(\tau_1, \tau_2; N) = \lim_{M \to 0} \int \prod_{\alpha,\beta=1}^M D G_{\alpha\beta} D\Sigma_{\alpha\beta}\ \e^{-N S[G_{\alpha\beta}, \Sigma_{\alpha\beta}]} G_{MM}(\tau_1, \tau_2)\,. 
\ee
We see that a quenched correlator is expressed in terms of replica correlators in the limit $M \to 0$. Note that this limit is taken \textit{before} the thermodynamic limit $N \to \infty$. 

\subsubsection{Replica-offdiagonal case}
\label{sec:RODD}

A common diagonostic of the spin glass physics is the \textit{quenched replica-offdiagonal} correlator \cite{Gur-Ari18,Sachdev15,Sachdev92,Georges00,Anninos16,ParisiBook,Fu16,SK}. Let us start with the two replicas $\alpha$ and $\beta$ of the given system. The offdiagonal quenched correlator is defined as 
\be
\mG_{\alpha\beta} (\tau_1, \tau_2; N) := \frac1N \sum_{i=1}^N \overline{\langle \psi_i^\alpha (\tau_1) \psi^\beta_i (\tau_2) \rangle} = \frac{1}{2!N} \sum_{i=1}^N \overline{\frac{1}{Z^{(2)}_{\bj}(\beta; 0)} \left.\frac{\delta}{\delta \eta^\alpha_{i}(\tau_1)} \frac{\delta}{\delta \eta^\beta_{i}(\tau_2)} Z_{\bj}^{(2)}(\beta; \eta^\alpha, \eta^{\beta}) \right|_{\eta = 0}}\,. 
\ee
Here $Z^{(2)}$ is the two-replica partition function. If the two replicas are independent, then  $Z^{(2)}_\bj(\beta; \eta^\alpha, \eta^\beta) = Z_\bj (\beta; \eta^\alpha) Z_\bj (\beta; \eta^\beta)$ and the expression under line can factorizes into product of two one-point functions: 
\be
\overline{\frac{1}{Z_{\bj}(\beta; 0)^2} \left.\frac{\delta}{\delta \eta^\alpha_{i}(\tau_1)} \frac{\delta}{\delta \eta^\beta_{i}(\tau_2)} Z_{\bj}(\beta; \eta^\alpha)  Z_{\bj}(\beta; \eta^\beta)\right|_{\eta = 0}} = \overline{\langle \psi_{i}^\alpha(\tau_1) \rangle \langle \psi_{i}^\beta(\tau_1) \rangle }\,.
\ee
In the fermionic theory, like SYK, the one-point functions vanish, which, assuming $N$ is finite, gives that in total $\mG_{\alpha\beta} (\tau_1, \tau_2)=0$\footnote{This argument does not generally work in the thermodynamic limit $N \to \infty$.}. This is a general argument that is often presented to support the absense of spin glass phase in fermionic models such as variations of SY and SYK \cite{Gur-Ari18,Sachdev92,Georges00,Anninos16,Fu16}. 

Let us now see what this tells us about the replica structure of the $G$, $\Sigma$ saddle points. We again define the replica bilocal field correlator 
\bea
\fG_{\alpha\beta} (\tau_1, \tau_2; N,M)&=&\label{repfGab} \int \prod_{\alpha,\beta = 1}^M D G_{\alpha\beta} D\Sigma_{\alpha\beta}\ \e^{-N S[G_{\alpha\beta}, \Sigma_{\alpha\beta}]} G_{\alpha\beta}(\tau_1, \tau_2)\,, \quad \alpha \neq \beta\,.
\eea
Note that, formally speaking, $\fG (\tau_1, \tau_2; N,M) = \fG_{MM} (\tau_1, \tau_2; N,M)$ (see (\ref{repfG})). Analogously to the diagonal case (\ref{quenched-diag}), one can derive the bilocal field representation for the quenched offdiagonal correlator: 
\be
\mG_{\alpha\beta} (\tau_1, \tau_2; N) = \lim_{M \to 0} \fG_{\alpha\beta} (\tau_1, \tau_2; N,M)\,. \label{quenched-offdiag}
\ee

The proof goes as follows. Proceeding analogously to the derivation of the expression (\ref{quenched-diag}) above, we write 
\be 
\mG_{\alpha\beta} (\tau_1, \tau_2; N) = \frac{1}{N} \overline{\lim_{M \to 0} Z_{\bj}(\beta; 0)^{M-2}\int D\psi^\alpha D\psi^\beta\ \e^{-S_{\text{SYK}}[\bj, \psi^\alpha]-S_{\text{SYK}}[\bj, \psi^\beta]} \psi_i^\alpha(\tau_1) \psi_i^\beta (\tau_2) }\,,
\ee
or in terms of the disorder-averaged replica bilocal field theory we obtain 
\be
\mG_{\alpha\beta}(\tau_1, \tau_2; N) = \lim_{M \to 0} \int \prod_{\gamma,\delta=1}^M D G_{\gamma\delta} D\Sigma_{\gamma\delta}\ \e^{-N S[G_{\gamma\delta}, \Sigma_{\gamma\delta}]} G_{\alpha\beta}(\tau_1, \tau_2)\,. 
\ee
This formula gives representation for non-diagonal disorder correlator in terms of the
disorder-averaged replica bilocal field theory. We once again see that the quenched correlator is governed by the replica structure in the limit $M \to 0$. 

The replica factorization and vanishing of the fermionic one-point functions at large (but finite) $N$ seems to tell that there should be no replica-nondiagonal saddle points, but in fact this is not conclusive. It is worth noting that the factorization argument holds at finite $N$, and it does not exclude possible dominant non-diagonal saddle points in the thermodynamic limit $N \to \infty$. The factorization adds a nontrivial nonperturbative constraint on the full $1/N$ expansion of the right hand side of (\ref{quenched-offdiag}). In particular, at some orders of $1/N$ one expects replica-nondiagonal contributions in perturbations even over the replica-diagonal saddle \cite{Anninos16,Gu16,Kitaev17}. The factorization in the fermionic case implies that all such perturbative corrections should resum to zero, together with any nonperturbative corrections (that come e.g. from subleading saddles). Of course, with only finite amount of the asymptotic $1/N$ series taken into account, one does not expect zero. This will be more relevant to our results in the case of annealed correlators. 
%

\subsection{Annealed quantities}

The quantities, which are obtained by performing annealed averaging, treat the disorder on equal footing with other microscopic degrees of freedom of the model. In the general case, one can also consider annealed quantities that deal with $M$ copies of the initial model by construction. The simplest example is the replica partition function $\overline{Z(\beta)^M}$, which we consider throughout the paper, taken with an integer $M$. 
As we show in sections \ref{sec:q=2} and \ref{sec:Numerics}, there are nontrivial exact replica-nondiagonal saddle points, contributing to this quantity in case of a finite replica number. We also show that these saddle points are suppressed at large $N$, giving nonperturbative contributions to the $1/N$-expansion that are suppressed exponentially like $\e^{-N S}$. 

Another closely related example of an annealed quantity is the spectral form factor  $\frac{1}{\overline{Z(\beta)^2}}\overline{Z(\beta + i T) Z(\beta - iT)}$, which is essential in studies of quantum chaos and quantum gravity \cite{Cotler16,Garcia-garcia16,Saad18}. The study of replica-nondiagonal saddle points of this and related quantities is the topic of the future work. 

Below we consider the annealed correlators. The annealed correlation functions are computed by performing the average over the disorder in the normalization and in the nominator of (\ref{correlators}) separately. 

\subsubsection{Single-replica correlators}

If the fields all belong to the same one replica of the theory initially, the correlator is written
\be
\langle \psi_{i_1}(\tau_1)\dots \psi_{i_k} (\tau_k) \rangle_{\text{annealed}}= \frac{1}{\overline{Z_{\bj}(\beta; 0)}} \frac{1}{k!} \left.\frac{\delta}{\delta \eta_{i_1}(\tau_1)} \dots \frac{\delta}{\delta \eta_{i_k}(\tau_k)} \overline{Z_{\bj}(\beta; \eta)}\right|_{\eta = 0}\,. \label{annealed}
\ee
The integral over the disorder is a straightforward Gaussian integral in both nominator and denominator and can be taken right away. The single-replica annealed correlators do not have any replica structure. 

In terms of the bilocal fields $G$, $\Sigma$ the annealed two-point function is written as 
\be
\frac1N \sum_{i=1}^N \langle \psi_{i}(\tau_1)\psi_{i} (\tau_2) \rangle_{\text{annealed}}= \frac{\int D G D \Sigma\ \e^{-N S[G, \Sigma]} G(\tau_1, \tau_2)}{\int D G D \Sigma\ \e^{-N S[G, \Sigma]}}\,.
\ee

\subsubsection{Replica-offdiagonal correlators}

Finally, one can consider the replica-offdiagonal annealed correlators of the form
\be
\frac1N \sum_i \langle \psi_i^\alpha (\tau_1) \psi^\beta_i (\tau_2) \rangle_{\text{annealed}} = \frac{1}{2!N} \sum_{i=1}^N \frac{1}{\overline{Z_{\bj}(\beta; 0)^2}} \left.\frac{\delta}{\delta \eta^\alpha_{i}(\tau_1)} \frac{\delta}{\delta \eta^\beta_{i}(\tau_2)} \overline{Z_{\bj}(\beta; \eta^\alpha)  Z_{\bj}(\beta; \eta^\beta)}\right|_{\eta = 0}\,. \label{annealed-offdiagonal}
\ee

\paragraph{Factorization.} Note that this correlator does not factorize into product of one-point functions due to the fact that disorder averaging introduces interaction between replicas. However, the fermionic path integral still factorizes at finite $N$. Taking the derivatives in the right hand side of (\ref{annealed-offdiagonal}), one obtains under the averaging line the product of two copies of 
\be
\int D\psi\ \e^{-S_{\text{SYK}}[\bj, \psi]}\,.
\ee
This is zero at finite $N$ due to parity, and therefore also nullifies the correlator (\ref{annealed-offdiagonal}). 

\paragraph{Bilocal field representation and $1/N$-expansion.} In terms of the bilocal replica fields, the correlator (\ref{annealed-offdiagonal}) reads 
\bea
\frac1N \sum_{i=1}^N \langle \psi_i^\alpha (\tau_1) \psi^\beta_i (\tau_2) \rangle_{\text{annealed}} &=& \frac{1}{\overline{Z(\beta)^2}} \fG_{\alpha\beta} (\tau_1, \tau_2; N,2) \label{annealed-offdiag}\\ &=&  \frac{ \int \prod_{\gamma,\delta=1}^2 D G_{\gamma\delta} D\Sigma_{\gamma\delta}\ \e^{-N S[G_{\gamma\delta}, \Sigma_{\gamma\delta}]} G_{\alpha\beta}(\tau_1, \tau_2)}{\int \prod_{\gamma,\delta=1}^2 D G_{\gamma\delta} D\Sigma_{\gamma\delta}\ \e^{-N S[G_{\gamma\delta}, \Sigma_{\gamma\delta}]}}\,,\nn
\eea
where $M=2$. Note that there is a normalization factor, unlike the quenched case. This correlator requires two replicas, and can detect replica-nondiagonal saddle points, obtained in sections \ref{sec:q=2} and \ref{sec:Numerics}. We have shown that those saddle points are suppressed at large $N$, thus their effect can only appear at finite $N$.

However, at finite $N$ the factorization of the fermionic path integral dictates that (\ref{annealed-offdiagonal}) exactly should be zero. But here one can repeat the same arguments as in the disorder case in the end of sec.\ref{sec:RODD}. We can also expect that at finite $N$ the replica-nondiagonal saddle points may give a nonperturbatively small contribution to annealed $4$- and higher-point nondiagonal correlators.

\end{document}